\documentclass{article}
\usepackage[a4paper,margin=2cm]{geometry}
\usepackage{graphicx}
\usepackage{subcaption}
\usepackage{amsmath} 
\usepackage{algorithm}
\usepackage{algpseudocode}
\usepackage{amssymb}
\usepackage{chngcntr}
\usepackage{float}
\usepackage{hyperref}

\counterwithout{figure}{section}
\title{The Space-Clock Elevator}
\vspace{-1.5cm}

\title{Space-Clock Elevator: Multi-Stage Orbital Transport via Rotating Tethers and Elliptical Nodes}

\author{
Maksim Kazanskii\\
Independent Researcher\\
\texttt{mkazanskii@gmail.com}
}

\begin{document}

\maketitle

Rotating space tethers have long been investigated as 
momentum-exchange devices for orbital transportation, 
including multi-stage architectures based on coordinated 
mass exchange between orbital facilities. In this work, 
we develop a mathematical and computational framework 
for the synthesis of multi-stage rotating tether systems 
connected through intermediate Keplerian transfer 
trajectories, termed elliptical nodes. The framework combines distributed-mass tether dynamics, 
systematic mechanical feasibility mapping, inverse 
kinematic reconstruction of neighboring tether 
configurations, and near-commensurability constraints 
between rotational and orbital frequencies. Numerical 
analysis further reveals an approximately separable 
structure of the residual rotational dynamics, enabling 
a reduced-order surrogate representation of tether 
trajectories. The construction of multi-stage transportation networks 
is formulated as a mixed discrete-continuous search 
problem and addressed using stochastic candidate 
generation, beam search, and receding-horizon evaluation. 
The numerical results demonstrate the construction of 
candidate transportation chains extending from low Earth 
orbit towards geostationary orbit while satisfying 
prescribed mechanical and nominal synchronization 
constraints. The resulting framework, referred to as the Space-Clock 
Elevator, provides a systematic computational approach 
to the design and evaluation of coordinated orbital 
transportation networks based on rotating tethers.

\section{Introduction}

Orbital transportation involves both mechanical energy 
transfer and the satisfaction of trajectory, momentum, 
and rendezvous constraints. The mechanical energy 
difference between orbital states provides a fundamental 
measure of the energetic requirements of orbital 
transportation.
For a body of mass \( m \) on a circular orbit of radius \( r \), the total mechanical energy is
\begin{equation}
E(r) = -\frac{G M_{\oplus} m}{2r},
\end{equation}
where \( G \) is the universal gravitational constant and \( M_{\oplus} \) is the mass of Earth.
The specific orbital energy for a circular orbit increases monotonically with orbital radius, implying that, in the absence of non-conservative losses, the energy required for orbital transfer depends only on the initial and final orbital radii through the difference in specific orbital energy.

\begin{figure}[t]
\centering
\includegraphics[width=0.85\linewidth]{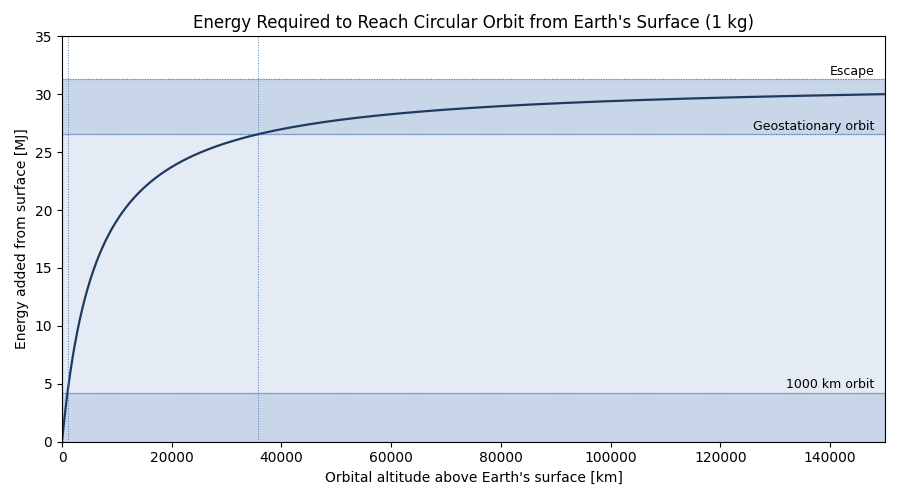}
\caption{Specific orbital energy as a function of altitude above Earth’s surface for circular orbits.}
\label{fig:orbital_energy}
\end{figure}

This energy scaling has important implications for 
orbital transportation. As illustrated in 
Fig.~\ref{fig:orbital_energy}, the specific mechanical 
energy of circular orbits increases with altitude, 
while its radial gradient decreases.

Neglecting Earth's rotation, the ideal mechanical 
energy required to transport a 1~kg payload from 
Earth's surface to a circular orbit at an altitude 
of 1000~km is approximately 35.5~MJ. Raising the 
same payload from this orbit to geostationary orbit 
requires an additional 22.3~MJ.

Although the initial injection into low Earth orbit 
accounts for the larger fraction of the ideal 
mechanical energy budget, transportation between 
low Earth orbit and higher orbital regimes still 
requires substantial energy expenditure.

This observation motivates reusable orbit-based 
transportation systems in which mechanical energy 
and momentum can be redistributed between ascending 
and descending masses. Such systems could potentially 
reduce the propellant requirements associated with 
repeated orbital transportation.

Throughout this work, an altitude of approximately 
\(1000\,\mathrm{km}\) is adopted as a representative 
lower orbital boundary. Atmospheric drag and other 
non-Keplerian perturbations are neglected within 
the idealized dynamical model.

\subsection{Space tether concepts.}
Space tether systems have long been investigated as enabling technologies for orbital manipulation, attitude stabilization, and momentum exchange \cite{CartmellMcKenzie2008}.
Classical studies of satellite attitude dynamics established
the importance of gravity-gradient torques in determining
rotational stability and libration behavior
\cite{Beletsky1966}.
These principles were subsequently developed within
the broader analytical framework of space tether dynamics
\cite{BeletskyLevin1993}.
Subsequent investigations examined the nonlinear dynamics
and control of tethered satellite systems, including
coupling between orbital, rotational, and internal
degrees of freedom
\cite{MisraModi1992,Misra2008,ModiLakshmananMisra1992,
Kalantzis1998}.
These investigations laid the theoretical groundwork for modern tether dynamics and remain central references in the field.

Beyond passive gravity-gradient stabilization, rotating tethers have been proposed as active momentum-
exchange devices capable of transferring payloads between orbits without onboard propulsion
\cite{Moravec1977,Hoyt2000}.
The historical origins of space elevator concepts can
be traced to early proposals for Earth-to-space
structures \cite{Tsiolkovsky1895}.
Later investigations developed tensile orbital tower
and space elevator architectures
\cite{Pearson1975,Edwards2000,vanPelt2009}.

While terrestrial space elevators remain beyond current
engineering capabilities, shorter orbital tethers have
been investigated as potential momentum-exchange devices
for in-space transportation
\cite{Moravec1977,CartmellMcKenzie2008,Bangham1998}.

\noindent
\subsection{From single tethers to coordinated systems.}
Previous studies have investigated momentum-exchange
tether systems for multi-stage orbital transportation,
including architectures designed to distribute the
required velocity increments across multiple rotating
tether facilities
\cite{Bangham1998,Hoyt2000,Landgraf2025}.
Particularly relevant to the present study are the 
investigations of Sidorov 
\cite{Sidorov2000,Sidorov2002,Sidorov2004}, 
which examined coordinated groups of orbital tethers 
and exchange masses for transportation and energy 
transformation without continuous propellant 
expenditure.

In particular, Sidorov~\cite{Sidorov2004} investigated 
multi-stage payload transportation between rotating 
tethers, configurations involving circular and 
elliptical orbits, near-zero-relative-velocity 
rendezvous, and bidirectional mass exchange for 
restoring the mechanical energy of the system. 
The transportation configurations presented in that 
work constitute direct precedents for the physical 
architecture considered here.

Related investigations by Ivanov et al.~
\cite{Ivanov2010} addressed the determination of tether 
parameters for zero-relative-velocity rendezvous 
in circular and elliptical orbital configurations.

The present study therefore does not claim novelty 
in the general concept of multi-stage orbital 
transportation using rotating tethers, elliptical 
transfer trajectories, or bidirectional mass exchange.

Instead, the objective is to develop a systematic 
computational framework for constructing candidate 
multi-stage transportation networks under simultaneous 
kinematic, synchronization, and mechanical constraints.

The methodological emphasis is placed on the inverse 
reconstruction of neighboring tether configurations 
from prescribed inertial transfer states, the 
identification of near-commensurate frequency 
relationships, the systematic mapping of mechanically 
admissible configurations, and their integration 
into a mixed discrete-continuous network synthesis 
procedure.

In this paper, we develop a computational synthesis 
framework for a class of modular orbital transportation 
architectures, hereafter referred to as the 
\emph{Space-Clock Elevator}.
The system consists of multiple rotating tethers deployed on circular orbits spanning from a lower orbit (e.g., LEO) to a higher orbit (e.g., near GEO).
Adjacent tethers interact via active transfer platforms, termed \emph{elliptical nodes}, which temporarily carry payloads along Keplerian elliptical trajectories between neighboring orbits.
By synchronizing tether rotation with the orbital motion of these elliptical nodes, payloads can be passed upward through the system while compensating mass moves downward.
Under idealized conservative dynamics, the mechanical 
energy required for outward payload transportation can 
be supplied through the inward transportation of 
compensating masses. Repeated operation requires 
appropriate balancing of energy and angular momentum 
exchange, together with compensation for non-ideal 
dynamical effects.
The term ``space-clock'' reflects the fact that the operation of the system relies on rational frequency relationships between tether rotation and orbital motion.
These near-commensurability conditions provide a means 
of identifying candidate recurrence intervals for 
payload exchange. Actual rendezvous additionally 
requires simultaneous satisfaction of inertial 
position and velocity matching conditions.

\subsection{Scope and contributions.}

The primary objective of this work is to develop a 
mathematical and computational framework for the 
synthesis of mechanically constrained multi-stage 
orbital transportation networks based on rotating 
tethers.

Building upon established momentum-exchange tether 
concepts and classical tether dynamics, the present 
study makes the following methodological contributions:

\begin{enumerate}

\item \textbf{Mechanical feasibility mapping.}
A distributed-mass dynamical model is employed to 
systematically characterize the admissible parameter 
space of rotating tethers in terms of orbital radius, 
tether length, and relative endpoint velocity, subject 
to maximum tension and tension-variation constraints.

\item \textbf{Reduced-order representation of rotational 
dynamics.}
An approximately parameter-independent normalized 
profile of the residual rotational trajectory is 
identified numerically within the investigated 
parameter domain. This observation enables the 
reconstruction of rotational trajectories using 
a surrogate model predicting the residual amplitude 
and rotation period.

\item \textbf{Inverse kinematic construction.}
An inverse formulation is developed for reconstructing 
candidate neighboring tether configurations from the 
known inertial position and velocity of an intermediate 
Keplerian transfer node. For a prescribed center-of-mass 
orbital radius, the construction reduces to a quadratic 
equation in the reciprocal tether rotation rate.

\item \textbf{Frequency-based synchronization constraints.}
Near-commensurate relationships between tether rotational 
frequencies and node orbital frequencies are incorporated 
into the candidate selection procedure, allowing nominal 
recurrence intervals to be evaluated under prescribed 
phase-error tolerances.

\item \textbf{Multi-stage network synthesis.}
The inverse construction, mechanical feasibility 
constraints, and frequency-based synchronization 
criteria are integrated into a mixed discrete-continuous 
search procedure employing stochastic candidate 
generation, beam search, and receding-horizon evaluation.

\end{enumerate}

The resulting numerical configurations illustrate 
the applicability of the framework to the construction 
of candidate transportation chains extending from 
low Earth orbit towards geostationary orbit.

The emphasis of the present study is on computational 
network synthesis rather than the introduction of 
a new physical mechanism for orbital transportation.

\section{Single Tether System}

\subsection{General details}

\begin{figure}[t]
  \centering
  \includegraphics[width=0.8\linewidth]{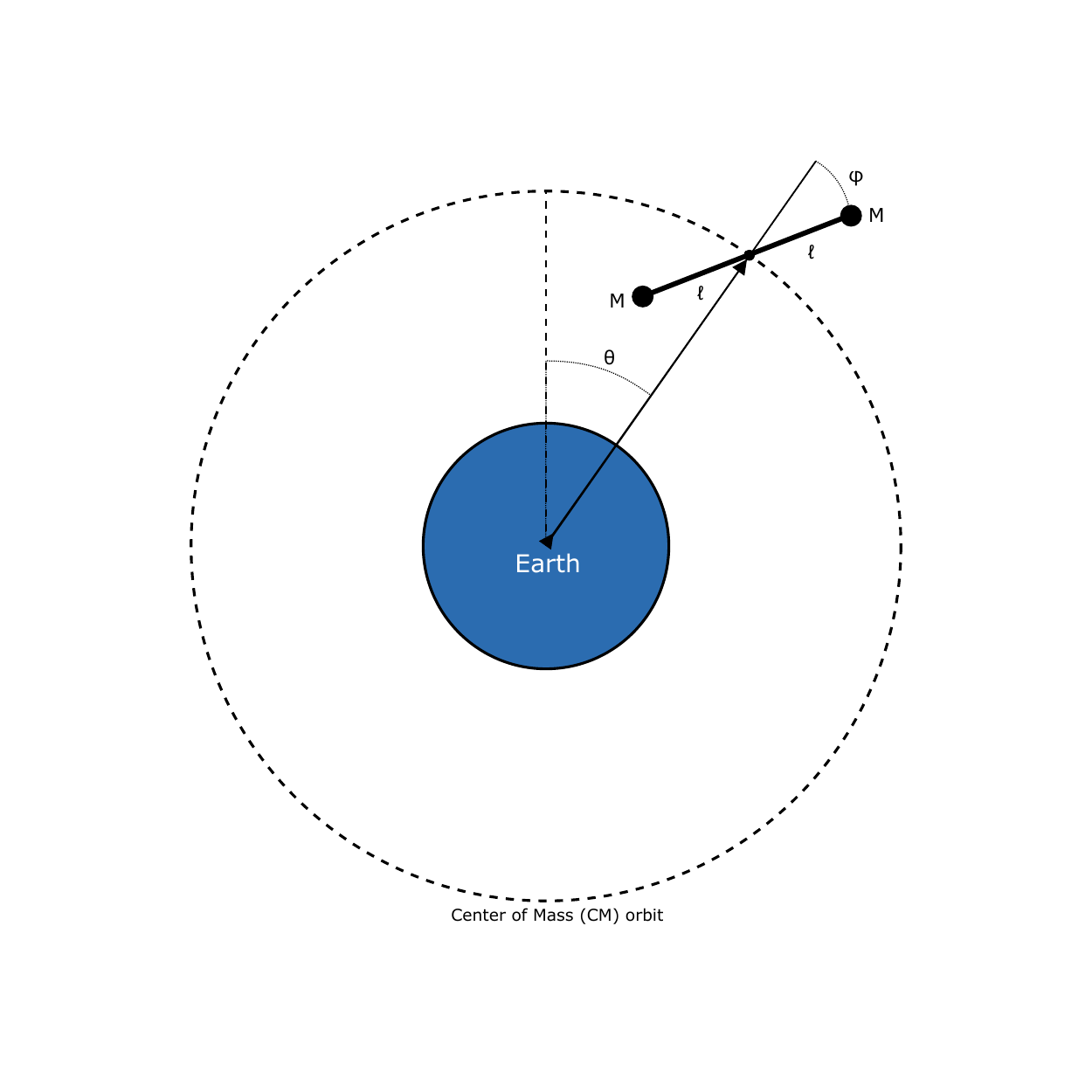}
  \caption{Tether geometry.}
  \label{fig:tether_geometry}
\end{figure}

For simplicity, the tether system is modeled as an inextensible rotating cable with two identical point masses \(M\) attached at its ends. The motion of the tether is assumed to be planar and confined to the orbital plane. Such a configuration represents a minimal yet physically meaningful model that captures the essential dynamics of a massive tether system while allowing for analytical tractability. Elastic deformation of the cable is neglected and the tether is therefore treated as inextensible. The implications of this assumption are discussed in Subsection~\ref{subsec:inext}. The mass of the cable is taken into account through a constant linear mass density \(\rho\). 
The center of mass of the tether is assumed to move along a circular orbit around the Earth. 
This assumption is consistent with a nominal operational regime in which orbital perturbations and radial oscillations are neglected. 
Accordingly, the velocity of the center of mass is always perpendicular to the Earth-centered radius vector. 
The magnitude of the velocity is given by 
\begin{equation}
v_{\mathrm{cm}} = R \dot{\theta},
\end{equation}
where \(R\) is the orbital radius of the center of mass and \(\theta\) denotes the orbital angle. A schematic illustration of the tether configuration is shown in Fig.~\ref{fig:tether_geometry}.

A material point of the tether is parameterized by the coordinate 
\(\sigma \in [-\ell,\ell]\), measured along the tether from its center of mass. 
The total length of the tether is therefore \(2\ell\).
The corresponding infinitesimal mass element of the cable is expressed as
\begin{equation}
\mathrm{d}m = \rho\,\mathrm{d}\sigma.
\end{equation}

\subsection{Single Tether Kinematics}

The kinematics of the tether are described in the orbital plane using a polar coordinate system centered at the Earth, following standard formulations in orbital and tether dynamics \cite{BeletskyLevin1993}.
Let \(\mathbf{e}_r\) and \(\mathbf{e}_\theta\) denote the radial and tangential unit vectors, respectively.
The position vector of the center of mass is given by
\begin{equation}
\mathbf{R}_{\mathrm{cm}} = R\,\mathbf{e}_r.
\end{equation}
The orientation of the tether is described by the angle \(\phi\), defined with respect to the local radial direction.
This angle fully characterizes the rotational state of the tether relative to the orbital frame, as commonly adopted in gravity-gradient and rotating tether models \cite{BeletskyLevin1993}.
The unit vector along the tether is therefore
\begin{equation}
\mathbf{e}_t = \cos\phi\,\mathbf{e}_r + \sin\phi\,\mathbf{e}_\theta.
\end{equation}

The position of a generic material point of the tether is then
\begin{equation}
\mathbf{r}(\sigma,t) = \mathbf{R}_{\mathrm{cm}} + \sigma\,\mathbf{e}_t,
\end{equation}
which is consistent with standard distributed-mass representations in analytical dynamics \cite{Greenwood1988}.

Differentiation with respect to time yields the velocity of a tether element,
\begin{equation}
\dot{\mathbf{r}}(\sigma,t)
= R\dot{\theta}\,\mathbf{e}_\theta
+ \sigma(\dot{\phi}+\dot{\theta})\,\mathbf{e}_n,
\end{equation}
where \(\mathbf{e}_n\) is the unit vector perpendicular to the tether in the orbital plane.
This expression highlights the combined contribution of orbital motion and tether rotation to the local velocity field, in agreement with classical kinematic treatments of rotating tether systems \cite{BeletskyLevin1993}.
\subsection{Kinetic energy}

The kinetic energy of the two endpoint masses is given by
\begin{equation}
T_{\mathrm{m}} = \frac{1}{2}M\left[\dot{\mathbf{r}}^2(\ell,t)
+ \dot{\mathbf{r}}^2(-\ell,t)\right].
\end{equation}

The kinetic energy of the cable is obtained by integrating the kinetic energy density over the tether length,
\begin{equation}
T_{\mathrm{c}} = \frac{1}{2}\int_{-\ell}^{\ell}
\rho\,\dot{\mathbf{r}}^2(\sigma,t)\,\mathrm{d}\sigma.
\end{equation}

After performing the integration, the total kinetic energy of the system can be written in compact form as
\begin{equation}
T = \frac{1}{2}J_\theta\,\dot{\theta}^2
+ \frac{1}{2}J_\phi\,(\dot{\phi}+\dot{\theta})^2,
\end{equation}
where the effective inertia coefficients are defined as
\begin{equation}
J_\theta = 2\bigl(M+\rho\ell\bigr)R^2,
\qquad
J_\phi = 2\left(M\ell^2 + \frac{\rho\ell^3}{3}\right).
\end{equation}

\subsection{Gravitational potential energy}

The gravitational potential energy of the two endpoint masses is
\begin{equation}
V_{\mathrm{m}} = -\mu M
\left(
\frac{1}{\lvert\mathbf{r}(\ell,t)\rvert}
+
\frac{1}{\lvert\mathbf{r}(-\ell,t)\rvert}
\right),
\end{equation}
where \(\mu\) is the Earth’s gravitational parameter.

The contribution of the cable is obtained by integrating the gravitational potential over its length,
\begin{equation}
V_{\mathrm{c}} = -\mu \int_{-\ell}^{\ell}
\frac{\rho\,\mathrm{d}\sigma}
{\lvert\mathbf{r}(\sigma,t)\rvert}.
\end{equation}

The total gravitational potential energy of the system is therefore
\begin{equation}
V(\phi) = V_{\mathrm{m}} + V_{\mathrm{c}}.
\end{equation}
\

\subsection{Lagrangian formulation}

The dynamics of the tether system are derived using the Lagrangian formalism, which provides a natural framework for systems with coupled rotational and translational degrees of freedom.
The Lagrangian is defined as
\begin{equation}
\mathcal{L} = T - V
= \frac{1}{2}J_\theta\,\dot{\theta}^2
+ \frac{1}{2}J_\phi\,(\dot{\phi}+\dot{\theta})^2
- V(\phi).
\end{equation}

The equations of motion for the generalized coordinates \(\theta\) and \(\phi\) are then obtained from the corresponding Euler-Lagrange equations.

\subsection{Equations of motion}

The equations of motion are obtained by applying the Euler-Lagrange equations
\begin{equation}
\frac{\mathrm{d}}{\mathrm{d}t}
\left(
\frac{\partial \mathcal{L}}{\partial \dot{q}_i}
\right)
-
\frac{\partial \mathcal{L}}{\partial q_i}
= 0,
\end{equation}
where \(q_i \in \{\theta,\phi\}\) are the generalized coordinates.

\vspace{0.5em}
\noindent
\textit{Orbital angle \(\theta\).}  
Since the Lagrangian does not explicitly depend on \(\theta\), this coordinate is cyclic and the corresponding generalized momentum is conserved. The conjugate momentum associated with \(\theta\) is given by
\begin{equation}
p_\theta
=
\frac{\partial \mathcal{L}}{\partial \dot{\theta}}
=
J_\theta \dot{\theta}
+
J_\phi (\dot{\phi}+\dot{\theta}).
\end{equation}

Conservation of \(p_\theta\) implies
\begin{equation}
\frac{\mathrm{d}p_\theta}{\mathrm{d}t} = 0,
\end{equation}
which yields the first equation of motion,
\begin{equation}
(J_\theta+J_\phi)\,\ddot{\theta}
+
J_\phi\,\ddot{\phi}
=
0.
\label{eq:theta_eom}
\end{equation}

\vspace{0.5em}
\noindent
\textit{Tether orientation angle \(\phi\).}  
The Euler-Lagrange equation associated with \(\phi\) takes the form
\begin{equation}
\frac{\mathrm{d}}{\mathrm{d}t}
\left[
J_\phi(\dot{\phi}+\dot{\theta})
\right]
+
\frac{\partial V(\phi)}{\partial \phi}
=
0,
\end{equation}
which leads to the second equation of motion,
\begin{equation}
J_\phi(\ddot{\phi}+\ddot{\theta})
+
\frac{\partial V(\phi)}{\partial \phi}
=
0.
\label{eq:phi_eom}
\end{equation}

Equations~\eqref{eq:theta_eom} and~\eqref{eq:phi_eom} form a coupled system describing the rotational dynamics of the tethered system. 
Since the orbital angle \(\theta\) is a cyclic coordinate, the associated generalized momentum \(p_\theta\) is conserved. 
This conservation law provides a first integral of motion, allowing \(\dot{\theta}\) to be expressed as a function of \(\dot{\phi}\).
Substitution of this relation into Eq.~\eqref{eq:phi_eom} reduces the system to a single second-order differential equation governing the evolution of the tether orientation angle \(\phi(t)\).

\subsection{Modelling}
\label{sec:feasibility_map}

The equations of motion were integrated numerically using a fixed-step fourth-order Runge-Kutta scheme, which is a standard for the time integration of coupled nonlinear ordinary differential equations arising in mechanical systems \cite{Press2007}. The primary objective of the numerical simulations was to construct feasibility maps based on the maximum tether tension and the post-transient tension variation over a wide range of orbital and rotational parameters. A detailed discussion of the tether tension characteristics is provided in Section \ref{subsec:tether-tension}.
The angular velocity of the tether relative to the orbital frame, the tether radius, and the center-of-mass orbital radius were systematically varied in order to assess their combined influence on the maximum tension levels and tension variability.

The derivative of the gravitational potential with respect to the tether orientation angle was evaluated numerically at each integration step by explicitly accounting for the distributed mass of the cable.
The continuous mass distribution was discretized uniformly along the tether using $N_\sigma = 400$ points, and all spatial integrals were computed using trapezoidal quadrature. The influence of the number of discretization points $N_\sigma$ on the solution accuracy is also investigated.

Since the orbital angle $\theta$ is a cyclic coordinate, the associated generalized momentum was conserved and used to reconstruct the orbital angular velocity $\dot{\theta}(t)$ algebraically at each time step.
The orbital angle was then obtained by direct time integration.
The conserved generalized momentum, associated with
the cyclic coordinate $\theta$ in the Lagrangian
formulation \cite{Greenwood1988}, was used as a
consistency check throughout the numerical simulations. The internal tension was evaluated at the tether center ($\sigma=0$), which corresponds to the maximum tensile load for symmetric configurations. All tension-related integrals were evaluated using the same spatial discretization as in the torque computation to ensure numerical consistency. Each simulation was propagated for two orbital periods using a fixed integration time step $\Delta t = 0.1~\mathrm{s}$.
To exclude transient effects associated with the initial conditions, only data obtained after half an orbital period were retained for post-processing.

The orbital radius $R$, tether half-length $\ell$ (or tether radius), and initial relative tangential velocity $v_{\mathrm{rel}}$ were systematically varied to assess their combined influence on the maximum tension and its post-transient variation. The orbital radius was varied from $7200$ to $42000~\mathrm{km}$ in steps of $1000~\mathrm{km}$, the tether half-length from $20$ to $2300~\mathrm{km}$ in steps of $40~\mathrm{km}$, and the initial relative tangential velocity from $100$ to $2500~\mathrm{m\,s^{-1}}$ in steps of $100~\mathrm{m\,s^{-1}}$. In addition to the grid-based parameter sweep, $1500$ additional simulations were performed using randomly sampled parameter combinations within the same ranges. These simulations were used for off-grid evaluation in order to assess the robustness of the numerical results with respect to parameter values not located on the predefined grid.
\begin{figure*}[t]
    \centering
    \includegraphics[width=\textwidth]{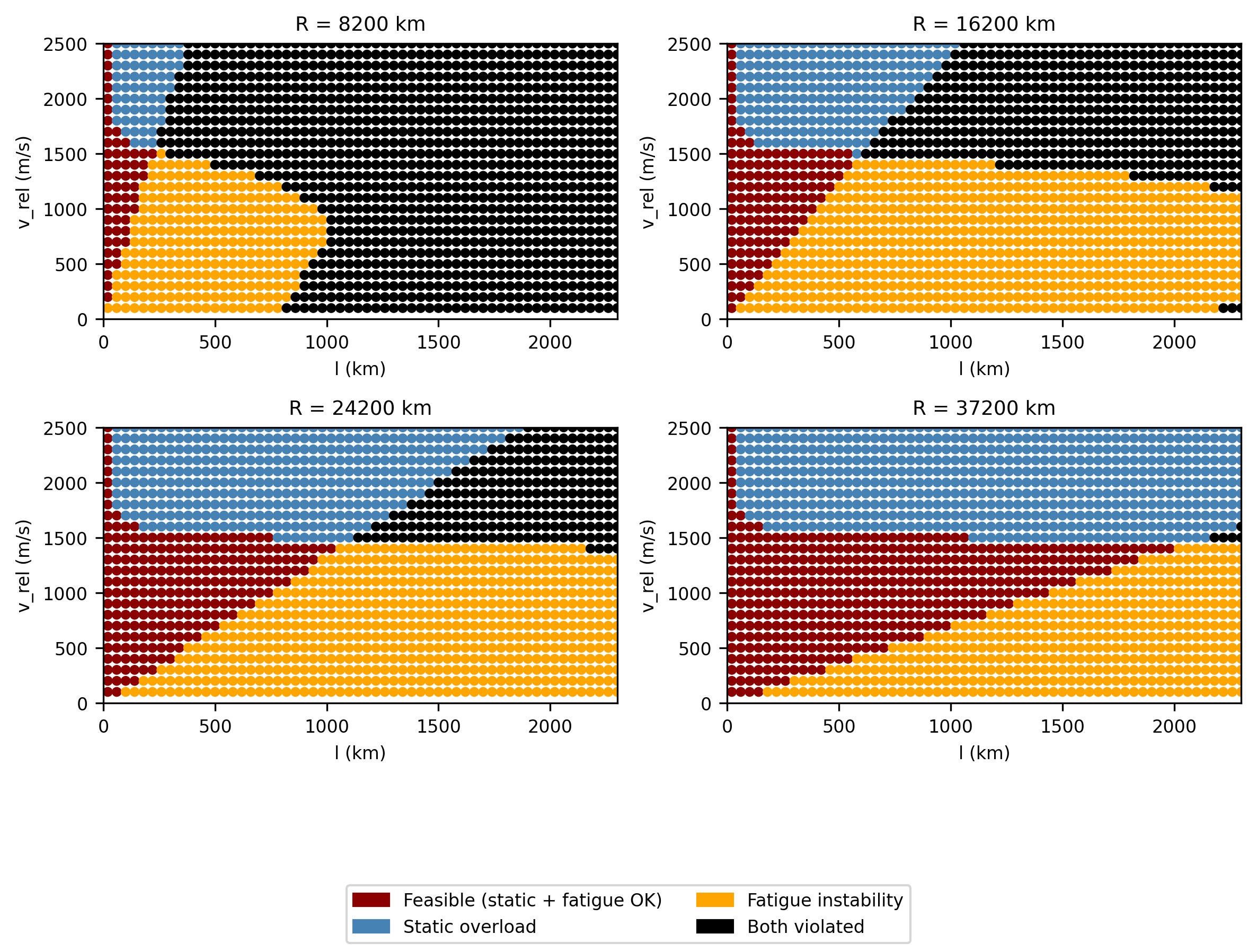}
    \caption{Feasibility maps for the tether system in the $(\ell, v_{\mathrm{rel}})$ parameter space at representative orbital radii.
    Each panel corresponds to a fixed orbital radius and shows the classification of dynamic responses according to static strength and fatigue criteria.
    Red regions indicate configurations satisfying both static and fatigue limits, blue regions correspond to static overload, orange regions indicate fatigue instability, and black regions denote cases where both constraints are violated.}
    \label{fig:feasibility_panels}
\end{figure*} 
For each parameter combination, the system was initialized with a small libration angle $\phi_0 = 0.1~\mathrm{rad}$ and an orbital angular velocity corresponding to a circular orbit,
$\dot{\theta}_0 = \sqrt{\mu/R^3}$.
The initial tether angular velocity $\dot{\phi}_0$ was selected to satisfy the prescribed relative velocity between the tether tip and the orbital motion.
For each simulation, the maximum center tension $T_{\max}$ and the tension variation for a full spin of the tether $\Delta T$ were extracted and recorded.
The resulting dynamic response was classified according to static strength and fatigue criteria based on the allowable Zylon cable stress (please see Section~\ref{subsec:tether-tension} for more details). The outcomes of the feasibility analysis are summarized in Fig.~\ref{fig:feasibility_panels}, which presents two-dimensional feasibility maps in the $(\ell, v_{\mathrm{rel}})$ plane for representative orbital radii. Each panel corresponds to a fixed orbital radius and indicates regions that satisfy both static and fatigue constraints, violate either constraint individually, or violate both simultaneously. The resulting dynamic response was classified according to static strength and fatigue criteria based on the allowable Zylon cable stress.

In the figure legend, feasibility is determined by two conditions: the maximum center tension remaining below the allowable limit, $T_{\max} < T_{\mathrm{allow}}$, and the normalized post-transient tension variation satisfying $\Delta T / T_{\max} < \Delta T_{\mathrm{crit}}$. Cases meeting both conditions are classified as feasible, while violations of either or both criteria are indicated by distinct colors. The physical motivation and detailed justification of these thresholds are discussed in Section~\ref{subsec:tether-tension}.

\subsection{Numerical Convergence Analysis}
A fixed integration time step of $\Delta t = 0.1~\mathrm{s}$ was used for all feasibility sweeps.
To verify that this choice does not influence the qualitative feasibility regions, a temporal convergence study was performed using ten random parameter combinations $(R,\ell,v_{\mathrm{rel}})$.
For each case, the endpoint position and tether tension over one orbital revolution were compared against a reference solution obtained with a smaller time step $\Delta t_{\mathrm{ref}} = 10^{-3}~\mathrm{s}$.
\begin{figure}[!t]

\centering

\noindent
\begin{minipage}[t]{0.48\textwidth}
    \centering
    \includegraphics[width=\linewidth]{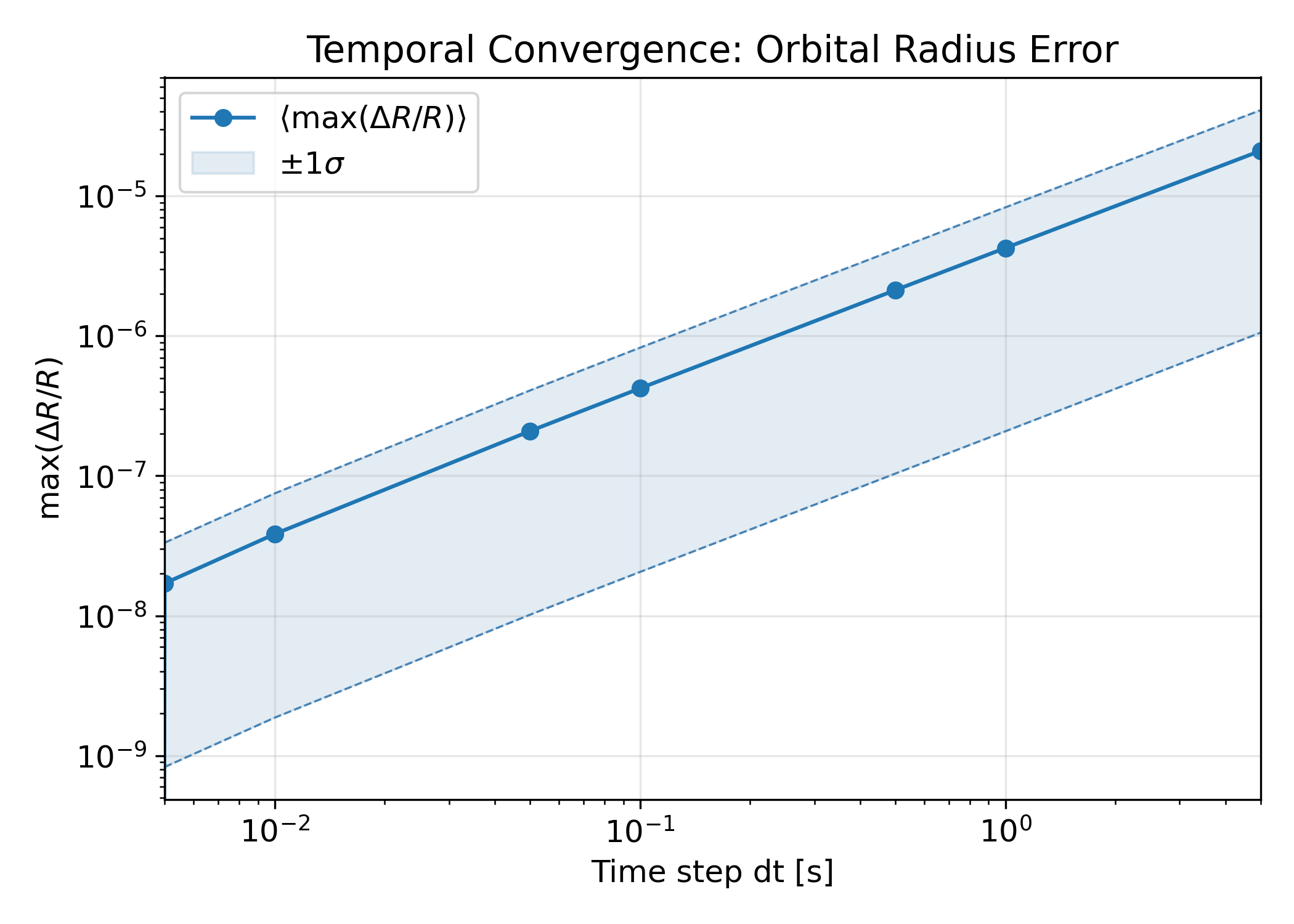}
\end{minipage}
\hfill
\begin{minipage}[t]{0.48\textwidth}
    \centering
    \includegraphics[width=\linewidth]{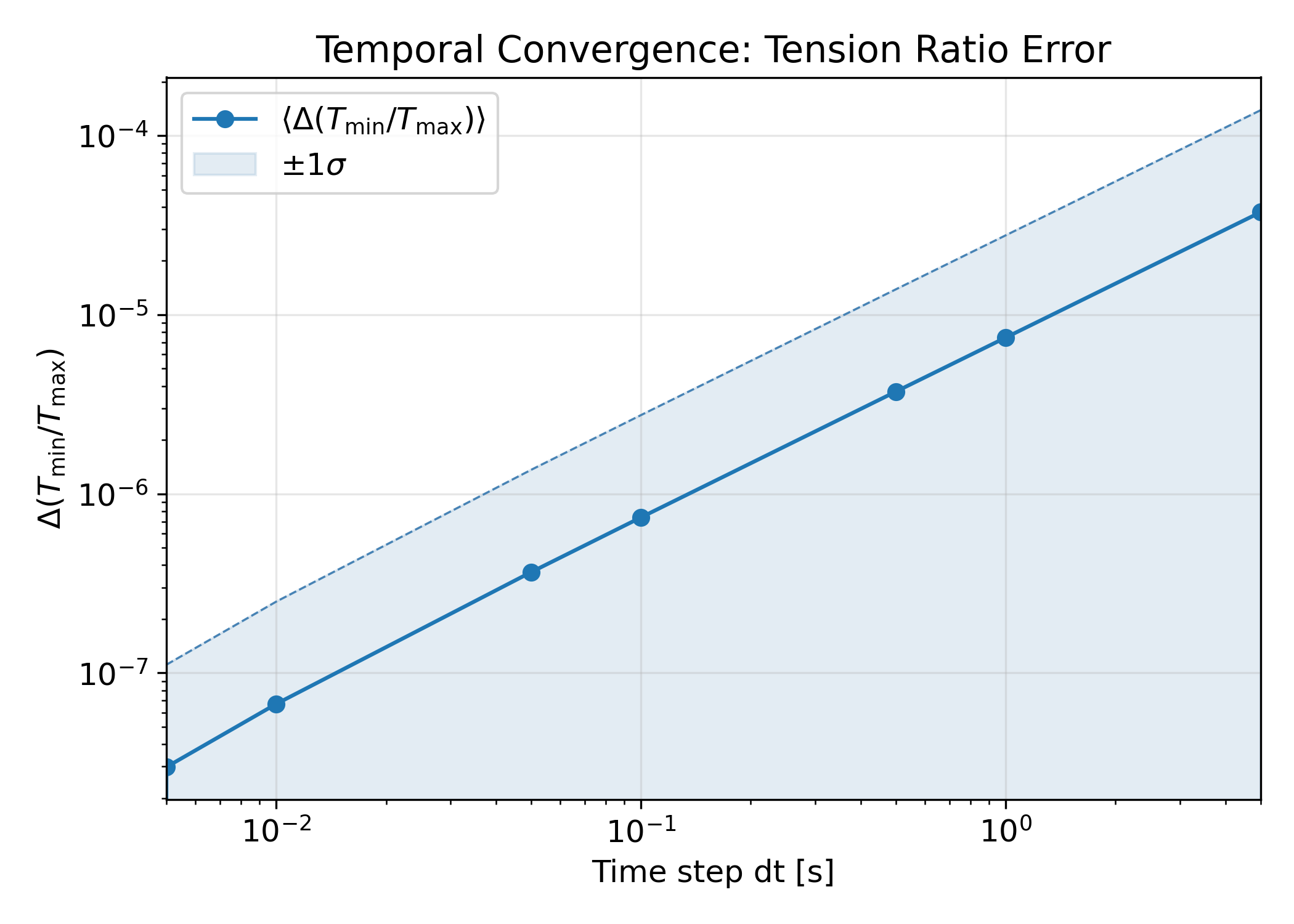}
\end{minipage}

\vspace{0.5em}

\noindent
\begin{minipage}[t]{0.48\textwidth}
    \centering
    \includegraphics[width=\linewidth]{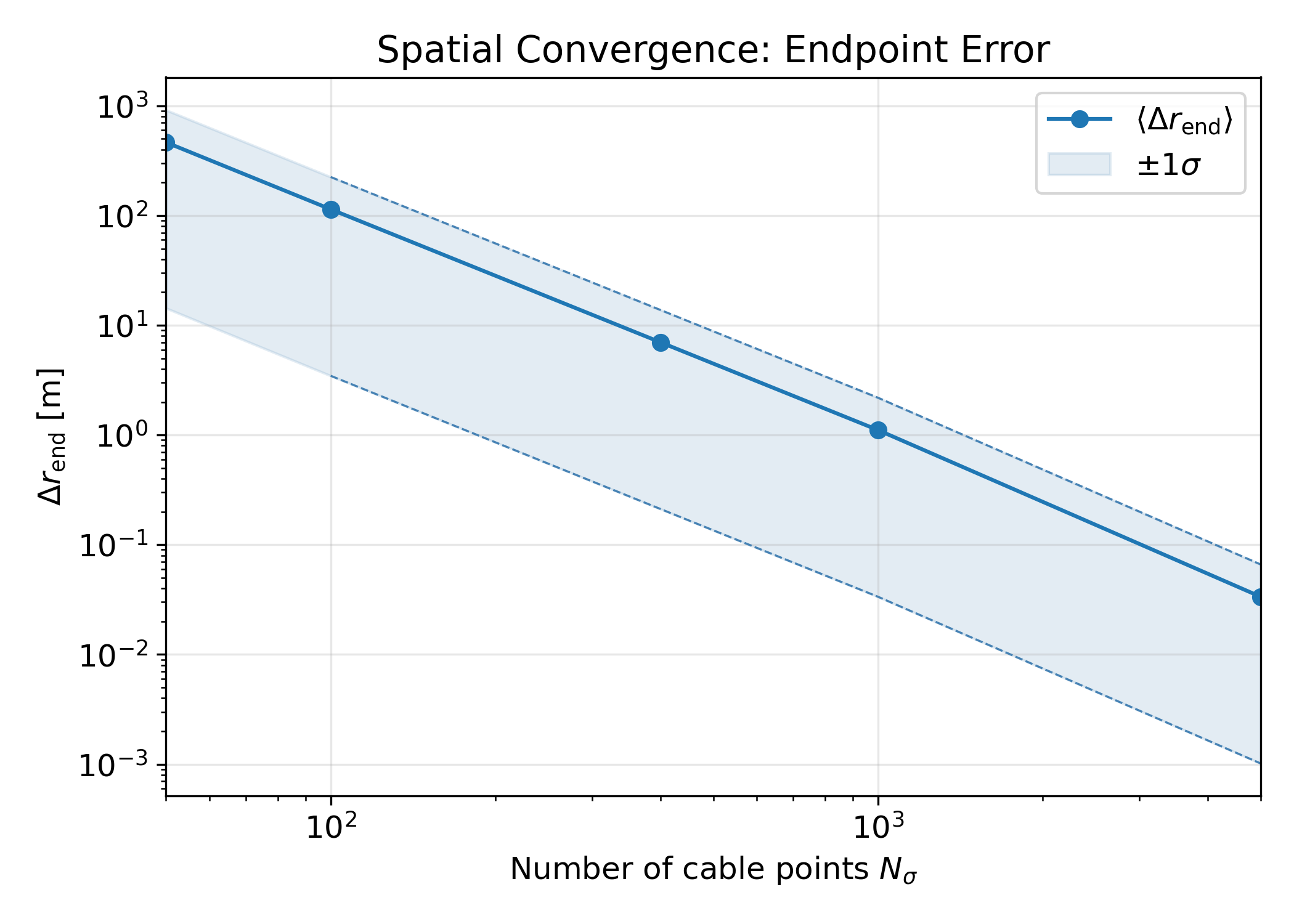}
\end{minipage}
\hfill
\begin{minipage}[t]{0.48\textwidth}
    \centering
    \includegraphics[width=\linewidth]{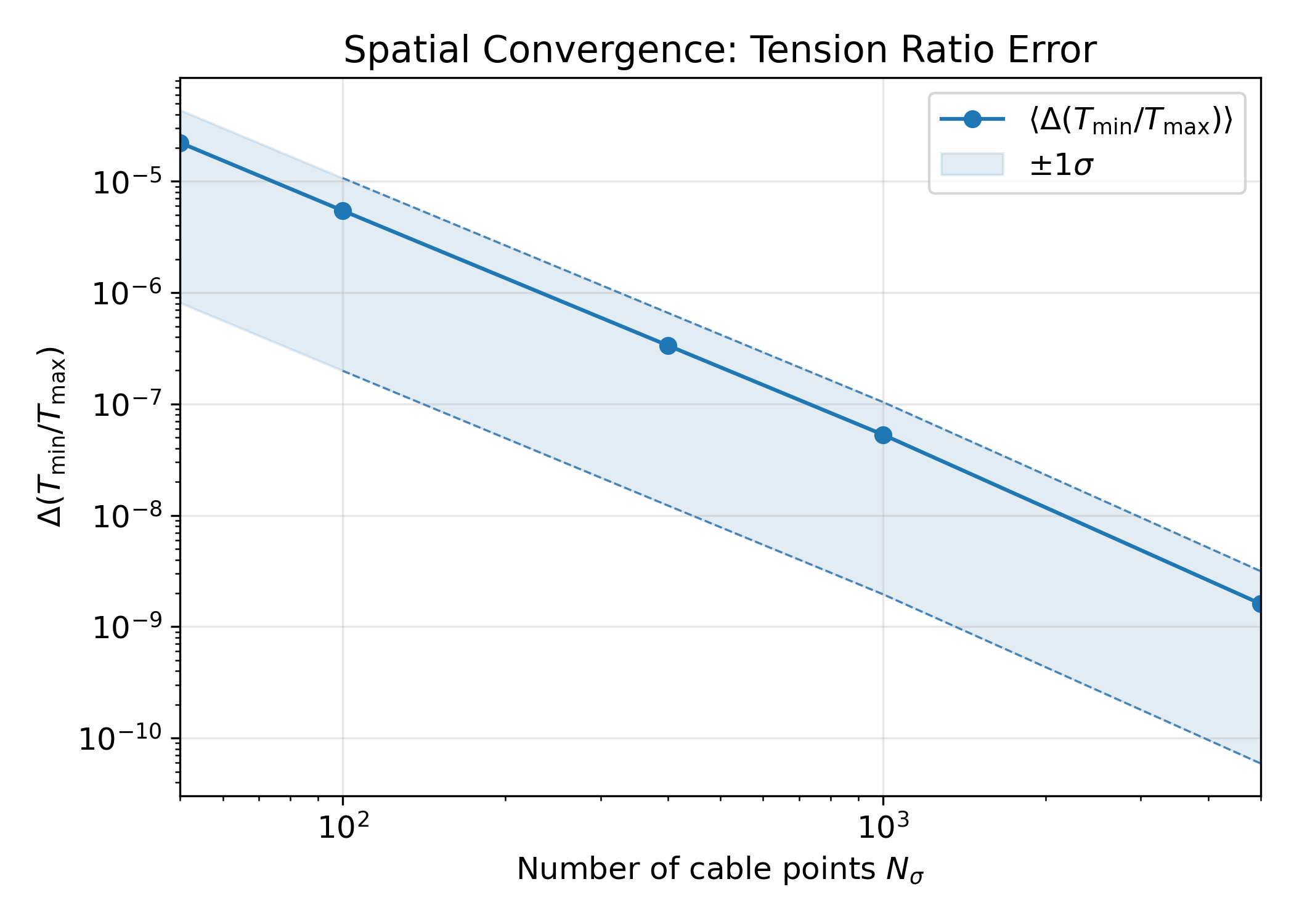}
\end{minipage}

\caption{
Numerical convergence diagnostics for the tether-orbit dynamics.
Top left: mean endpoint radial error $\langle \max(\Delta r_{\mathrm{end}}/R) \rangle$ as a function of the integration time step $\Delta t$.
Top right: mean error in the tension ratio $\langle \Delta(T_{\min}/T_{\max}) \rangle$ as a function of $\Delta t$.
Bottom left: mean endpoint position error $\langle \Delta r_{\mathrm{end}} \rangle$ as a function of the number of cable integration points $N_\sigma$.
Bottom right: mean error in the tension ratio $\langle \Delta(T_{\min}/T_{\max}) \rangle$ as a function of $N_\sigma$.
Solid curves denote averages over ten random orbital and tether parameter combinations, while shaded regions indicate $\pm1\sigma$ variability across cases.}
\label{fig:num_conv}
\end{figure} 
To enable meaningful comparison across different orbital configurations, endpoint position discrepancies were reported in normalized form as $\max(\Delta r_{\mathrm{end}}/R)$, while tension discrepancies were quantified using the ratio $\Delta(T_{\min}/T_{\max})$.
For the selected time step $\Delta t = 0.1~\mathrm{s}$, the mean normalized endpoint position error was
\[
\langle \max(\Delta r_{\mathrm{end}}/R) \rangle \approx 1.5\times10^{-7}.
\]
For the largest orbital radius considered ($R \approx 4.2\times10^{7}~\mathrm{m}$), this corresponds to a conservative endpoint displacement of order
\[
\Delta r_{\mathrm{end}} \sim \mathcal{O}(10~\mathrm{m})
\]
per orbit.

The numerical convergence diagnostics for the endpoint position and tether tension as functions of the integration time step are shown in the top row of Fig.~\ref{fig:num_conv}.
For $\Delta t = 0.1~\mathrm{s}$, the resulting numerical errors are several orders of magnitude smaller than the characteristic system length scales, namely the orbital radius and the tether length, and therefore do not affect the feasibility classification.
Moreover, the corresponding errors in the tension ratio $\Delta(T_{\min}/T_{\max})$ are found to be negligibly small compared to the tolerance levels implicit in the feasibility conditions discussed above, indicating that the numerical time discretization does not alter the assessment of tether tension limits.
Consequently, the numerical integration error associated with $\Delta t = 0.1~\mathrm{s}$ does not influence the synchronization feasibility assessments presented in the following sections or the resulting feasibility maps.

The spatial convergence curves shown in the bottom row of Fig.~\ref{fig:num_conv} appear strictly linear in logarithmic scale because the distributed cable gravity term is evaluated using a fixed-order trapezoidal quadrature applied to a smooth integrand. For such integrals, the discretization error scales asymptotically as $O(N_{\sigma}^{-2})$, independent of the specific orbital configuration, with case-to-case variability affecting only the prefactor of the error. In the present study, each coarse discretization is compared against a high-resolution reference solution computed with the same time integration scheme, causing temporal discretization errors to cancel in the difference and isolating the spatial quadrature error. As a result, all tested cases exhibit the same second-order power-law dependence on $N_{\sigma}$, and ensemble averaging over multiple parameter combinations preserves this exponent while reducing noise, yielding nearly perfect straight lines in log–log representations. Based on these results, a discretization of $N_{\sigma}=400$ cable integration points is sufficient to achieve the required numerical precision for all subsequent simulations.

\subsection{Trajectory}
In addition to the global dynamical quantities described above, the detailed rotational trajectory of the tether was reconstructed for each simulated configuration. 
In particular, the evolution of the tether orientation angle $\phi(t)$ was analyzed in order to characterize the rotational dynamics of the system over a complete spin cycle. 
Here $\phi(t)$ denotes the instantaneous orientation angle of the tether with respect to the local radial direction $\mathbf{e}_r$ in the orbital frame.

The instantaneous tether orientation angle $\phi(t)$ was first unwrapped to remove $2\pi$ discontinuities, yielding a continuous angular signal $\phi_{\mathrm{cont}}(t)$. 
To exclude transient behavior associated with the initial conditions, only rotations occurring after half of an orbital period were considered. 
A complete tether revolution was identified when the continuous angle increased by $2\pi$, and the final two such crossings were used to isolate one steady-state rotational cycle. 
The angle trajectory over this cycle was then uniformly resampled at $360$ points using linear interpolation, producing a normalized angular signal $\phi_{360}$ representing the evolution of the tether orientation over one full rotation. 
The corresponding rotation period was also recorded.

The residual rotational dynamics of the tether were also analyzed by removing the dominant linear component of the angle evolution over a single rotation cycle. 
After extracting the mean angular trend corresponding to the average rotation rate, the remaining signal represents periodic deviations from uniform rotation. 
These residuals provide a convenient way to visualize librational oscillations and other nonlinear effects in the tether motion. 

Despite large variations in orbital radius, tether length, and relative velocity across the parameter sweep, the residual dynamics exhibit a strongly periodic structure, indicating that the rotational motion remains close to uniform within the feasible region introduced above. 
This behavior is consistent with the tension feasibility criteria imposed in the simulations: within the admissible parameter domain, the variation of the tether tension remains limited, preventing large dynamical perturbations of the rotational motion. 
As a result, the tether rotation deviates only slightly from uniform rotation, leading to a stable periodic residual signal.

We observed that, after subtraction of the linear component corresponding to the mean angular velocity, the residual rotational signal exhibits a nearly invariant functional profile across the explored parameter space. In particular, the shape of the residual function remains essentially unchanged, while only its amplitude varies with the system parameters. Within the investigated mechanically admissible 
parameter domain, this property allows the residual 
dynamics to be approximated by a common normalized 
profile scaled by a parameter-dependent amplitude. 
The resulting approximate separability provides a 
reduced-order representation of the nonlinear 
rotational dynamics.

We define the phase-wrapped tether orientation over one rotation cycle and separate the dominant uniform-rotation component from the remaining periodic modulation. 
Let $\phi(t)$ denote the unwrapped tether angle, and consider a single steady-state rotation interval $t\in[t_0,t_0+T_{\mathrm{rot}}]$, where $T_{\mathrm{rot}}$ is the measured rotation period. 
Introduce the normalized phase
\begin{equation}
\tau \;=\; \frac{t-t_0}{T_{\mathrm{rot}}}\in[0,1].
\end{equation}
The linear component associated with the mean angular velocity is
\begin{equation}
\phi_{\mathrm{lin}}(t)
\;=\;
\phi(t_0) + \omega_{\mathrm{rot}}(t-t_0),
\qquad
\omega_{\mathrm{rot}}=\frac{2\pi}{T_{\mathrm{rot}}}.
\end{equation}
The residual angle is then defined as
\begin{equation}
\phi_{\mathrm{res}}(t)
\;=\;
\phi(t)-\phi_{\mathrm{lin}}(t).
\label{eq:phi_res_def}
\end{equation}
Numerically, we observe that $\phi_{\mathrm{res}}(t)$ admits an approximately separable representation of the form
\begin{equation}
\phi_{\mathrm{res}}(t;R,\ell,v_{\mathrm{rel}})
\;\approx\;
A(R,\ell,v_{\mathrm{rel}})\, f(\tau),
\label{eq:phi_res_factorization}
\end{equation}
where $A(R,\ell,v_{\mathrm{rel}})$ is a scalar amplitude 
depending on the system parameters and $f(\tau)$ is 
an approximately parameter-independent normalized 
profile over one rotation within the investigated 
mechanically admissible parameter domain.

A convenient normalization is obtained by defining the amplitude
\begin{equation}
A \;=\; \max_{\tau\in[0,1]} \bigl|\phi_{\mathrm{res}}(\tau)\bigr|,
\end{equation}
and the corresponding normalized residual profile
\begin{equation}
f(\tau)
\;=\;
\frac{\phi_{\mathrm{res}}(t_0+\tau T_{\mathrm{rot}})}{A},
\qquad
\max_{\tau\in[0,1]}|f(\tau)|=1.
\label{eq:f_def}
\end{equation}
With this decomposition, the full angular evolution over one cycle can be reconstructed as
\begin{equation}
\phi(t)
\;\approx\;
\phi(t_0)+\omega_{\mathrm{rot}}(t-t_0)
+
A(R,\ell,v_{\mathrm{rel}})\, f\!\left(\frac{t-t_0}{T_{\mathrm{rot}}(R,\ell,v_{\mathrm{rel}})}\right),
\qquad t\in[t_0,t_0+T_{\mathrm{rot}}].
\label{eq:phi_reconstruct}
\end{equation}

Based on this observation, the full rotational trajectory $\phi(t)$ can be
reconstructed from two quantities: the residual amplitude $A$ and the
rotation period $T_{\mathrm{rot}}$. To obtain these quantities for arbitrary
$(R,\ell,v_{\mathrm{rel}})$, a regression model was trained using the
simulation data generated in the parameter sweep. Such surrogate approximations are commonly used to replace computationally
expensive simulations with fast analytical predictors \cite{Forrester2008}.

To enable rapid evaluation of the rotational dynamics across the feasible parameter space, a surrogate regression model was constructed to approximate the mapping
\[
(R,\ell,v_{\mathrm{rel}}) \rightarrow (A,T_{\mathrm{rot}}),
\]
where \(A\) is the residual oscillation amplitude and \(T_{\mathrm{rot}}\) is the rotation period extracted from the numerical simulations. The model is trained on the simulation dataset generated in the parameter sweep and uses dimensionless input variables based on orbital scaling. This surrogate allows the rotational characteristics of the tether to be estimated without performing a full dynamical integration. The detailed formulation of the regression model, including feature normalization, local polynomial regression, and the reconstruction of the angular trajectory \(\phi(t)\), is provided in Appendix~\ref{sec:regression_model}.

\section{Multi-Tether System}

\subsection{General details} Following established concepts of orbital transportation 
through coordinated mass exchange 
\cite{Sidorov2002,Sidorov2004}, we consider a system 
composed of multiple rotating tethers operating at 
different orbital radii.

The fundamental operational principle is the 
redistribution of mechanical energy and angular 
momentum through controlled mass exchange between 
neighboring orbital facilities. Under idealized 
conservative dynamics, outward payload transportation 
can be supported by the inward transportation of 
compensating masses.

This physical mechanism has important precedents 
in earlier orbital tether transportation concepts. 
The emphasis of the present work is instead on 
the computational construction of candidate 
multi-stage configurations satisfying prescribed 
kinematic, mechanical, and synchronization constraints.

Repeated operation requires appropriate balancing 
of mechanical energy and angular momentum, together 
with compensation for non-ideal dynamical effects.

\begin{figure}[t]
  \centering
  \includegraphics[width=\linewidth]{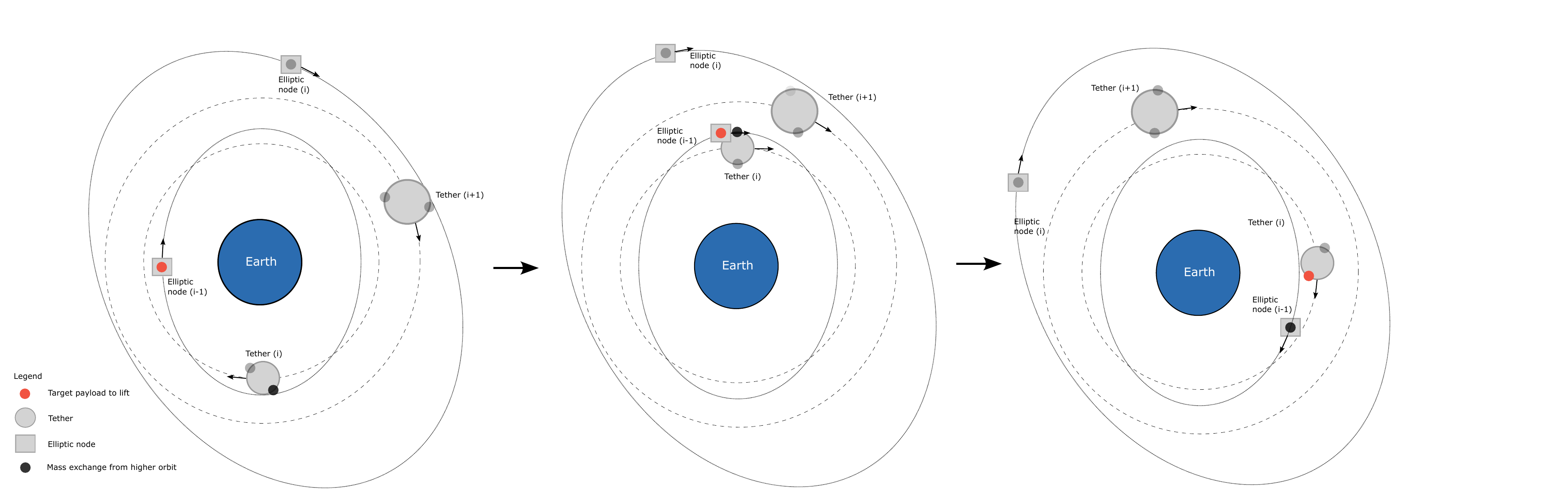}
  \caption{Multitether scheme.}
  \label{multi_tether}
\end{figure}
\subsection{Elliptical nodes}
The use of intermediate exchange masses following 
orbital trajectories between neighboring tether 
facilities has important precedents in the work 
of Sidorov~\cite{Sidorov2004}. Within the present computational synthesis framework, 
we consider a multi-tether configuration in which 
neighboring tethers are connected through intermediate 
transfer platforms, hereafter referred to as 
\emph{elliptical nodes}. A schematic illustration of the multi-tether system is shown in Fig.~\ref{multi_tether}. The elliptical nodes act as active transfer platforms that mediate controlled mass exchange between adjacent tethers. During each exchange episode, the payload mass carried by the endpoint of a tether and the mass associated with the corresponding elliptical node are exchanged at the same spatial location and share an identical instantaneous velocity vector. As a result, the detachment and subsequent attachment processes occur without discontinuities in either position or velocity (within predetermined tolerance). This kinematic continuity ensures conservation of linear momentum at the exchange point and avoids impulsive forces, allowing the transfer process to be modeled, in a first approximation, as quasi lossless exchange. For further discussion about the lossless property of the system please refer to Section~\ref{subsec:lossless}.  Within the proposed computational framework,
the elliptical node provides an intermediate
Keplerian transfer trajectory connecting candidate
release and capture configurations.

Exact position and velocity matching constitutes
an idealized rendezvous condition rather than an
intrinsic property of the elliptical node. Perfect alignment in position and velocity is not fully guaranteed due to perturbations and finite tolerances. However, small mismatches could be compensated by propulsion units located on the elliptical nodes, where the dynamics are simpler and active correction is feasible. In this chapter, active correction is not taken into account, and the analysis is limited to the nominal dynamical trajectories. For more details on the active correction please refer to Section~\ref{subsec:active-correction}.

Following detachment from the tether endpoint, the elliptical node carries the payload and continues its motion along the original trajectory, with the velocity imparted at the moment of exchange. Since this velocity generally differs from the local circular orbital velocity at the release radius, the subsequent motion of the elliptical node follows a general Keplerian trajectory which may be elliptical or parabolic, depending on the velocity magnitude. In particular, the trajectory is elliptical if the node velocity remains below the local escape velocity and parabolic if it reaches the escape threshold. In the present analysis, we restrict attention to elliptical trajectories. Accordingly, the velocity at the moment of detachment is required to remain below the local escape velocity corresponding to the release radius. This condition guarantees that the subsequent motion of the elliptical node is bounded and that the payload follows a closed Keplerian orbit, enabling repeatable capture by an adjacent tether.

\subsection{Velocity conditions}
To put the construction analytically, each tether \(i\) is modeled as an inextensible rotating cable with two terminal platforms located at its ends. The inextensibility limitations are explained in Section~\ref{subsec:inext}. These platforms are capable of attaching and detaching payloads at prescribed phases of the tether rotation. The center of mass of tether \(i\) is assumed to move on a circular orbit of radius
\begin{equation}
R_{\mathrm{cm},i},
\end{equation}
which is chosen such that the minimum radial distance reached by the tether endpoints satisfies a prescribed safety constraint. 

Let tether \(i\) rotate with angular velocity \(\omega_i\) relative to the local orbital frame. 
The motion of the tether is confined to the orbital plane, and all vectors are expressed in an inertial Earth-centered frame.

The position of the tether center of mass is
\begin{equation}
\mathbf{R}_{\mathrm{cm},i}(t) = R_{\mathrm{cm},i}\,\mathbf{e}_r(t),
\end{equation}
where \(\mathbf{e}_r(t)\) is the radial unit vector rotating with the orbital angular velocity
\begin{equation}
\Omega_i = \sqrt{\frac{GM_{\oplus}}{R_{\mathrm{cm},i}^3}} .
\end{equation}

The position of a tether endpoint relative to the center of mass is denoted by the vector
\begin{equation}
\mathbf{s}_i(t) = l_i
\left(
\cos\phi_i(t)\,\mathbf{e}_r(t)
+
\sin\phi_i(t)\,\mathbf{e}_\theta(t)
\right),
\end{equation}
where \(l_i\) is the tether half-length, \(\phi_i\) is the tether orientation angle measured in the orbital frame, and \(\mathbf{e}_\theta(t)\) is the tangential unit vector. The total Earth-centered position of the endpoint is therefore
\begin{equation}
\mathbf{r}_i(t) = \mathbf{R}_{\mathrm{cm},i}(t) + \mathbf{s}_i(t).
\end{equation}

Differentiating \(\mathbf{R}_{\mathrm{cm},i}(t)\) with respect to time yields the inertial velocity of the tether center of mass,
\begin{equation}
\dot{\mathbf{R}}_{\mathrm{cm},i}
=
\Omega_i\,\mathbf{J}\mathbf{R}_{\mathrm{cm},i},
\end{equation}
where \(\mathbf{J}\) denotes the planar rotation operator by \(+\pi/2\).

The time derivative of the endpoint offset \(\mathbf{s}_i(t)\) contains two contributions: rotation of the orbital frame at rate \(\Omega_i\), and rotation of the tether relative to that frame at rate \(\omega_i\). Thus,
\begin{equation}
\dot{\mathbf{s}}_i
=
(\Omega_i + \omega_i)\,\mathbf{J}\mathbf{s}_i.
\end{equation}

Combining the two contributions, the instantaneous inertial velocity of the tether endpoint is
\begin{equation}
\mathbf{V}_i
=
\dot{\mathbf{r}}_i
=
\dot{\mathbf{R}}_{\mathrm{cm},i}
+
\dot{\mathbf{s}}_i
=
\Omega_i \,\mathbf{J}\mathbf{R}_{\mathrm{cm},i}
+
(\Omega_i + \omega_i)\,\mathbf{J}\mathbf{s}_i .
\end{equation}

The elliptical node is an active transfer platform that detaches from the tether endpoint with velocity
\begin{equation}
\mathbf{V}^{(e)}_i = \mathbf{V}_i .
\end{equation}
As a result, the instantaneous inertial state of the elliptical node at the release point is identical to that of the tether endpoint. The subsequent trajectory of the elliptical node is therefore fully determined by this inertial position-velocity pair and follows a Keplerian orbit.

For the elliptical node to remain on a bound trajectory, its specific orbital energy must be negative, which imposes the condition
\begin{equation}
\lVert \mathbf{V}_i \rVert < \sqrt{\frac{2GM_{\oplus}}{\lVert \mathbf{r}_i \rVert}},
\end{equation}
where \(\mathbf{r}_i\) is the Earth-centered position vector of the release point. Under this condition, the elliptical node follows an elliptical Keplerian orbit whose parameters are uniquely determined by \(\mathbf{r}_i\) and \(\mathbf{V}_i\).

The elliptical trajectory of the node is designed to intersect the capture region of a neighboring tether at a later time. Capture requires that the endpoint of the neighboring tether pass through the same spatial location with an inertial velocity matching that of the elliptical node. In the numerical implementation, this velocity-matching condition is enforced by solving an inverse tether problem that reconstructs the parameters \((r_{i+1}, R_{\mathrm{cm},i+1}, \omega_{i+1})\) of the next tether from the instantaneous inertial state of the node.

The motion of each elliptical node between successive attachment events follows a classical Keplerian elliptical trajectory. The corresponding orbital elements are obtained from the relation,
\begin{equation}
a_i = \left( \frac{2}{\lVert \mathbf{r}_i \rVert} - \frac{\lVert \mathbf{V}_i \rVert^2}{GM_{\oplus}} \right)^{-1},
\end{equation}
with specific angular momentum
\begin{equation}
h_i = \lVert \mathbf{r}_i \times \mathbf{V}_i \rVert ,
\end{equation}
and eccentricity
\begin{equation}
e_i = \sqrt{1 - \frac{h_i^2}{a_i GM_{\oplus}}} .
\end{equation}
The corresponding perigee radius is
\begin{equation}
r_{p,i} = a_i(1 - e_i).
\end{equation}
Only trajectories satisfying
\begin{equation}
r_{p,i} \ge R_{\oplus} + R_{\mathrm{safe}}
\end{equation}
are retained. Here \(R_{\mathrm{safe}}\) denotes a prescribed safety altitude above the Earth's surface that excludes atmospheric drag, thermal loading, and ground intersection effects. For more details please refer to Sec.~\ref{subsec:atm}. This constraint ensures that all elliptic-node trajectories remain strictly non-intersecting with the Earth and evolve under purely Keplerian dynamics throughout their motion.

Under this admissibility condition, the elliptical node follows a bound Keplerian orbit whose mean orbital angular frequency is
\begin{equation}
f_i = \sqrt{\frac{GM_{\oplus}}{a_i^3}} .
\end{equation}

\subsection{Inverse solutions}
The goal is to determine the parameters of the neighboring tether $i+1$
such that one of its endpoints becomes dynamically compatible with the
elliptical endpoint released from tether $i$.
Specifically, the tether must be positioned so that, at the capture time
$t_c$, its endpoint coincides with this elliptical endpoint in physical space
and has an identical inertial velocity (within a prescribed tolerance).
This requirement defines an inverse tether construction problem, in which
the geometric and rotational parameters of the capturing tether are
reconstructed from the known inertial state of the elliptical endpoint.

Let the elliptical endpoint released from tether $i$ propagate ballistically
to a capture time $t_c$.
Denote its inertial state at capture by
\begin{equation}
\mathbf{r}_{e,i} \equiv \mathbf{r}_{e,i}(t_c),
\qquad
\mathbf{v}_{e,i} \equiv \dot{\mathbf{r}}_{e,i}(t_c).
\end{equation}

The neighboring tether $i+1$ has unknown parameters
\begin{equation}
\left( l_{i+1},\, R_{\mathrm{cm},i+1},\, \omega_{i+1} \right),
\end{equation}
and a corresponding orbital angular velocity
\begin{equation}
\Omega_{i+1} =
\sqrt{\frac{GM_{\oplus}}{R_{\mathrm{cm},i+1}^3}} .
\label{eq:Omega_i+1}
\end{equation}

For the purpose of constructing a feasible capture configuration,
the tether rotation is approximated with a constant angular
velocity $\omega_{i+1}$.
Under this approximation, the inertial velocity of the tether endpoint is

\begin{equation}
\mathbf{v}_{\mathrm{end}}
=
\mathbf{v}_{\mathrm{cm}}
+
\boldsymbol{\omega}_{i+1}
\times
\mathbf{r}_{\mathrm{end}},
\end{equation}

which allows the inverse construction problem to be solved by matching
the endpoint position and velocity to the state
$(\mathbf{r}_{e,i}, \mathbf{v}_{e,i})$ at the capture time $t_c$.

The constant-$\omega$ approximation is used only at the feasibility
stage to determine a geometrically consistent tether configuration.
This assumption makes the inverse construction analytically tractable,
since the endpoint position and velocity can then be expressed in closed
form in terms of the tether parameters, as shown further. Without this simplification the
inverse problem would require solving the full nonlinear tether dynamics,
which is not analytically feasible.

To assess the validity of the constant-$\omega$ approximation used at
this stage, the endpoint trajectory obtained under the assumption
$\dot{\phi}(t)=\bar{\omega}=\mathrm{const}$ was compared with the
endpoint trajectory reconstructed from the full simulated nonuniform
evolution $\phi(t)$.
For each case, the position and velocity mismatches
\[
\Delta r(t)=\|\mathbf r_{\mathrm{full}}(t)-\mathbf r_{\mathrm{const}}(t)\|,
\qquad
\Delta v(t)=\|\mathbf v_{\mathrm{full}}(t)-\mathbf v_{\mathrm{const}}(t)\|
\]
were evaluated over the capture interval.
The resulting normalized errors remained small, with average values
$\Delta r/l \approx 2.7\times10^{-3}$ and
$\Delta v/v \approx 4.1\times10^{-3}$,
which correspond to average absolute errors of about
$2.5\,\mathrm{km}$ in position and $4\,\mathrm{m\,s^{-1}}$ in velocity.
Although these discrepancies are small relative to 
the characteristic dimensions and velocities of the 
system, they are not negligible for physical docking 
operations.

The constant-angular-velocity approximation is 
therefore used as a candidate generation mechanism 
rather than as a complete rendezvous solution.

The resulting position and velocity mismatches 
would require further trajectory refinement or 
active correction at the elliptical node. The 
design and validation of such correction procedures 
remain outside the scope of the present analysis.

Let $\mathbf{e}_r(t_c)$ and $\mathbf{e}_\theta(t_c)$ denote the local orbital basis vectors of tether $i+1$ at time $t_c$.
The center-of-mass position of the tether is
\begin{equation}
\mathbf{R}_{\mathrm{cm},i+1}(t_c)
=
R_{\mathrm{cm},i+1}\,\mathbf{e}_r(t_c).
\end{equation}

The endpoint offset relative to the center of mass is
\begin{equation}
\mathbf{s}_{i+1}(t_c)
=
r_{i+1}
\left(
\cos\phi_{i+1}(t_c)\,\mathbf{e}_r(t_c)
+
\sin\phi_{i+1}(t_c)\,\mathbf{e}_\theta(t_c)
\right),
\end{equation}
so that the total Earth-centered position of the endpoint is
\begin{equation}
\mathbf{r}_{i+1}(t_c)
=
\mathbf{R}_{\mathrm{cm},i+1}(t_c)
+
\mathbf{s}_{i+1}(t_c).
\end{equation}

Using the inertial velocity expression derived previously, the endpoint velocity is
\begin{equation}
\mathbf{V}_{i+1}(t_c)
=
\Omega_{i+1}\,\mathbf{J}\mathbf{R}_{\mathrm{cm},i+1}(t_c)
+
\left( \Omega_{i+1} + \omega_{i+1} \right)
\,\mathbf{J}\mathbf{s}_{i+1}(t_c).
\end{equation}

Capture of the elliptical node by tether $i+1$ requires the existence of a tether phase
$\phi_{i+1}(t_c)$ such that the endpoint position and velocity match those of the node:
\begin{equation}
\mathbf{r}_{i+1}(t_c) = \mathbf{r}_{e,i}(t_c),
\qquad
\mathbf{V}_{i+1}(t_c) = \mathbf{v}_{e,i}(t_c) .
\end{equation}

The conditions above therefore define the inverse capture problem: given the inertial state
$(\mathbf r_{e,i}(t_c), \mathbf v_{e,i}(t_c))$ at time $t_c$, determine whether there exists a
rigidly rotating tether whose endpoint can instantaneously pass through this state.
The matching requirement corresponds to the idealized
zero-relative-velocity rendezvous condition considered
in momentum-exchange tether transportation systems
\cite{Moravec1977,Bangham1998,Sidorov2004,Ivanov2010}.
In this limit, the payload and tether endpoint coincide
in position and inertial velocity at the exchange instant,
thereby avoiding impulsive momentum transfer.

Within the present computational synthesis framework,
these matching conditions are formulated as an inverse
kinematic problem in which candidate geometric and
rotational parameters of the capturing tether are
reconstructed directly from a prescribed inertial
transfer state.

The resulting quadratic formulation provides an
analytically tractable candidate generation mechanism
for sequential multi-stage network synthesis.

The endpoint offset relative to the center of mass of the tether is therefore
\begin{equation}
\mathbf{s}
=
\mathbf{r}_{e,i}(t_c)
-
\mathbf{R}_{\mathrm{cm},i+1}(t_c).
\end{equation}

Rigid-body kinematics then require that the inertial velocity of the
endpoint be expressible as
\begin{equation}
\mathbf{v}_{e,i}(t_c)
=
\Omega_{i+1}\,\mathbf{J}\mathbf{R}_{\mathrm{cm},i+1}(t_c)
+
\left( \Omega_{i+1} + \omega_{i+1} \right)
\mathbf{J}\mathbf{s}.
\end{equation}

Substituting, rearranging and multiplying by  $\mathbf{J}^{-1}$ yields
\begin{equation}
\mathbf{J}^{-1}\mathbf{v}_{e,i}(t_c)
-
(\Omega_{i+1}+\omega_{i+1})\mathbf{r}_{e,i}(t_c)
=
-\omega_{i+1}\mathbf{R}_{\mathrm{cm},i+1}(t_c),
\end{equation}
and therefore the required center-of-mass position must satisfy
\begin{equation}
\mathbf{R}_{\mathrm{cm},i+1}(t_c)
=
\left(1+\frac{\Omega_{i+1}}{\omega_{i+1}}\right)
\mathbf{r}_{e,i}(t_c)
-
\frac{\mathbf{J}^{-1}\mathbf{v}_{e,i}(t_c)}{\omega_{i+1}}.
\end{equation}

Introduce the state-dependent vector
\begin{equation}
\mathbf{q} \equiv \mathbf{J}^{-1}\mathbf{v}_{e,i}(t_c),
\end{equation}
and define the radius-dependent mismatch vector
\begin{equation}
\mathbf{d}(R_{\mathrm{cm},i+1})
\equiv
\mathbf{q}
-
\Omega_{i+1}\,\mathbf{r}_{e,i}(t_c).
\end{equation}

Then the center-of-mass position can be written in the equivalent form
\begin{equation}
\mathbf{R}_{\mathrm{cm},i+1}(t_c)
=
\mathbf{r}_{e,i}(t_c)
-
\frac{1}{\omega_{i+1}}\,
\mathbf{d}(R_{\mathrm{cm},i+1}).
\end{equation}

Imposing the circular-orbit constraint
\begin{equation}
\left|
\mathbf{R}_{\mathrm{cm},i+1}(t_c)
\right|
=
R_{\mathrm{cm},i+1}
\end{equation}
yields the scalar condition
\begin{equation}
\left|
\mathbf{r}_{e,i}(t_c)
-
\frac{1}{\omega_{i+1}}\,
\mathbf{d}(R_{\mathrm{cm},i+1})
\right|^2
=
R_{\mathrm{cm},i+1}^2 .
\end{equation}

Introducing $u=1/\omega_{i+1}$ gives
\begin{equation}
\left|
\mathbf{r}_{e,i}(t_c)
-
u\,\mathbf{d}(R_{\mathrm{cm},i+1})
\right|^2
=
R_{\mathrm{cm},i+1}^2,
\end{equation}
which expands to the quadratic equation
\begin{equation}
D(R_{\mathrm{cm},i+1})\,u^2
-
2\left(\mathbf{r}_{e,i}(t_c)\cdot
\mathbf{d}(R_{\mathrm{cm},i+1})\right)u
+
\left(
\mathbf{r}_{e,i}(t_c)\cdot
\mathbf{r}_{e,i}(t_c)
-
R_{\mathrm{cm},i+1}^2
\right)
=0,
\end{equation}
where
\begin{equation}
D(R_{\mathrm{cm},i+1})
\equiv
\mathbf{d}(R_{\mathrm{cm},i+1})\cdot
\mathbf{d}(R_{\mathrm{cm},i+1}).
\end{equation}

Thus, for any prescribed center-of-mass orbit radius $R_{\mathrm{cm},i+1}$
(and hence $\Omega_{i+1}$ from~\eqref{eq:Omega_i+1}),
the matching conditions reduce to a single quadratic equation in $u$.
Consequently, for a fixed $R_{\mathrm{cm},i+1}$ there are at most two real
solutions $u_{\pm}$ (corresponding to two branches of $\omega_{i+1}$),
or no real solutions.
If a real solution $u$ exists, the corresponding angular rate is
\begin{equation}
\omega_{i+1}=\frac{1}{u},
\end{equation}
the center-of-mass position at capture is obtained from
\begin{equation}
\mathbf{R}_{\mathrm{cm},i+1}(t_c)
=
\mathbf{r}_{e,i}(t_c) - u\,\mathbf{d}(R_{\mathrm{cm},i+1}),
\end{equation}
and the required tether half-length (endpoint offset magnitude) follows as
\begin{equation}
l_{i+1} = \left\|\mathbf{r}_{e,i}(t_c)-\mathbf{R}_{\mathrm{cm},i+1}(t_c)\right\|.
\end{equation}

\subsection{Synchronization}

Candidate recurrence intervals between an elliptical
node and its neighboring tethers are identified using
near-commensurate relationships between tether
rotational and node orbital frequencies.

Within the present framework, these relationships
serve as computational screening criteria rather
than sufficient conditions for physical rendezvous.
Let $\omega_i$ denote the rotational angular velocity of tether $i$ and $f_i$ the mean
orbital angular frequency (mean motion) of the associated elliptical node.
Rather than requiring exact commensurability, we require that the ratios admit sufficiently accurate
rational approximations,
\begin{equation}
\left| \frac{\omega_i}{f_i} - \frac{p_i}{q_i} \right| \le \delta_i,
\qquad
\left| \frac{\omega_{i+1}}{f_i} - \frac{l_i}{m_i} \right| \le \delta_i,
\end{equation}
where $p_i,q_i,l_i,$ and $m_i$ are coprime positive integers, and $\delta_i$ is a
stage-dependent synchronization tolerance.
The value of $\delta_i$ is not prescribed a priori in the frequency domain, but is instead
determined by a chosen admissible spatial mismatch at the exchange point.
Specifically, $\delta_i$ is selected such that the accumulated phase drift over the
corresponding synchronization interval results in a relative displacement not exceeding
a prescribed geometric tolerance.

These relations imply that tether $i$ executes approximately $p_i$ rotations during
$q_i$ orbital revolutions of the elliptical node, while tether $i+1$ executes approximately
$l_i$ rotations during $m_i$ node revolutions. The integers $q_i$ and $m_i$ therefore set
candidate repeat intervals of the relative configurations, up to a bounded phase error.

Let $\phi_{\mathrm{node}}(t)$ denote the inertial angular phase of the elliptical node and
$\phi_i(t)$ and $\phi_{i+1}(t)$ those of tethers $i$ and $i+1$. The relative phases are
\begin{equation}
\Delta\phi_i(t) = \phi_i(t) - \phi_{\mathrm{node},i}(t),
\qquad
\Delta\phi_{i+1}(t) = \phi_{i+1}(t) - \phi_{\mathrm{node},i}(t).
\end{equation}
Mass exchange requires that, at the exchange point, tether $i$ presents its endpoint outward
while tether $i+1$ presents its endpoint inward, corresponding to relative phase offsets differing by $\pi$,
\begin{equation}
\Delta\phi_{i+1} = \Delta\phi_i + \pi \quad (\mathrm{mod}\; 2\pi),
\end{equation}
which can be absorbed into the initial phase constants and therefore does not affect the synchronization period.

The relative phase evolution is
\begin{equation}
\Delta\phi_i(t) = (\omega_i - f_i)t + \Delta\phi_i(0).
\end{equation}
A repeat of the required geometric configuration after $N_{\mathrm{node}}$ node revolutions
occurs when the accumulated phase drift is within a prescribed spatial tolerance at the exchange radius.
Writing $N_{\mathrm{node}} = f_i T/(2\pi)$, we enforce the geometric congruence condition
\begin{equation}
\Delta s_i
=
r_{\mathrm{ex},i}\,
\Bigl(2\pi N_{\mathrm{node}}\Bigr)\,
\left| \frac{\omega_i}{f_i} - \frac{p_i}{q_i} \right|
\le \Delta s_{\mathrm{tol}},
\end{equation}
where $r_{\mathrm{ex},i}$ is the radial distance of the exchange point and
$\Delta s_{\mathrm{tol}}$ is the maximum allowable spatial mismatch.
An analogous condition is imposed for tether $i+1$ with integers $(l_i,m_i)$.

Under these constraints, tether $i$ admits an exchange configuration approximately repeating
after $q_i$ node revolutions (within the tolerance), while tether $i+1$ does so after $m_i$ node revolutions.
For both tethers to be simultaneously aligned with the elliptical node, the number of node revolutions must be a common multiple of
$q_i$ and $m_i$. The minimal joint synchronization interval is therefore
\begin{equation}
N_{\mathrm{node}}^{(i)} = \operatorname{lcm}(q_i,m_i),
\end{equation}
where $\operatorname{lcm}$ denotes the least common multiple.

The corresponding synchronization time for stage $i$ is
\begin{equation}
T_i = \operatorname{lcm}(q_i,m_i)\,\frac{2\pi}{f_i}.
\end{equation}

Assuming sequential transfers across $n$ stages, the cumulative synchronization-based
propagation time is
\begin{equation}
T_{\mathrm{eff}}
= \sum_{i=0}^{n-1} T_i
= \sum_{i=0}^{n-1} \operatorname{lcm}(q_i,m_i)\,\frac{2\pi}{f_i}.
\end{equation}
If the node frequencies $f_i$ are rational multiples of a common reference frequency,
the multi-tether system admits a strictly periodic solution and $T_{\mathrm{eff}}$
coincides with an exact cycle time. Otherwise, the dynamics are quasi-periodic and
$T_{\mathrm{eff}}$ should be interpreted as an effective propagation time. Because synchronization is enforced only up to finite phase tolerances, residual phase
drift generally prevents individual exchange events from occurring at exactly these times.
Consequently, $T_{\mathrm{eff}}$ is a nominal (lower-bound) estimate of propagation time,
with actual transfers occurring within bounded temporal offsets.

\subsection{Problem definition and experiment methods}

We consider the problem of constructing a dynamically feasible chain of rotating orbital tethers connected by elliptical transfer nodes in order to determine whether sustained outward payload transport can be achieved under Keplerian orbital dynamics, synchronization constraints, and finite geometric tolerances.

The system is built sequentially in stages. At stage $i$, a tether $i$, characterized by half-length $l_i$, center-of-mass orbital radius $R_{\mathrm{cm},i}$, and angular velocity $\omega_i(t)$, releases a payload from one of its endpoints onto an elliptical transfer trajectory determined by the release position and velocity. The payload subsequently follows this trajectory until it is intercepted by the tether of the next stage.

The total number of stages is not prescribed \textit{a priori}. Instead, new stages are added sequentially until the inner endpoint radius of the final tether satisfies

\begin{equation}
R_{\mathrm{cm},n} - l_n \ge R_{\mathrm{target}}.
\end{equation}

here $R_{\mathrm{target}}$ denotes the target high orbital radius, chosen in this work as the geostationary orbit for specificity.

The construction is governed by continuous parameters describing the tether geometry and rotation, namely the half-length $l_i$ and angular velocity $\omega_i(t)$. For a given tether configuration, the corresponding elliptical node trajectory is uniquely determined by the release state, and its orbit frequency $n_i$ follows from the resulting Keplerian orbit. Synchronization properties are then analyzed by approximating the ratios between the relevant angular frequencies by rational numbers. The numerical construction proceeds through the following sequence of steps.

\paragraph{Step 1. Construction of the initial tether family.}

The algorithm begins by generating a family of admissible base tethers. For each candidate tether half-length $l$, the center-of-mass orbital radius is chosen such that the inner endpoint of the tether lies at the safe orbital radius,

\begin{equation}
R_{\mathrm{cm}} - l = R_{\mathrm{safe}},
\end{equation}

\begin{equation}
R_{\mathrm{cm}} = R_{\mathrm{safe}} + l,
\end{equation}

where

\begin{equation}
R_{\mathrm{safe}} = R_{\oplus} + R_{\mathrm{safe,alt}}.
\end{equation}

In this work we take $R_{\mathrm{safe,alt}} = 1000~\mathrm{km}$, thus ensuring negligible atmospheric drag (see Section \ref{sec:lowest_tether_drag}).

The tether half-length is scanned over the interval

\begin{equation}
R_{\min} \le l \le R_{\max},
\end{equation}

with

\begin{equation}
R_{\min} = 20~\mathrm{km}.
\end{equation}

The search is performed on a discrete grid with step size

\begin{equation}
\Delta R = 100~\mathrm{m}
\label{eq:grid_step}
\end{equation}

The upper bound $R_{\max}=500~\mathrm{km}$ is introduced only as a numerical search limit and does not reflect constraints related to material strength or tether feasibility.

The center of mass of the tether moves on a circular Keplerian orbit with angular velocity $\Omega$. 
To parameterize the initial tether configurations, the base tether rotation rate is fixed to a representative value

\begin{equation}
\omega_{\mathrm{fixed}} = 0.02~\mathrm{rad\,s^{-1}} .
\end{equation}

This value is not a fundamental constraint of the model but serves as a practical reference point for generating the initial set of feasible configurations. 
In principle, the tether rotation frequency may take any value within the mechanically admissible range of the system. 
However, both excessively small and excessively large angular velocities lead to undesirable dynamical behavior.

If the rotation rate is too small, the tangential velocity at the tether endpoint becomes insufficient to produce energetically viable transfer trajectories. 
In this regime the released payload remains close to the circular orbit of the tether center of mass, resulting in elliptical trajectories with low apoapsis and limited outward propagation capability. 
Moreover, weak release velocities increase the risk of trajectories with unsafe perigee distances, potentially violating operational safety constraints.

Conversely, excessively large rotation rates produce endpoint velocities approaching the local escape velocity. 
In this case the resulting trajectories become highly eccentric and approach parabolic escape conditions, making resonant synchronization with subsequent tether stages increasingly rare. 
From an engineering perspective, very large rotation rates are also undesirable because they require large stored rotational energy and impose substantial structural loads on the tether. The selected value  represents a moderate operating regime that generates sufficiently energetic release trajectories while remaining compatible with realistic mechanical constraints.

Candidate tethers must also satisfy mechanical feasibility constraints encoded in the feasibility map described in Section~\ref{sec:feasibility_map}. This map restricts admissible combinations of $R_{\mathrm{cm}}$, tether length $l$, and endpoint relative velocity.
\paragraph{Step 2. Generation of elliptical transfer nodes.}

For each tether retained at stage $i$, the payload release state is evaluated for multiple tether phases $\phi$ uniformly distributed over the interval $[0,2\pi)$ using 1200 discretization steps. The inertial position and velocity of the released payload are computed from the tether geometry, the center-of-mass orbital motion, and the tether rotation.

From each release state, the Keplerian orbit of the detached payload is reconstructed. This imposes the elliptical admissibility condition on the full energy: 

\begin{equation}
\varepsilon < 0,
\end{equation}

together with the safety requirement

\begin{equation}
r_{p,i} \ge R_{\mathrm{safe}},
\end{equation}

which ensures that the perigee of the elliptical orbit does not intersect the Earth.

\paragraph{Step 3. Synchronization between tether rotation and elliptic-node motion.}

For every dynamically admissible elliptical node, the ratio between the node orbital frequency $n_i$ and the inertial rotation frequency of the emitting tether $\omega_i^{\mathrm{inert}}$ is examined. Near-rational relations
\begin{equation}
\frac{n_i}{\omega_i^{\mathrm{inert}}} \approx \frac{p_i}{q_i}
\end{equation}
are identified by enumerating integer pairs satisfying
\begin{equation}
1 \le p_i,q_i \le p_{\max},
\qquad
\operatorname{lcm}(p_i,q_i) \le \operatorname{lcm}_{\max}.
\end{equation}

Exact resonance is not required. Instead, approximate synchronization is accepted if the accumulated phase drift corresponds to a calculated spatial misalignment below a prescribed tolerance,
\begin{equation}
\Delta s_i \le \Delta s_{\mathrm{tol}}
\label{eq:res_tol}
\end{equation}

In this work we adopt
\begin{equation}
\Delta s_{\mathrm{tol}} = 500~\mathrm{m}
\label{eq:res_tol_value}
\end{equation}

To prevent impractically long synchronization intervals, the integer search is restricted by
\begin{equation}
\operatorname{lcm}(p_i,q_i) \le \operatorname{lcm}_{\max}.
\end{equation}
In the conceptual formulation we take
\begin{equation}
{p_{max} =200},\qquad \operatorname{lcm}_{\max} = 2000
\end{equation}

The parameters $p_{\max}$, $\operatorname{lcm}_{\max}$, and $\Delta s_{\mathrm{tol}}$ could be treated as hyperparameters of the synchronization search. The limits $p_{\max}$ and $\operatorname{lcm}_{\max}$ are chosen empirically to ensure that the approximate time between synchronization events does not become excessively large, while still allowing a sufficient number of candidate resonances to be identified.

The spatial tolerance $\Delta s_{\mathrm{tol}}$ represents the admissible misalignment between the tether emission phase and the orbital node. 
It is selected to be small enough that the resulting deviation can be corrected through active control without significant energy expenditure, yet large enough to allow practical resonance detection within the discrete search space. In this work, $\Delta s_{\mathrm{tol}} = 500\,\mathrm{m}$ is adopted. 
This value is also consistent with the numerical accuracy of the Runge–Kutta integration scheme introduced earlier, ensuring that the tolerance remains not smaller than the expected numerical error of the trajectory propagation.
\paragraph{Step 4. Propagation of elliptical nodes and search for receiving tethers.}

Each resonant elliptical node is propagated along its Keplerian trajectory. At discrete points along the orbit, the simulation searches for tethers whose endpoint position and velocity match the node state. This step enforces the velocity compatibility condition: the inertial velocity of the elliptical node must coincide with that of the receiving tether endpoint at the docking moment. Candidate receiving tethers are constructed analytically and retained only if they satisfy both the geometric docking condition and the mechanical feasibility constraints of the feasibility map. 

In the calculations reported here, the propagation step along the trajectory was set to 5000~m due to computational limitations. Consequently, the procedure detects only a subset of the full set of feasible receiving tethers, and a smaller step size would be expected to identify additional solutions. Nevertheless, the obtained results are sufficient to demonstrate the existence of feasible transfer configurations.

\paragraph{Step 5. Synchronization between elliptical node and receiving tether.}

For every candidate receiving tether, the ratio of elliptical-node frequency and the inertial rotation frequency of the tether is evaluated,
\begin{equation}
\frac{n_i}{\omega_{i+1}^{\mathrm{inert}}} \approx \frac{l_i}{m_i}.
\end{equation}
To enable an analytical construction of the receiving tether trajectory, the average inertial angular velocity $\omega_{i+1}^{\mathrm{inert}}$ over the synchronization interval is used. This approximation slightly modifies the instantaneous position and velocity of the tether endpoint relative to the exact time-dependent motion; however, the resulting discrepancy is small and can be compensated by a corresponding adjustment of the elliptical node along its orbit.

As before, integer pairs $(l_i,m_i)$ are used to approximate the ratio, and the configuration is accepted only if the resulting spatial misalignment remains below the prescribed tolerance
\begin{equation}
\Delta s_i \le \Delta s_{\mathrm{tol}}.
\end{equation}

The associated synchronization interval is
\begin{equation}
T_i = \frac{2\pi m_i}{\omega_{i+1}^{\mathrm{inert}}}.
\end{equation}

Additionally, the candidate tether must produce outward radial progress relative to the previous stage,
\begin{equation}
\Delta R_i =
\bigl(R_{\mathrm{cm},i+1}-r_{i+1}\bigr)
-
\bigl(R_{\mathrm{cm},i}-r_i\bigr) > 0.
\end{equation}

In the first experiment, the search over candidate configurations was restricted to a finite radial increment grid, with $\Delta R_i$ sampled within the range $5\times10^5$ to $5\times10^6$ m. In the second experiment, this restriction was removed, allowing the radial propagation to extend up to $4.5\times10^7$ m in order to investigate the emergence of large radial jumps and long-range transfer links within the tether chain.

\paragraph{Step 6. Selection based on propagation efficiency.}

Among all admissible receiving tethers, configurations are ranked by the effective radial propagation speed
\begin{equation}
v_{\mathrm{prop},i} = \frac{\Delta R_i}{T_i}.
\end{equation}

Here $T_i$ denotes the interval between successive synchronization opportunities rather than the actual physical travel time of the spacecraft. It therefore represents the characteristic waiting time required for the next docking configuration to occur. The metric penalizes solutions that achieve outward transport only through extremely long synchronization intervals. To maintain computational feasibility, the search is performed using a beam search strategy. At each stage, the best candidates are retained according to the propagation efficiency metric, with the beam width limited to 30 configurations.

To avoid redundancy arising from discretization of the search space, near-duplicate tether configurations are removed prior to ranking. Specifically, candidate tethers are compared in the parameter space $(R_{\mathrm{cm}}, r, w)$, and any pair of candidates $(i,j)$ satisfying
\begin{equation}
\left|R_{\mathrm{cm}}^{(i)} - R_{\mathrm{cm}}^{(j)}\right| < \delta_R, \quad
\left|r^{(i)} - r^{(j)}\right| < \delta_r, \quad
\left|w^{(i)} - w^{(j)}\right| < \delta_w
\end{equation}
is considered redundant. Within each such cluster, only a single representative configuration is retained, while the remaining candidates are discarded. In the present implementation, the tolerance values are chosen to correspond to sub-kilometer differences in 
$(R_{\mathrm{cm}})$ and ($r$), and small relative deviations in angular velocity, ensuring that merged configurations are physically indistinguishable.
In practice, the tolerances are chosen as $\delta_R \approx 500~\mathrm{m}$ and $\delta_r \approx 500~\mathrm{m}$, consistent with the spatial congruence threshold used in the synchronization condition. The angular velocity tolerance is defined in relative terms, with $\delta_w / w \sim 10^{-4}$, ensuring that only configurations with negligible dynamical differences are merged. 

The configuration maximizing $v_{\mathrm{prop},i}$ within the filtered candidate set is selected as the next stage.

\paragraph{Step 7. Iteration of the staged construction.}

Once the optimal receiving tether is identified, it forms the basis of the next stage. The procedure then repeats: release phases are scanned again, elliptical nodes are generated, synchronization conditions are evaluated, and candidate receiving tethers are searched.

The algorithm terminates when one of the following conditions occurs:

\begin{itemize}
\item no dynamically admissible elliptical nodes are generated,
\item resonant elliptical nodes exist but no feasible receiving tether can be constructed,
\item the target orbital radius is reached,
\begin{equation}
R_{\mathrm{cm},n}+l_n \ge R_{\mathrm{target}}.
\end{equation}
\end{itemize}

The cumulative transfer time is determined by the synchronization intervals of the accepted stages,
\begin{equation}
T = \sum_{i=0}^{n} T_i .
\end{equation}

Thus both the number of stages $n$ and the total transfer time emerge naturally from the admissibility structure of the staged construction rather than being specified a priori. A pseudocode description of the algorithm is provided  in Appendix~B.

\subsection{Simulation results}

\paragraph{Resonant tether chain construction.}

\begin{figure*}[t]
\centering

\begin{subfigure}{0.48\textwidth}
\centering
\includegraphics[width=\linewidth]{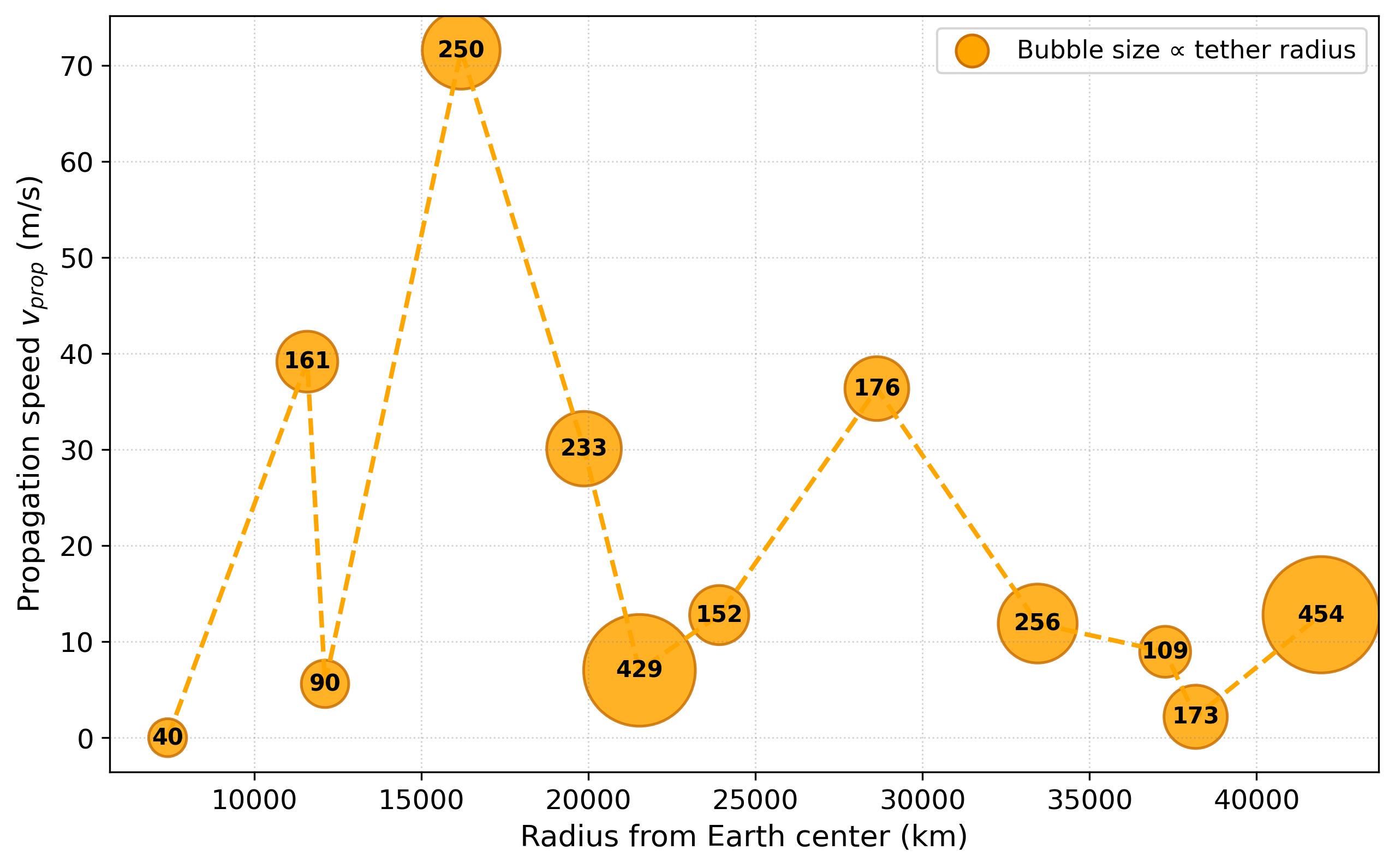}
\caption{Propagation speed between successive resonant tether stages as a function of orbital radius. Each bubble corresponds to a transfer between two stages of the elevator chain. The bubble size and the number inside represents the tether half-length (km).}
\end{subfigure}
\hfill
\begin{subfigure}{0.48\textwidth}
\centering
\includegraphics[width=\linewidth]{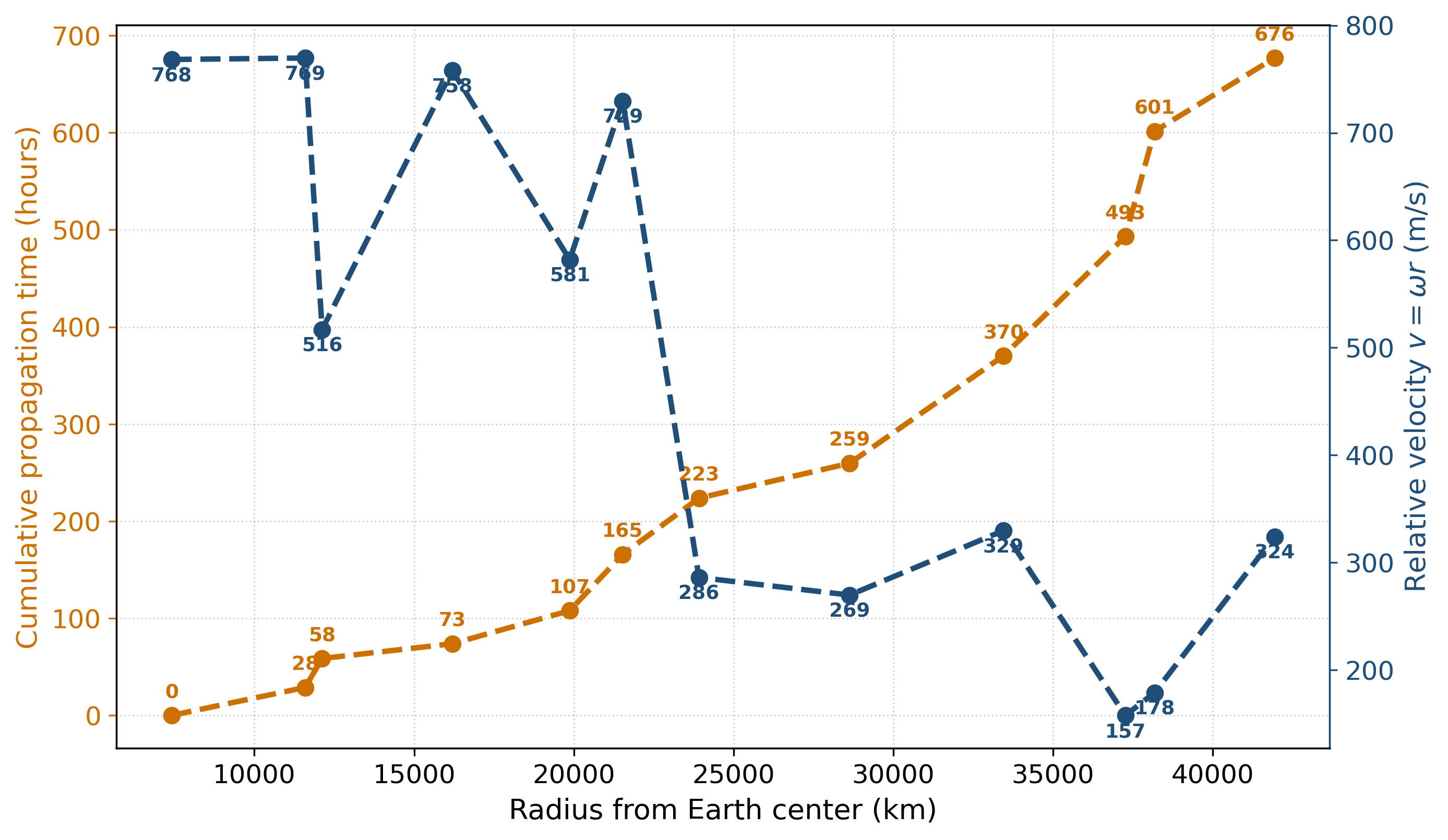}
\caption{Cumulative propagation time required to reach increasing orbital radii through successive resonant transfers, together with the rotational velocity relative to the tether’s center of mass. The time curve represents the total synchronization time accumulated across all preceding tether stages.}
\end{subfigure}

\vspace{0.4cm}

\begin{subfigure}{0.48\textwidth}
\centering
\includegraphics[width=\linewidth]{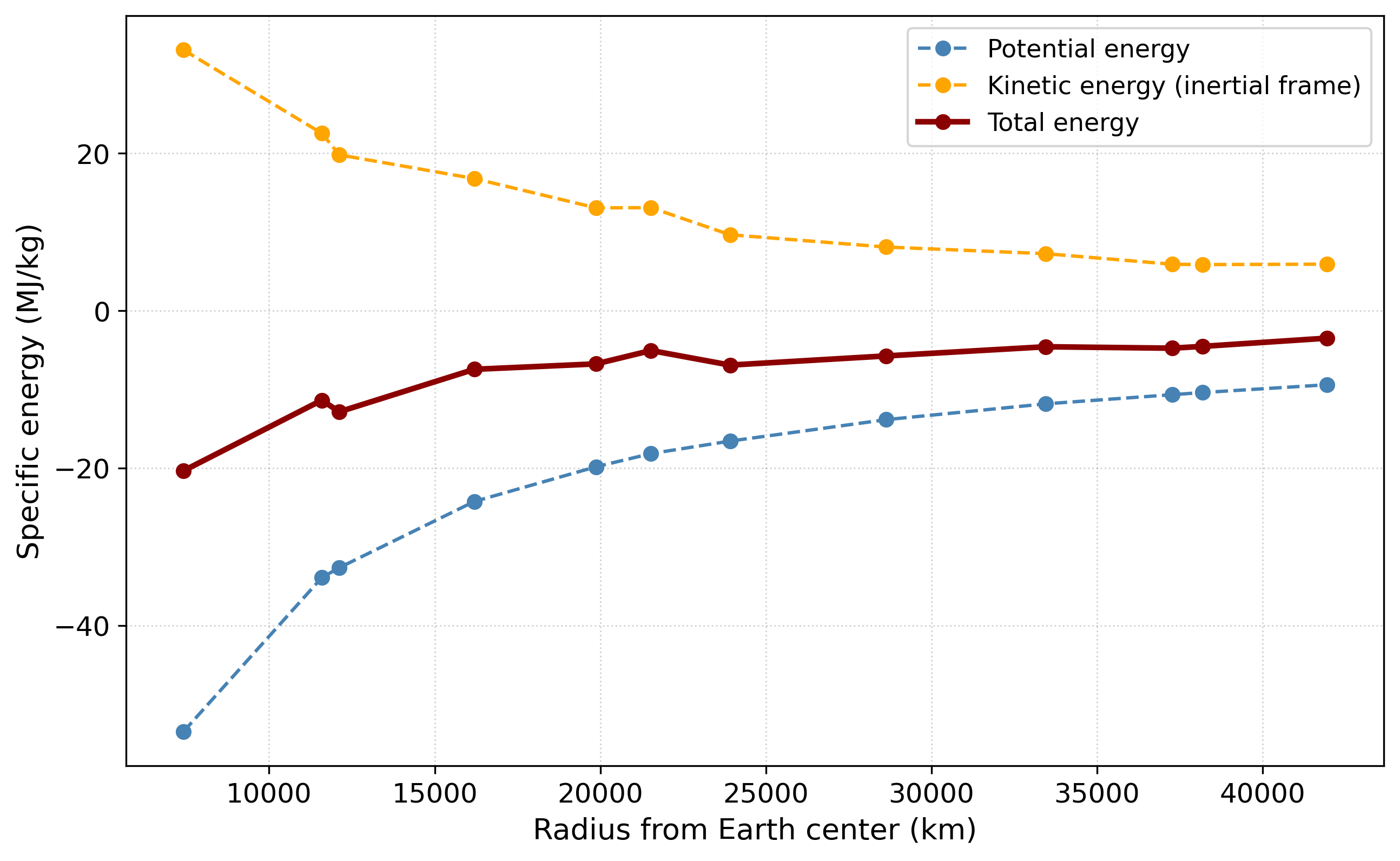}
\caption{Specific orbital energy components along the elevator chain. The gravitational potential energy increases toward zero with radius, while the kinetic energy decreases due to lower orbital velocity at larger radii.}
\end{subfigure}
\hfill
\begin{subfigure}{0.48\textwidth}
\centering
\includegraphics[width=\linewidth]{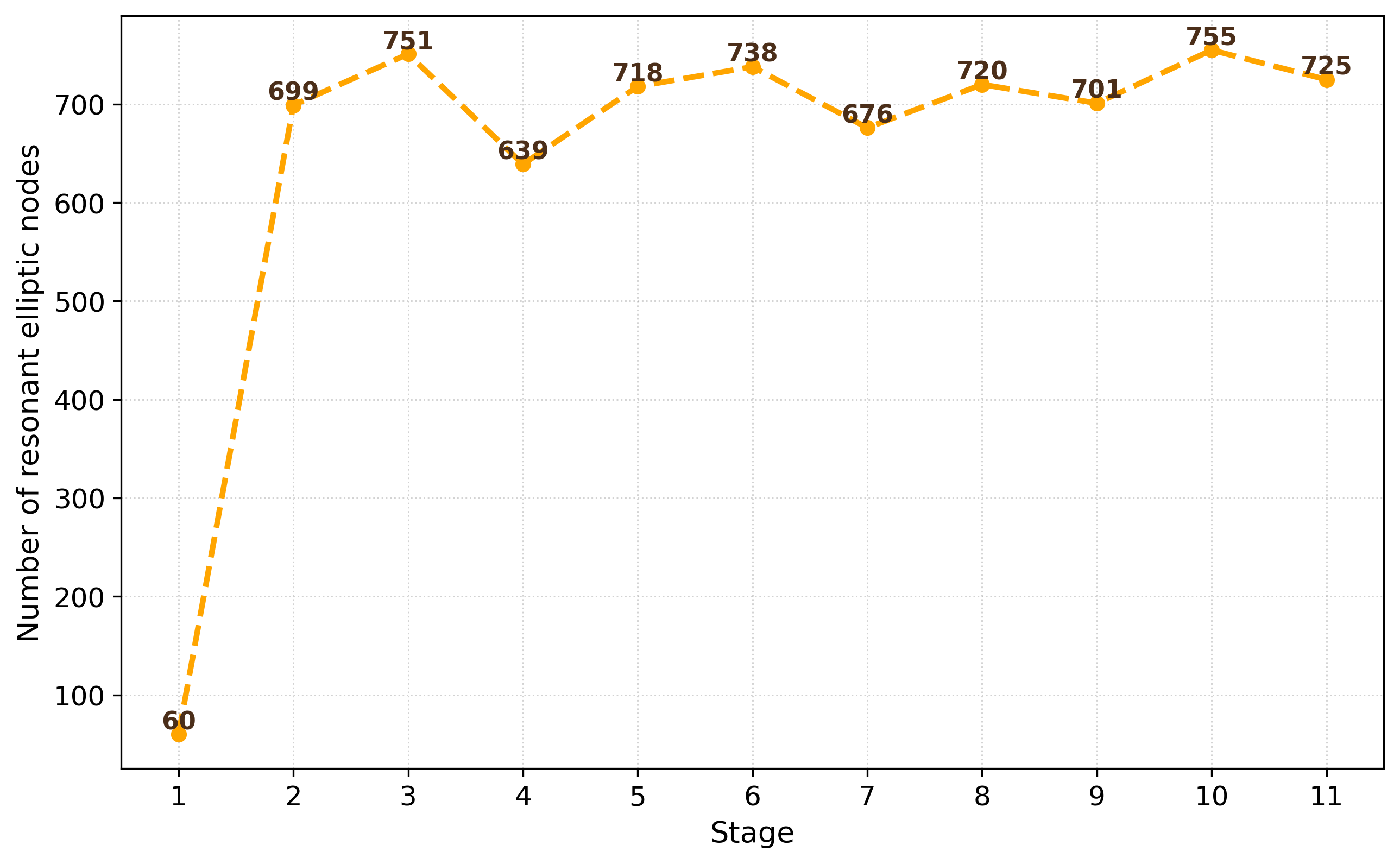}
\caption{Number of resonant elliptical nodes identified at each construction stage of the tether chain. The large number of available resonant nodes demonstrates the abundance of dynamically compatible transfer opportunities between stages. Stage 0 is omited.}
\end{subfigure}

\caption{Resonant multi-stage orbital elevator architecture obtained from the discrete resonance search algorithm. The panels illustrate the propagation dynamics between successive tether stages, cumulative synchronization time required for outward transport, the corresponding variation of specific orbital energy, and the number of available resonant elliptical nodes per stage. }
\label{fig:elevator_results}
\end{figure*}

Using a beam search with beam width equal to $15$, we constructed a continuous chain of resonant tethers extending from low Earth orbit toward geostationary radius. The resulting sequence contains 12 stages and provides a dynamically consistent pathway through successive resonant transitions. The propagation structure and the effective transfer speeds between successive stages are illustrated in Fig.~\ref{fig:elevator_results}(a). In this experiment, the maximum allowed radial increment between successive stages was constrained to $\Delta R \leq 5{,}000$ km. The resulting chain shows that most transitions occur with significantly smaller radial steps. The cumulative propagation time along the chain is approximately 676 hours, as shown in Fig.~\ref{fig:elevator_results}(b). This quantity does not represent the physical travel time of a payload but rather the sum of synchronization periods required for resonant alignment between orbital motion and tether rotation at each stage. These synchronization intervals determine when a transfer between successive tether stages becomes dynamically feasible. The number of available resonant elliptical nodes exhibits a strongly non-monotonic dependence on the stage number, as shown in Fig.~\ref{fig:elevator_results}(d). The initial stage contains a relatively small number of nodes (60), which is primarily due to the limited size of the initial tether family. In the subsequent stages, the number of nodes increases sharply to approximately 700–750 and then fluctuates around this level with moderate variations between stages. Overall, the results indicate that the solution space remains sufficiently rich across all stages, without a systematic depletion of resonant configurations at larger radii.

An important observation is that very large tethers are not required to maintain the chain. Although the search algorithm allowed tether half-lengths up to $500\,\mathrm{km}$, most selected tethers remain well below $250\,\mathrm{km}$. This indicates that moderate tether sizes could be sufficient to sustain resonance propagation through the system. 

As a sanity check, we also evaluated the specific orbital energy along the chain (Fig.~\ref{fig:elevator_results}(c)). The computed gravitational potential energy increases toward zero with radius, while the kinetic energy correspondingly decreases, consistent with the expected behavior of orbital mechanics in an Earth-centered inertial frame. The smooth variation of these quantities confirms the dynamical consistency of the generated tether sequence. Notably, the total specific energy increases along the chain, indicating a net upward energy transfer, consistent with mass and momentum exchange mechanisms.

Overall, these results demonstrate that a multi-stage resonant tether architecture can provide a dynamically feasible pathway from low Earth orbit to near-geostationary radii using moderate tether lengths and a sequence of resonance-driven transfers. The abundance of resonant elliptical nodes and the relatively short cumulative synchronization time suggest that such architectures may offer a promising framework for scalable orbital elevator systems.
\begin{figure*}[t]
\centering

\begin{subfigure}{0.48\textwidth}
\centering
\includegraphics[width=\linewidth]{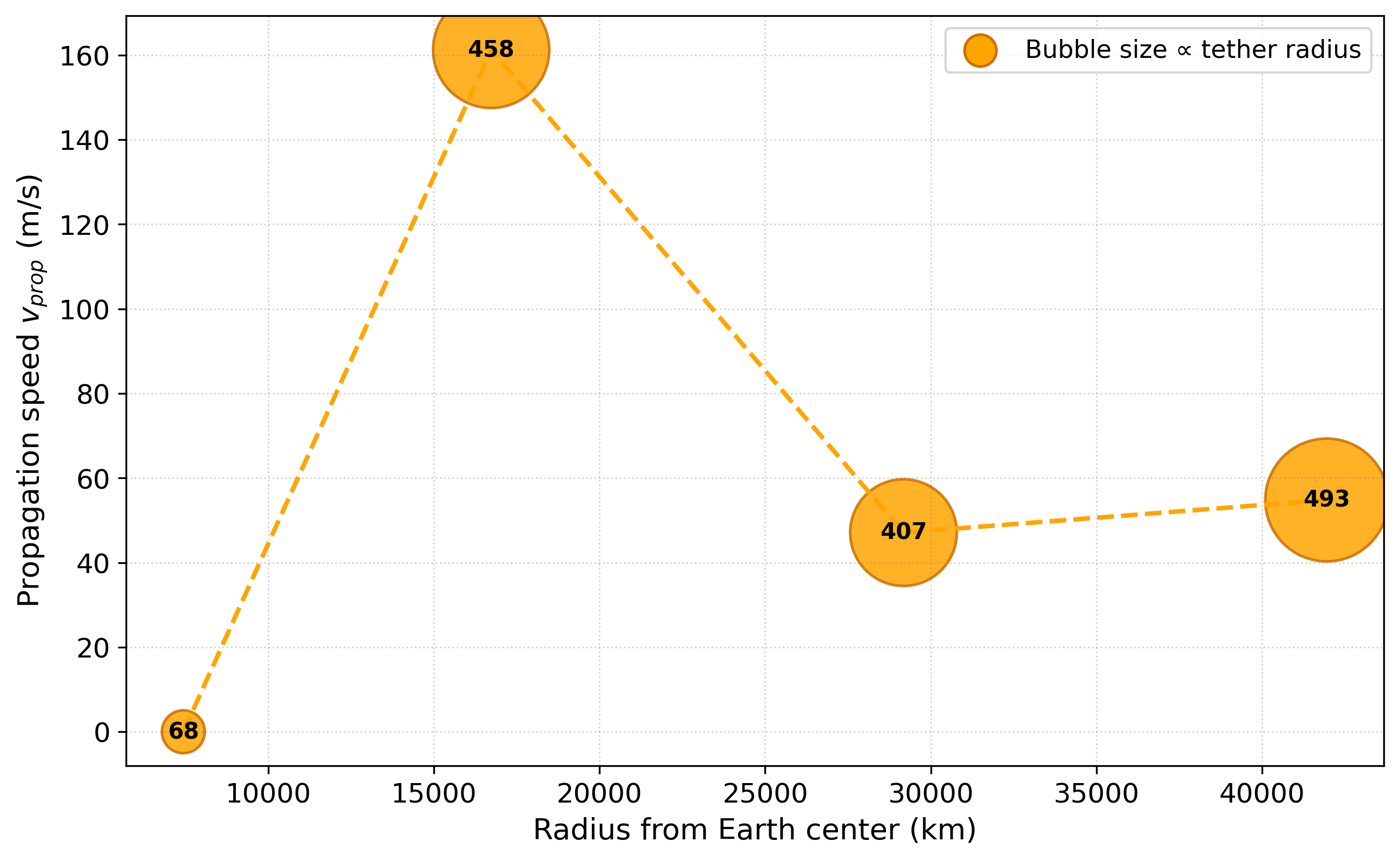}
\caption{Propagation speed between successive resonant tether stages as a function of orbital radius. Each bubble corresponds to a transfer between two stages of the elevator chain. The bubble size and the number inside represents the tether half-length (km).}
\end{subfigure}
\hfill
\begin{subfigure}{0.48\textwidth}
\centering
\includegraphics[width=\linewidth]{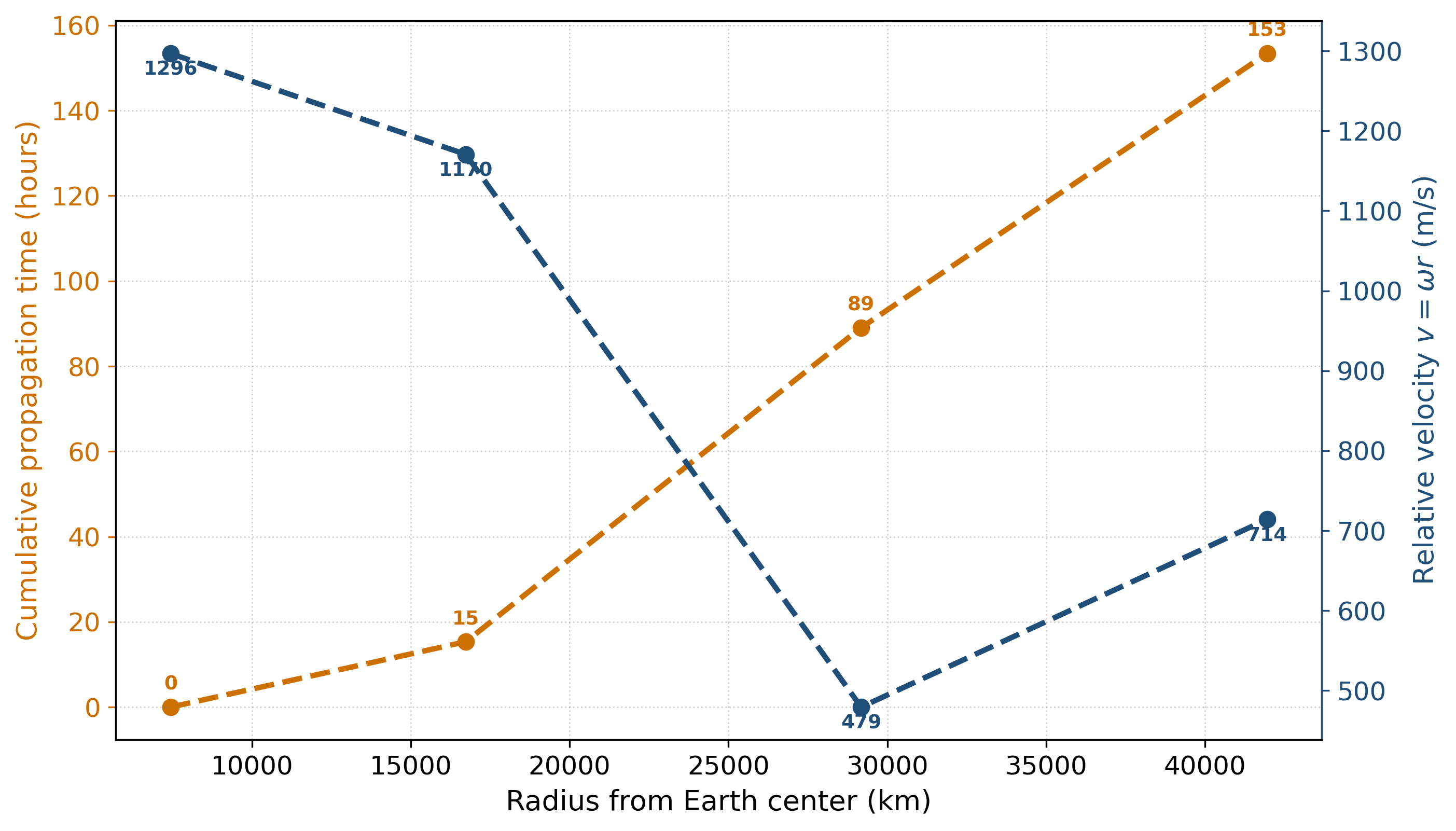}
\caption{Cumulative propagation time required to reach increasing orbital radii through successive resonant transfers, together with the rotational velocity relative to the tether’s center of mass. The time curve represents the total synchronization time accumulated across all preceding tether stages.}
\end{subfigure}

\vspace{0.4cm}

\begin{subfigure}{0.48\textwidth}
\centering
\includegraphics[width=\linewidth]{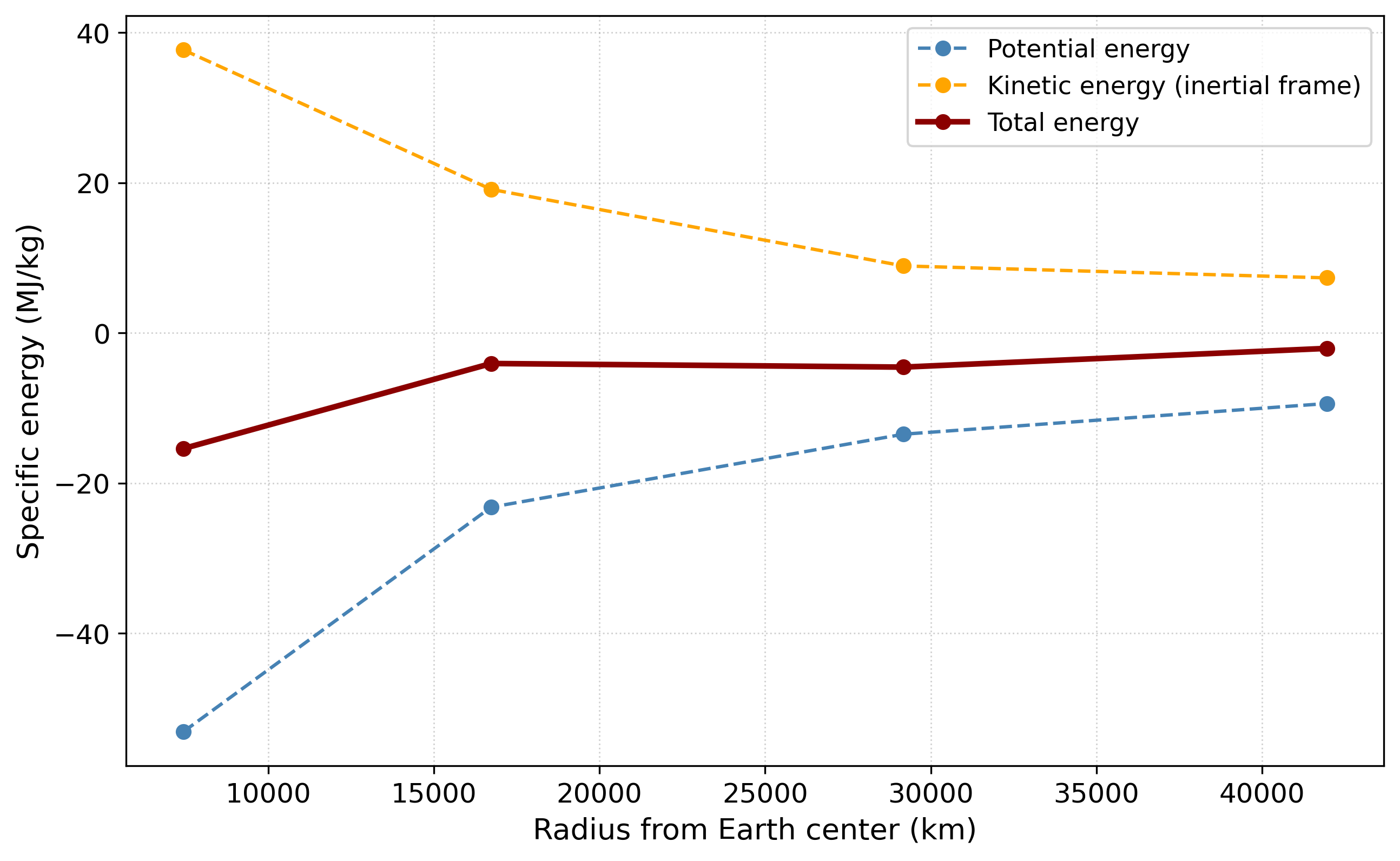}
\caption{Specific orbital energy components along the elevator chain. The gravitational potential energy increases toward zero with radius, while the kinetic energy decreases due to lower orbital velocity at larger radii.}
\end{subfigure}
\hfill
\begin{subfigure}{0.48\textwidth}
\centering
\includegraphics[width=\linewidth]{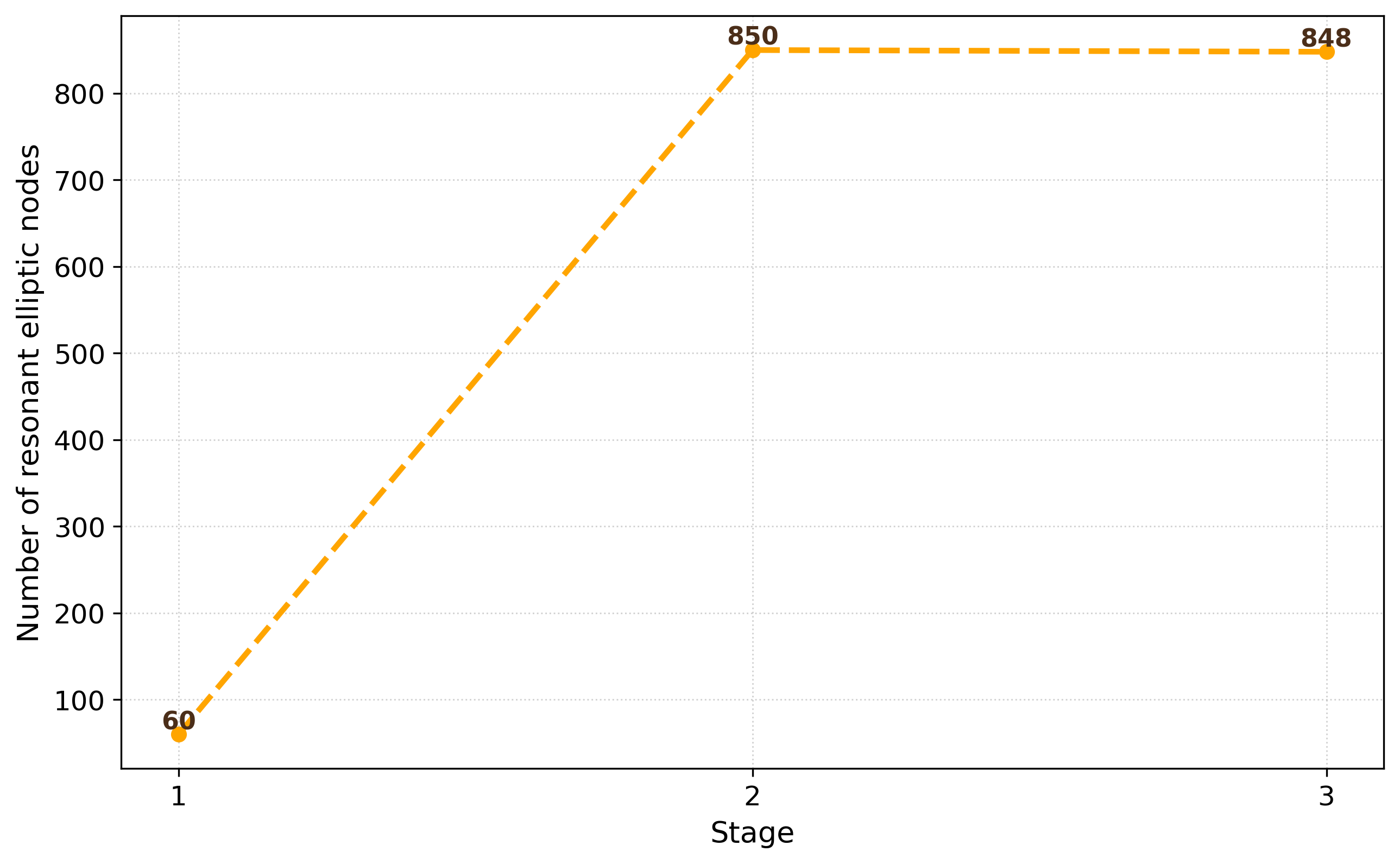}
\caption{Number of resonant elliptical nodes identified at each construction stage of the tether chain. The large number of available resonant nodes demonstrates the abundance of dynamically compatible transfer opportunities between stages. Stage 0 is omited.}
\end{subfigure}

\caption{Resonant multi-stage orbital elevator architecture obtained from the discrete resonance search algorithm. The panels illustrate the propagation dynamics between successive tether stages, cumulative synchronization time required for outward transport, the corresponding variation of specific orbital energy, and the number of available resonant elliptical nodes per stage. }
\label{fig:elevator_results_nolimit}
\end{figure*}
An important observation is that very large tethers are not required to maintain the chain. Although the search algorithm allowed tether half-lengths up to $500\,\mathrm{km}$, most selected tethers remain well below $250\,\mathrm{km}$. This indicates that moderate tether sizes are sufficient to sustain resonance propagation through the system. The increment limit unlimited.

\begin{figure}[t]
\centering
\begin{subfigure}[t]{0.48\textwidth}
\centering
\includegraphics[width=\linewidth]{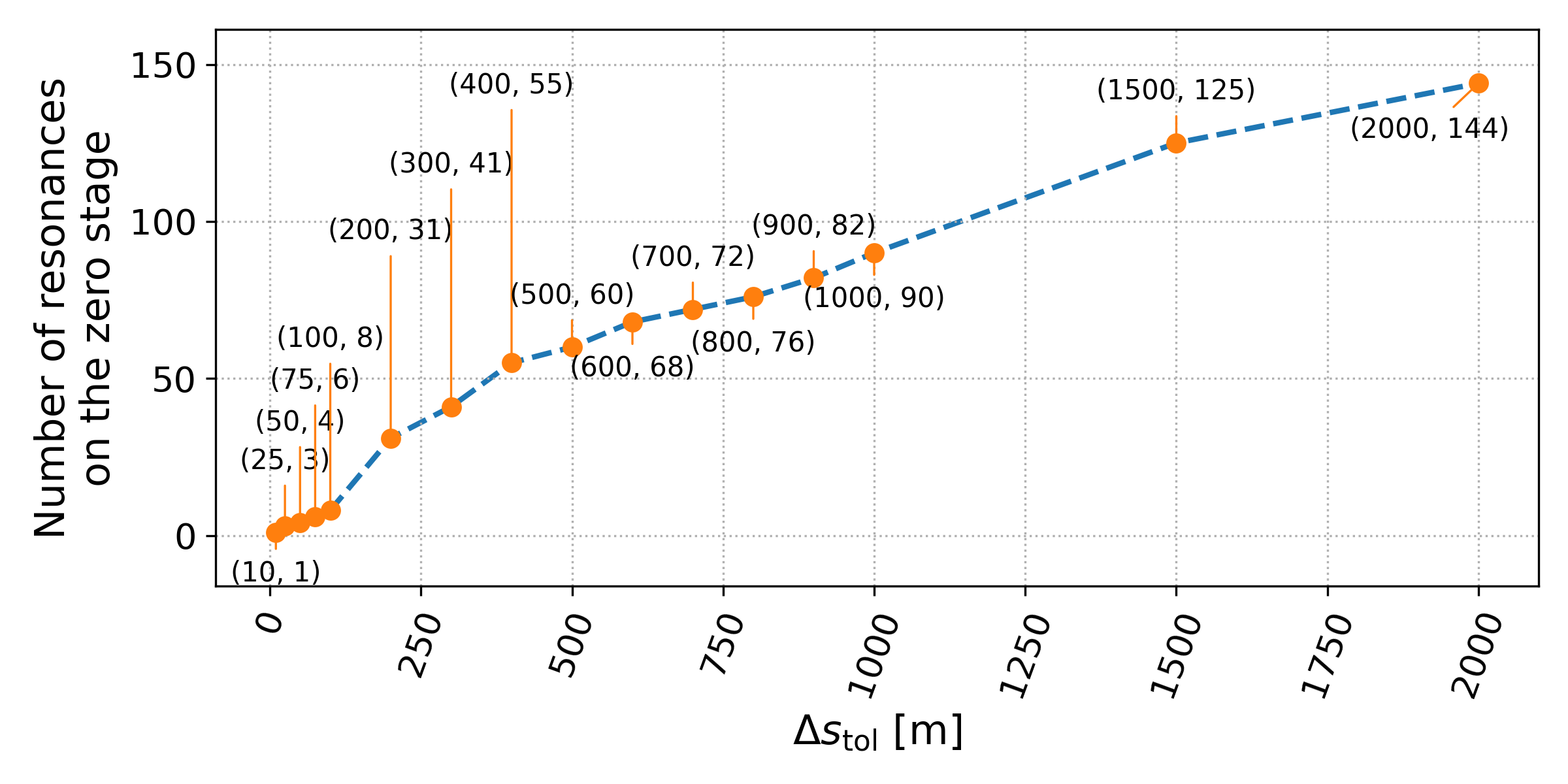}
\caption{Dependence of the number of resonances on the synchronization tolerance $\Delta s_{\mathrm{tol}}$.The left value is the tolerance in meters, the right value is the number of elliptical resonances.}
\label{fig:ablation_tol}
\end{subfigure}
\hfill
\begin{subfigure}[t]{0.48\textwidth}
\centering
\includegraphics[width=\linewidth]{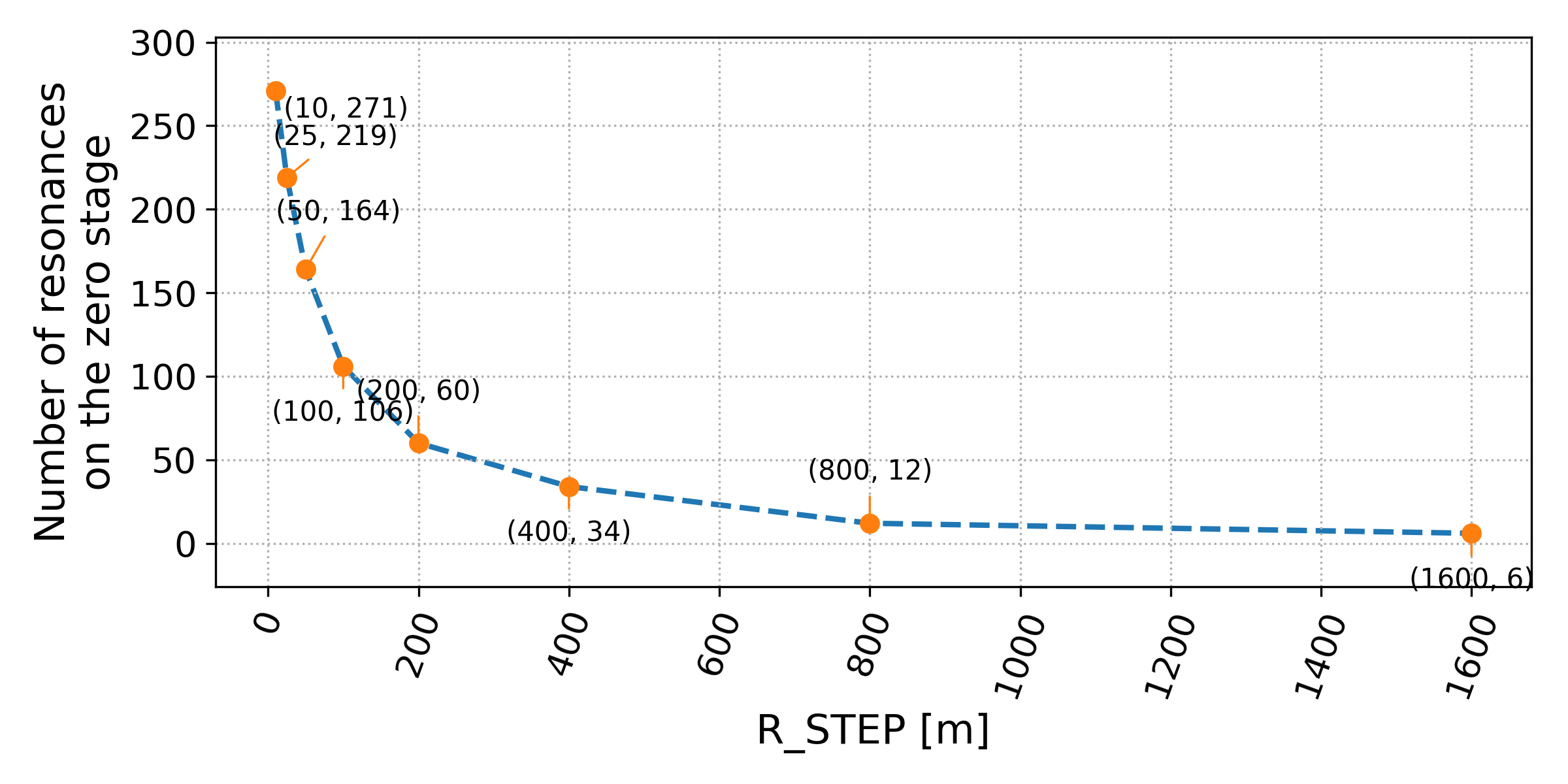}
\caption{Dependence of the number of resonances on the radial grid step $\Delta R$. The left number is the $\Delta R$ in meters, the right number is the number of elliptical resonances.}
\label{fig:ablation_step}
\end{subfigure}
\caption{Sensitivity analysis of the discrete synchronization search. The plots show how the number of detected resonances at the initial stage depends on the key discretization parameters of the algorithm.}
\label{fig:ablation}
\end{figure}
A notable limitation of the proposed architecture is the relatively low effective propagation speed of the payload along the tether chain. The average propagation speed obtained in the simulations is on the order of magnitude  $10$--$100\,\mathrm{m\,s^{-1}}$, which is significantly lower than the velocities typically achieved by conventional rocket-based transfer systems. As a result, the total propagation time across the full chain reaches hundreds of hours. It is important to emphasize that this time does not correspond to continuous propulsion but rather represents the cumulative synchronization time required for successive resonant transfers between rotating tethers.  It inherently results in longer delivery times compared to traditional chemical propulsion systems. A detailed numerical example of the constructed tether chain, including the parameters of the rotating tethers and the corresponding elliptical synchronization nodes, is provided in Appendix C.

\paragraph{Resonance tether chain construction with unlimited increments.}

In the second experiment, a maximum increment of $\Delta R = 37{,}000$ was used. Thus, it becomes possible to construct a terminal solution with a single tether. The results are consistent with those obtained in the previous experiments and are shown in Figure~\ref{fig:elevator_results_nolimit}. However, we observe that the number of tethers required is significantly smaller, while the terminal rotational velocity is higher. The reduction in tether count is expected, since the absence of a strict constraint on $\Delta R$ allows the algorithm to place tethers at larger radial distances in a single step. This effectively removes the need for intermediate configurations. More interestingly, the increase in terminal velocity is non-trivial. It can be explained by the fact that larger radial placements enable more efficient angular momentum accumulation. When tethers are placed farther from the center, the tangential velocity is inherently higher, and each placement contributes a larger increment to the system’s total angular momentum. In contrast, constrained step sizes force the system to evolve through a sequence of locally suboptimal configurations. A detailed numerical
example of the constructed tether chain for unlimited increment, including the parameters of the rotating tethers and the corresponding elliptical synchronization nodes, is provided in Appendix C.

\paragraph{Computational Timing.}

The computational cost of the propagation and tether search procedure is substantial due to the combinatorial nature of the candidate space. The full procedure requires approximately $\sim 40$ hours for the case of unlimited increments and $\sim 20$ hours for limited increments (for $R_{\max} = 500\,\text{km}$) on a single CPU core (Apple M1). At the level of a single stage, the runtime is on the order of several hours, ranging from $\sim 2$ hours for limited increments up to $\sim 10$ hours for the unlimited increment case. 

The dominant computational cost arises from evaluating candidate configurations at each stage, which are generated through the propagation of elliptical trajectories. This is followed by a significant reduction step involving resonance filtering, feasibility checks, and duplicate removal. Typically, on the order of $10^6$--$10^7$ elliptical orbits are identified (depending on the stage), which are then reduced to approximately $10^3$ unique resonant candidates. It is worth noting that the current simulations are not globally optimal and retain only a beam size of $15$ candidate tethers (with the highest propagation velocity) for further simulation; therefore, the obtained solutions are generally suboptimal.

The computational cost exhibits approximately linear scaling with the number of candidate nodes and inverse scaling with respect to the propagation step size. Consequently, reducing the step size or increasing angular resolution leads to a substantial increase in runtime. In the present implementation, an elliptical propagation step of $5000\,\text{m}$, an angular discretization of $1200$ samples in $\phi$, and most importantly a beam size of $15$ were used to maintain tractability. A finer discretization would likely reveal additional feasible configurations, at the expense of significantly increased computational cost.
\paragraph{Sensitivity analysis.}
To verify that the existence of feasible transport chains is not an artifact of the synchronization tolerance, additional tests were performed using a substantially stricter spatial tolerance of $10\,\mathrm{m}$, which is comparable to the numerical resolution of the trajectory integration. Even under this stringent condition, valid synchronization configurations were still identified. This observation indicates that the existence of solutions does not rely on the larger tolerance value used in the main search procedure. The tolerance $\Delta s_{\mathrm{tol}} = 500\,\mathrm{m}$ is therefore introduced primarily as an operational and computational margin rather than as a requirement for dynamical feasibility. The results are presented in Figure~\ref{fig:ablation}(a).

The sensitivity analysis was conducted to assess how the number of discrete synchronization
resonances detected at the initial stage depends on the key hyperparameters of the search procedure.
In particular, we varied the tether grid step size $\Delta R$ (introduced in Eq.~\ref{eq:grid_step}) and the
resonance congruence tolerance $\Delta s_{\mathrm{tol}}$ defined by the synchronization condition
in Eq.~\ref{eq:res_tol}.  The results show that the number of detected resonances increases monotonically with $\Delta s_{\mathrm{tol}}$,
as larger tolerance values admit a broader class of near-resonant configurations satisfying the synchronization condition.
This growth is approximately sublinear, indicating diminishing returns at larger tolerance values.  Conversely, increasing the grid step size $\Delta R$ significantly reduces the number of detected resonances,
since a coarser discretization of the tether configuration space limits the number of candidate configurations
capable of satisfying the resonance constraints. A similar dependence is expected at subsequent stages of
the propagation process, as each stage relies on the similar discrete synchronization conditions between
orbital motion and tether rotation. The results are presented in Figure~\ref{fig:ablation}(b).

\section{Discussion}

\subsection{Requirement for a large mass on the final orbit}
\label{subsec:atm}
The most significant limitation of the proposed system is the requirement for a
large mass reservoir in a high Earth orbit, whose  energy
is exploited to enable sustained upward payload transfer.
In the present formulation, this mass reservoir is assumed to be effectively unlimited,
rendering the concept primarily theoretical at current technological readiness
levels.

One possible mitigation strategy is the emplacement
of asteroid-derived material into a bound Earth orbit
or into a suitable orbit within the Earth--Moon system.

Previous studies have investigated the feasibility
of retrieving near-Earth asteroid material using
long-duration low-thrust propulsion \cite{Brophy2012},
as well as the accessibility of asteroid resources
in near-Earth space \cite{SanchezMcInnes2011}.

The availability of suitable asteroid resources
depends on their composition, abundance, and
orbital accessibility \cite{Elvis2014,SanchezMcInnes2011}.

Solar electric propulsion represents a potential
technology for gradual orbital modification
\cite{Landis2015}.

However, the feasibility of establishing a mass
reservoir of the scale required by the proposed
architecture remains an open engineering question
and is not established by the present analysis. An alternative mitigation strategy is to exploit the Moon itself as a long-term
mass reservoir, where the Moon offers an effectively vast and accessible source of material.
The stability, control, and station-keeping of large mass reservoirs
remain nontrivial challenges, as do the logistical and energetic costs associated
with sustained lunar launch operations. 

A key distinction of lunar-based implementations is the absence of an atmosphere,
which substantially relaxes several constraints that dominate Earth-based tether
systems.
Atmospheric drag and aerodynamic heating are entirely
absent, allowing tether endpoints to be placed significantly closer to the lunar
surface.
This, in turn, enables alternative mass-transport mechanisms that do not rely on
vertical rocket launch.
In principle, material could be accelerated predominantly in the horizontal
direction using surface-based electromagnetic launchers, mass drivers, or other
non-chemical propulsion systems, injecting payloads directly into
tether-compatible trajectories.
Related transportation concepts include lunar tether
systems and electromagnetic mass drivers
\cite{Forward1991,ONeill1981}.
The reduced lunar gravity and negligible atmosphere
provide a physical motivation for investigating
such non-chemical surface-to-orbit transportation
mechanisms.

While these approaches remain speculative and face substantial engineering
challenges, the combination of atmosphere-free operation, reduced gravity, and
continuous mass exchange suggests that lunar-based tether-elliptical transport
architectures may operate under fundamentally relaxed constraints than their
terrestrial counterparts.
As a result, the Moon may represent a natural anchor point for large-scale,
long-term orbital mass transfer systems.

\subsection{Placement of the lowest tether and atmospheric drag}
\label{sec:lowest_tether_drag}

The lowest tether in the proposed architecture cannot be placed arbitrarily close to Earth,
as aerodynamic drag introduces non-conservative energy losses that violate the idealized
Keplerian assumptions adopted in this work.
Although lowering the first tether would reduce the orbital energy gap between the surface
and the initial operational orbit, atmospheric friction increases rapidly with decreasing
altitude and imposes a practical lower bound on the usable regime.

For an object moving with velocity \(v\) relative to the atmosphere, the aerodynamic drag force
scales as \cite{NASAGlennDrag,BateMuellerWhite}
\begin{equation}
F_{\mathrm{d}} \propto \rho_{\mathrm{atm}}(h)\,v^{2},
\end{equation}
with an associated power dissipation
\begin{equation}
P_{\mathrm{d}} \propto \rho_{\mathrm{atm}}(h)\,v^{3},
\end{equation}
where \(\rho_{\mathrm{atm}}(h)\) is the atmospheric density at altitude \(h\).
In near-circular orbit, the orbital velocity of the
tether center of mass scales as
\(v \sim \sqrt{\mu/R}\)
\cite{BateMuellerWhite}, whereas reference atmospheric
density decreases rapidly with increasing altitude
\cite{USStdAtm1976}.

Actual thermospheric densities also depend on solar
and geomagnetic activity. The adopted minimum
altitude should therefore be interpreted as a
simplifying modeling constraint rather than a
universal boundary for negligible atmospheric drag.

As a consequence, drag-induced energy dissipation grows rapidly for low-altitude orbits,
and sustained motion below a few hundred kilometers requires continuous energy input to
maintain orbital altitude.
Such non-conservative effects are incompatible with the lossless mass-exchange concept
considered in the present analysis.
For this reason, we restrict attention to orbital radii above approximately
\(1000~\mathrm{km}\), where atmospheric density is sufficiently low that drag effects are
negligible over orbital timescales and purely gravitational dynamics provide an accurate
first-order description.

This altitude cutoff is introduced as a modeling assumption rather than an optimized design
choice and serves only to delimit the regime in which the dynamical feasibility analysis
remains valid.

\subsection{ Reliability and modularity considerations.}
The reliability of orbital tether systems depends on
environmental exposure, structural complexity, deployment
requirements, and operational conditions.
The hazards associated with orbital debris and collisions
in Earth orbit are well documented
\cite{Kessler1978,LiouJohnson2006,Johnson2010}.
For extended structures such as space tethers, collision
exposure represents an important engineering consideration.
Tether deployment, tension management, and dynamical
stability introduce additional engineering challenges
\cite{BeletskyLevin1993,CartmellMcKenzie2008}.

A modular architecture composed of multiple shorter
tethers distributed across adjacent orbits may offer
advantages in manufacturing, deployment, monitoring,
maintenance, and component replacement.
However, increasing the number of individual tethers
also introduces additional interfaces, synchronization
requirements, and potential failure modes.

Deploying several tethers on the same orbit with
controlled phase offsets could provide operational
redundancy, potentially allowing payload traffic
to be redistributed among the remaining units
if an individual tether becomes inoperable.
Such a configuration may facilitate scheduled
maintenance and improve fault tolerance, provided
that suitable synchronization and traffic
redistribution mechanisms are available.

A quantitative comparison of system-level reliability,
collision exposure, maintenance costs, and operational
availability is beyond the scope of the present study.

\subsection{Third-body gravitational perturbations}

The orbital dynamics of the tethers are primarily governed by Earth’s gravity; however, gravitational perturbations from other Solar System bodies, most notably the Moon and the Sun, introduce additional long-term effects. These third-body forces give rise to long-period variations in the orbital elements, including nodal precession and apsidal rotation, as well as small oscillations in eccentricity. In addition, differential third-body gravitational acceleration along the tether length may induce small but persistent tidal stresses that modulate the internal tension distribution.

In the present analysis, third-body perturbations were not explicitly included, as the focus was on
first-order feasibility and the internal dynamics of the multi-tether system under idealized Keplerian
conditions.
Nevertheless, the treatment of lunar and solar perturbations is well established within classical celestial
mechanics and can be incorporated systematically.

At the analytical level, third-body effects may be introduced through a disturbing function
\(\mathcal{R}_{3b}\), added to the gravitational potential of the Earth-centered two-body problem.
For a perturbing body of mass \(m_p\) at position \(\mathbf{r}_p\), the disturbing potential acting on a tether
element at position \(\mathbf{r}\) is
\begin{equation}
\mathcal{R}_{3b} =
G m_p
\left(
\frac{1}{|\mathbf{r}_p - \mathbf{r}|}
-
\frac{\mathbf{r}\cdot\mathbf{r}_p}{|\mathbf{r}_p|^3}
\right),
\end{equation}
where the second term removes the indirect acceleration of the Earth-centered reference frame.
Orbit-averaging of \(\mathcal{R}_{3b}\) over the fast orbital motion yields secular evolution equations for the
orbital elements, which can be treated using standard Lagrange planetary equations.

For near-circular and near-equatorial orbits, the dominant secular effects of lunar and solar perturbations
manifest as slow precession of the orbital plane, including variation of the longitude of the ascending node. 
The characteristic timescales of these variations are typically weeks to months for low Earth orbit and
months to years for medium and high Earth orbits.
These timescales are significantly longer than the characteristic rotational periods of the tethers and
the synchronization times of the elliptical nodes, implying a natural separation of timescales.

From a numerical perspective, third-body perturbations can be incorporated directly by extending the
equations of motion to include lunar and solar point-mass accelerations, using ephemerides for the
Moon and Sun.
Numerical propagation using Runge–Kutta integrators allows
accurate long-term simulation of the coupled orbital and rotational dynamics without modifying the
underlying tether model.

Operationally, the slow and predictable nature of third-body perturbations enables several mitigation
strategies.
First, orbital phasing can be selected such that nodal precession rates are synchronized or averaged across
neighboring tethers, reducing relative drift within the multi-tether system.
Second, modest active control such as small corrective torques applied through spin-rate adjustment of the tether system can compensate for accumulated phase errors.
Third, periodic orbit maintenance maneuvers at the level of the tether centers of mass can be employed to
re-circularize or re-phase the orbits over long operational intervals.

Importantly, because the Space-Clock Elevator architecture relies on relative synchronization between
neighboring tethers rather than absolute inertial pointing, global perturbations that affect all system elements in a similar manner are largely benign for the system architecture.
As long as third-body effects remain secular and slowly varying, they can be absorbed into the timing and
synchronization framework without compromising the lossless mass-exchange mechanism.

\subsection{Elliptical Nodes as Dynamic Buffers for Tether Operations}

Elliptical nodes can significantly improve the operational robustness of rotating tether systems by acting as intermediate dynamic elements between the primary tethers. One of their key advantages is that they relax the requirement of strict synchronization between multiple tethers. Perfect synchronization of rotating tethers is difficult in practice due to the need to simultaneously synchronize multiple dynamical parameters. Elliptical nodes provide a flexible interface that allows different tether systems to operate with slight phase offsets while still maintaining coordinated exchange operations.

In addition to mitigating synchronization constraints, elliptical nodes serve as the primary locations for active trajectory adjustment. Directly modifying the trajectory or phase of a rotating tether is mechanically complex and energetically costly, since the tether is a distributed system with significant inertia and elastic coupling. Elliptical nodes, by contrast, are comparatively compact structures whose orbital parameters can be adjusted more easily through small propulsion corrections. By performing most trajectory corrections at the elliptical nodes rather than on the tether itself, the overall system can maintain stable tether rotation while still allowing controlled adjustments of timing and geometry for payload transfer.

Another operational advantage arises during payload exchange. Capture and release events rarely occur at perfectly optimal instants, and even small timing deviations can generate reverse oscillations along the tether, producing transient tension spikes and destabilizing the rotational dynamics. Elliptical nodes act as dynamic buffers that moderate the propagation of these disturbances. This buffering capability makes it possible to perform tether-node exchange operations and the corresponding system-level payload exchange nearly simultaneously, without requiring extremely precise timing. The nodes therefore reduce sensitivity to phase mismatches and allow attachment or release events that are only approximately synchronized to occur without inducing large oscillatory responses.

Taken together, elliptical nodes function both as mechanical buffers and as operational control points. They decouple the strict timing requirements of tether rotation from the broader logistics of orbital transfer.

\subsection{Stability of mass exchange}

In the present formulation, mass exchange occurs directly between a tether endpoint
and an elliptical node.
At the exchange instant, the two coincide in position and inertial velocity, so that
attachment and detachment occur without impulsive momentum transfer.
The elliptical node subsequently accommodates differences in orbital radius, angular
velocity, and phase between neighboring tethers through its Keplerian motion, acting
as a dynamical buffer for the transfer process.
Nevertheless, even in the absence of impulsive momentum
transfer, the redistribution of mass at the tether
endpoint may modify its effective mechanical boundary
conditions.

The general framework for investigating the coupled
dynamics and control of tethered systems is established
in the literature
\cite{BeletskyLevin1993,Misra2008,
ModiLakshmananMisra1992,Kalantzis1998}.

Within this framework, changes in endpoint loading
may excite internal dynamical modes of the tether.
Such potential exchange-induced disturbances would
be distinct from the regular periodic tension
oscillations associated with gravity-gradient
and rotational dynamics.

Their amplitudes, frequencies, and propagation
characteristics would depend on the tether elasticity,
damping, endpoint mass distribution, and attachment
mechanism.

A detailed analysis of wave excitation, transient
stress amplification, and their interaction with
tether elasticity and damping requires an extensible
tether model and is therefore beyond the scope
of the present study.

\subsection{Near-lossless energy and momentum exchange}
\label{subsec:lossless}
The Space-Clock Elevator architecture is designed to enable orbital mass transfer with
minimal propulsive energy expenditure by exploiting controlled redistribution of
gravitational potential energy.
Nevertheless, the system is not strictly lossless.
All exchange processes are subject to finite geometric and kinematic tolerances, as
well as unavoidable perturbations, which necessitate active correction.

In the present formulation, synchronization between tether rotation and elliptic-node
motion is enforced only approximately.
Phase alignment is required to be satisfied within a prescribed spatial tolerance,
taken throughout this work as
\begin{equation}
\Delta s_{\mathrm{tol}} = 500~\mathrm{m}.
\end{equation}

Exchange events therefore occur within bounded spatial and temporal offsets rather
than at exact phase coincidences.
Reducing this tolerance requires increasingly accurate rational approximations between
tether rotation and node orbital frequencies, which in turn leads to longer
synchronization intervals and increased total delivery times.

Even in the limit of long synchronization times, exact zero tolerance cannot be
achieved in practice.
Residual phase drift arising from finite integer commensurability, numerical limits,
and external perturbations prevents perfect alignment.
As a consequence, continuous or intermittent adjustment of tether rotation rates and
phases is required to maintain repeated near-alignment, implying nonzero auxiliary
energy input.

In addition, the elliptical transfer nodes must actively compensate for residual mismatches
in position and velocity at attachment and detachment.
Although nominal trajectories are constructed to satisfy velocity-matching
conditions, small deviations due to synchronization tolerances, numerical
approximations, and perturbations require corrective maneuvers.
These corrections are naturally localized at the elliptical nodes, where control is
simplest, but they nevertheless contribute additional energy expenditure.

The system should therefore be regarded as \emph{near-lossless} rather than strictly
conservative.
The dominant energy exchange is provided by the controlled descent of counter-masses,
while auxiliary propulsion is required to maintain synchronization, enforce geometric
tolerances, and compensate phase drift.
Quantifying the magnitude of these correction costs relative to the transported
payload energy is an important topic for future work and lies beyond the scope of the
present feasibility analysis.
\subsection{Active correction under finite synchronization tolerance}
\label{subsec:active-correction}

Even in the absence of external perturbations, the Space-Clock Elevator architecture
cannot operate as a strictly passive or perfectly conservative system.
This limitation does not arise from deficiencies in the underlying equations of motion,
but rather from the necessity of enforcing repeated high-precision mass exchange events
under finite geometric and temporal tolerances.

Throughout this work, synchronization between tether rotation and elliptic-node motion
is enforced only approximately through near-rational frequency relationships.
For any finite integer bounds on the commensurability parameters
$(p_i,q_i,l_i,m_i)$, exact phase locking is impossible.
As a consequence, relative phase drift between tether endpoints and elliptical nodes
accumulates over time, even under ideal Keplerian dynamics.
Mass exchange is therefore required to occur within a prescribed spatial tolerance
$\Delta s_{\mathrm{tol}}$, rather than at exact phase coincidence.

This finite tolerance introduces an inherent requirement for active correction.
Although nominal trajectories are constructed such that endpoint position and inertial
velocity match at the exchange point to leading order, residual mismatches in both
position and velocity are unavoidable.
These mismatches arise from bounded synchronization error, numerical discretization,
and the finite resolution of rational frequency approximations.
As a result, purely ballistic attachment and detachment without any corrective action
cannot be sustained indefinitely.

Active correction is therefore an intrinsic component of the system architecture,
even under idealized two-body dynamics.
Because synchronization is enforced only within a finite tolerance, repeated exchange
events require small corrective actions to prevent secular phase drift and ensure
reliable capture.

Correction authority is distributed between both the elliptical transfer nodes and the
tethers.
The elliptical nodes provide local, event-level correction by applying small $\Delta v$
adjustments to trim their Keplerian trajectories and null residual relative velocity
at docking.
The tethers, by contrast, provide \emph{mandatory systematic correction} arising from
the finite synchronization tolerance: small but persistent phase errors accumulate over
successive exchange cycles and must be compensated through controlled adjustment of
tether spin rate, phase, and, when required, center-of-mass orbital motion.
Together, these mechanisms maintain long-term alignment within the prescribed tolerance
without requiring exact resonance.

The purpose of active correction of the tether is to eliminate accumulated phase error rather than
to correct velocity per se.
A finite synchronization tolerance corresponds to an admissible along-track phase
mismatch of order $\Delta s_{\mathrm{tol}}$, which must be actively removed prior to
exchange. Phase correction is achieved by temporarily modifying orbital or rotational parameters
(e.g., elliptic-node mean motion or tether spin rate), allowing the required phase
advance or delay to accumulate over time.
The associated velocity adjustments serve only as a control mechanism to generate
the desired phase drift and need not persist after alignment is restored.

The necessity of active correction does not undermine the core premise of the
Space-Clock Elevator.
The dominant energy transfer mechanism remains gravitational and conservative, with
active control serving only to maintain alignment within finite tolerances.
The system should therefore be regarded as near-lossless rather than strictly lossless.
Quantifying the cumulative correction cost and its impact on net transport efficiency
is an important topic for future work and lies beyond the scope of the present
feasibility analysis.

\subsection{Inextensibility}
\label{subsec:inext}

In the present feasibility analysis, the tether is treated as \emph{inextensible}, meaning
that its arclength is assumed constant and axial elasticity is neglected.
This approximation is justified when elastic strain variations induced by dynamical
tension fluctuations remain small, so that changes in tether length do not significantly
affect endpoint kinematics, phase synchronization, or docking geometry.
While a real tether undergoes finite static extension under load, only the \emph{relative}
length variations associated with tension oscillations are dynamically relevant for the
mass-exchange process considered here.

For a linearly elastic tether, the axial strain is related to the internal tension by
\begin{equation}
\varepsilon = \frac{\sigma}{E} = \frac{T}{A E},
\end{equation}
where \(E\) is Young’s modulus and \(A\) is the tether cross-sectional area.
A small tension variation \(\Delta T\) about the mean tension \(T_{\mathrm{mean}}\) therefore
induces a strain variation
\begin{equation}
\Delta \varepsilon = \frac{\Delta T}{A E}.
\end{equation}

Using the admissibility constraints adopted in this work,
\begin{equation}
T_{\mathrm{mean}} \le T_{\max}(A) = \sigma_{\mathrm{allow}} A,
\end{equation}
and
\begin{equation}
\frac{\Delta T}{T_{\mathrm{mean}}} \le \alpha,
\qquad
\alpha = 0.1,
\end{equation}
the maximum strain variation is bounded by
\begin{equation}
\Delta \varepsilon_{\max}
\le
\alpha\,\frac{\sigma_{\mathrm{allow}}}{E}.
\end{equation}

For high-modulus Zylon (PBO fiber), the nominal Young's
modulus is \(E \simeq 270~\mathrm{GPa}\)
\cite{Toyobo2005}.
Using the allowable tensile stress adopted in the present
study, \(\sigma_{\mathrm{allow}} = 1.16~\mathrm{GPa}\),
as defined in Section~\ref{subsec:tether-tension},
this yields

For a representative tether length of \(L = 1000~\mathrm{km}\),
the corresponding fluctuation-induced change in length is
\begin{equation}
\Delta L_{\max} = L\,\Delta \varepsilon_{\max}
\approx 4.3\times 10^{2}~\mathrm{m},
\end{equation}
which remains below the synchronization tolerance
\(\Delta s_{\mathrm{tol}} = 500~\mathrm{m}\) adopted in this study.

Importantly, this bound is independent of tether radius and mass, as the cross-sectional
area cancels when expressed in terms of the normalized tension variation
\(\Delta T/T_{\mathrm{mean}}\).
Consequently, under the imposed tension and fluctuation constraints, axial elasticity
introduces only small, bounded corrections to the tether geometry.
The inextensible-tether approximation is therefore self-consistent within the scope of
the present first-order feasibility analysis for tethers of the limited length.

Small variations in tether length modify the angular velocity distribution along the
tether and can therefore alter the precise release position and release velocity of the
payload. While these effects are expected to remain secondary at the level of the
present analysis, they may become relevant for high-precision synchronization and
should be examined in future studies using fully elastic tether models. Although the estimated length variations remain small relative to the total tether length and therefore do not affect the feasibility analysis presented here, they may become significant for the precise release trajectory of the payload. Variations in tether length modify the angular velocity distribution along the tether and can therefore introduce deviations in the release position and velocity. A more accurate analysis including tether extensibility will be addressed in future work.

\subsection{Tether tension and radius constraints}
\label{subsec:tether-tension}

Any rotating orbital tether is subject to finite tensile
strength and mass constraints, which together restrict
the admissible combinations of tether radius, length,
and rotation rate.

High-strength materials have been investigated in the
context of space elevators and orbital tether systems
\cite{Edwards2003,vanPelt2009}.

To quantify the mechanical constraints in the present
feasibility analysis, we adopt the nominal mechanical
properties of high-modulus Zylon (PBO fiber) as a
representative high-strength material reference
\cite{Toyobo2005}.

Throughout this section, Zylon parameters are used solely
as a numerical scale for defining bounded-stress criteria
in the feasibility maps. Their use does not imply
manufacturability, deployment feasibility, or operational
survivability of tethers at the corresponding dimensions.

According to the manufacturer's technical documentation
\cite{Toyobo2005}, the nominal material properties of
high-modulus Zylon fibers are
\begin{equation}
\sigma_{\mathrm{ult}} \approx 5.8~\mathrm{GPa},
\qquad
\rho \approx 1560~\mathrm{kg\,m^{-3}},
\qquad
E \approx 270~\mathrm{GPa},
\end{equation}
where $\sigma_{\mathrm{ult}}$ denotes the ultimate tensile
strength, $\rho$ the volumetric mass density, and $E$
the Young's modulus. These nominal fiber properties
provide reference values rather than experimentally
validated effective properties of a full-scale tether.

\paragraph{Static strength constraint.}
To introduce a nominal margin against material uncertainty,
manufacturing defects, and potential strength degradation,
we restrict the operational stress to a fraction of the
ultimate tensile strength,
\[
\sigma_{\mathrm{allow}} = \eta\,\sigma_{\mathrm{ult}},
\qquad
\eta = 0.2,
\]
corresponding to a nominal safety factor of five.
This value is adopted as a conservative numerical
screening parameter rather than a certified engineering
design limit. Environmental degradation of polymer-based
space tether materials, including Zylon, has been
documented experimentally \cite{Gittemeier2005}.
For a tether with circular cross-section $A=\pi r^{2}$, the corresponding maximum allowable
static tension is
\begin{equation}
T_{\max}(A) = \sigma_{\mathrm{allow}}\,A.
\label{eq:static_t_limit}
\end{equation}

This relation defines an upper bound on admissible internal tension for a given tether radius.
Increasing the radius raises the allowable tension linearly, but simultaneously increases
the tether mass per unit length and therefore the gravity-gradient and centrifugal loading.
The resulting trade-off between strength and self-weight is intrinsic to all rotating
tether systems and is naturally captured in the numerical feasibility maps.

\paragraph{Constraint on tension fluctuations.}
In addition to the absolute tension level, large cyclic variations in tension are dynamically
undesirable. To exclude strongly oscillatory solutions, we impose a bound on the normalized
post-transient tension variation,
\begin{equation}
\frac{\Delta T}{T_{\mathrm{mean}}} \le \alpha,
\qquad
\alpha = 0.1.
\label{eq:dT_ratio_limit}
\end{equation}

Experimental investigations of high-modulus PBO fibers
under cyclic tensile loading have examined the
relationship between maximum applied stress and
fatigue life, demonstrating a reduction in fatigue
strength following ultraviolet irradiation
\cite{Horikawa2012Fatigue}.

The threshold $\alpha=0.1$ is introduced here as a
phenomenological dynamic admissibility criterion rather
than an experimentally calibrated fatigue limit.
Its purpose is to exclude configurations exhibiting
large cyclic tension fluctuations and to identify
bounded-load solutions under idealized dynamics.

A quantitative prediction of fatigue life would require
material-specific stress--life relationships, loading
histories, environmental degradation models, and
experimental validation at the cable level. Such an
analysis is beyond the scope of the present study.

\paragraph{Role in feasibility maps.}
Equations \eqref{eq:static_t_limit} and \eqref{eq:dT_ratio_limit} together define the
mechanical admissibility criteria used to classify numerical solutions.
Parameter combinations satisfying both constraints are labeled as feasible in the maps
presented above. These maps therefore answer the question of \emph{existence} of bounded-stress
solutions under Zylon-scaled limits, rather than asserting engineering realizability.

\subsection{Results}

An important feature of the constructed architecture is the large number of feasible resonance families available at each stage of the propagation process. During the search procedure, typically more than one hundred candidate synchronization configurations were detected for a given stage. To control the combinatorial growth of the search tree, these candidates were filtered using a beam search strategy, retaining only the best $15$ configurations for further propagation.  Despite this strong pruning of the candidate set, the algorithm was still able to construct a complete transfer chain reaching the target orbital radius. This indicates that the solution does not rely on a single narrowly tuned resonance sequence but rather emerges from a large set of dynamically compatible resonance families. The present construction relies on initial conditions specified primarily from the inner boundary of the system, corresponding to the lowest operational tether. From this side, the sequence of admissible elliptical nodes and receiving tethers is generated through the synchronization and feasibility constraints described above. Consequently, the families of solutions obtained in the simulations are anchored by the dynamical conditions imposed at the lower orbital boundary.

The present construction relies on initial conditions specified primarily from the inner boundary of the system, corresponding to the lowest operational tether. From this side, the sequence of admissible elliptical nodes and receiving tethers is generated through the synchronization and feasibility constraints described above. Consequently, the families of solutions obtained in the simulations are anchored by the dynamical conditions imposed at the lower orbital boundary.

In contrast, the outer boundary of the architecture is currently treated in a more heuristic manner. The properties of the final tether, including its rotation rate and orbital radius, are selected empirically so as to produce sufficiently slow endpoint motion and to facilitate capture by the preceding stage. While this approach is adequate for demonstrating the existence of dynamically feasible transport chains, it does not yet impose a fully consistent boundary condition from the outermost orbit. A more rigorous formulation would incorporate constraints from both ends of the transport chain simultaneously, allowing the construction to converge toward solutions that satisfy boundary conditions at both the initial and final orbital radii.

Another important limitation concerns the estimation of the total transfer time. In the present formulation, the total time \(T\) is approximated as the sum of synchronization intervals between successive stages. This quantity represents the waiting time between admissible docking opportunities rather than the exact physical time of the payload trajectory. As a result, the estimated transfer duration should be interpreted as an approximate characteristic timescale of the staged transport process. Despite this approximation, the computed times remain sufficiently short to enable payload delivery across the entire tether chain. From a practical engineering perspective, the approximate nature of the timing estimate does not appear to represent a fundamental limitation of the proposed architecture.

Another aspect worth noting is the dependence of the solution families on the
maximum orbital radius \(R_{\max}\).
The simulations indicate that feasible transport chains exist over a wide range
of \(R_{\max}\) values. This suggests that the staged tether architecture does
not require extremely large outer orbits in order to sustain sequential
momentum exchange. Instead, dynamically admissible transport families appear
to persist even within relatively compact orbital domains.

The present construction of the resonant tether chain is initiated from the inner orbital boundary corresponding to the lowest operational tether altitude. From this starting point the sequence of admissible elliptical nodes and receiving tethers is generated iteratively using the synchronization and feasibility constraints described in the previous sections. Consequently, the resulting families of solutions are primarily determined by the dynamical conditions imposed at the inner boundary of the system. In the current formulation the outer boundary condition is not imposed explicitly; rather, the search procedure is allowed to propagate outward until the target orbital region is reached. As a result, the terminal tether configuration obtained in the simulations should be interpreted as one feasible realization of the architecture rather than a uniquely determined optimal endpoint. A more complete formulation could incorporate simultaneous constraints at both the inner and outer orbital boundaries and solve the construction problem as a two-sided boundary value problem. Investigating such bidirectional construction strategies may reveal additional families of solutions and constitutes an interesting direction for future work.

\section{Conclusion}

This work developed a mathematical and computational
framework for the synthesis of multi-stage orbital
transportation networks based on rotating tethers
and intermediate Keplerian transfer trajectories.

Building upon established momentum-exchange tether
transportation concepts
\cite{Sidorov2002,Sidorov2004,Ivanov2010},
the present study focused on the computational
construction and evaluation of candidate
transportation configurations under prescribed
mechanical, kinematic, and frequency-based
constraints.

For individual rotating tethers, a distributed-mass
dynamical model was employed to characterize
mechanically admissible regions of the parameter
space in terms of orbital radius, tether length,
and relative endpoint velocity.

Numerical analysis of the rotational dynamics
revealed an approximately separable residual
trajectory structure within the investigated
parameter domain. This property enabled the
construction of a reduced-order surrogate model
for estimating rotational trajectories without
repeated numerical integration of the full
dynamical equations.

For neighboring tether configurations, the
position and velocity matching conditions were
formulated as an inverse kinematic problem.
Under the adopted constant-angular-velocity
approximation, this construction reduced to
a quadratic equation in the reciprocal tether
rotation rate, providing an analytically
tractable candidate generation mechanism.

The inverse construction was subsequently
combined with mechanical feasibility constraints
and frequency-based recurrence criteria within
a mixed discrete-continuous network synthesis
procedure employing stochastic candidate
generation, beam search, and receding-horizon
evaluation.

The numerical experiments demonstrated the
construction of candidate multi-stage
transportation chains extending from low Earth
orbit towards geostationary orbit within the
adopted idealized model.

The primary contribution of the Space-Clock
Elevator framework lies in the integration of
these mathematical and computational components
rather than in the introduction of a new
physical mechanism for orbital transportation.

The present study remains limited by its
assumptions of tether inextensibility, nominal
orbital trajectories, simplified mass exchange,
and finite synchronization tolerances.

The resulting configurations do not constitute
a complete validation of repeated physical
rendezvous or sustained operational transportation.
Further investigations are required to address
trajectory refinement, active control, elastic
tether dynamics, long-term stability, and the
cumulative effects of repeated mass exchange.

Future work may also extend the computational
framework towards bidirectional network synthesis,
simultaneous inner and outer orbital boundary
constraints, and more comprehensive system-level
optimization.

The code of the project can be found at
\url{https://github.com/maksimkazanskii/spaceclock}.

\section*{Use of AI tools}

The large language model ChatGPT (version 5.2, OpenAI) was used solely for language editing and text polishing during the preparation of this manuscript. The model was not used to generate scientific ideas, perform analysis,  develop the methodology, or interpret results. All scientific content, algorithms, derivations, and conclusions were developed and verified entirely by the authors.

\clearpage
\bibliographystyle{IEEEtran}
\bibliography{bibliography}
\clearpage
\appendix
\renewcommand{\theequation}{\arabic{equation}}

\section{Regression Model for Residual Amplitude and Rotation Period}
\label{sec:regression_model}
\subsection{Problem formulation}
This appendix describes the regression model used to approximate the residual amplitude and the rotation period of the tether dynamics across the feasible parameter space.

The regression model is trained using the simulation dataset generated in the parameter sweep described in Section~\ref{sec:feasibility_map}. Each simulation produces a parameter triple $(R,\ell,v_{\mathrm{rel}})$ together with the corresponding residual amplitude $A$ and rotation period $T_{\mathrm{rot}}$ extracted from the angular trajectory analysis.

The regression model approximates the mapping

\begin{equation}
(R,\ell,v_{\mathrm{rel}}) \rightarrow (A,T_{\mathrm{rot}})
\end{equation}

allowing rapid estimation of the rotational characteristics of the tether without performing a full dynamical simulation.

\subsection{Dimensionless Feature Representation}

To improve generalization across different orbital radii, the input parameters are expressed in dimensionless form using characteristic orbital scales. Let $\mu$ denote the gravitational parameter of the central body.

Two dimensionless variables are defined as

\begin{equation}
x = \frac{\ell}{R}, \qquad
u = \frac{v_{\mathrm{rel}}}{\sqrt{\mu/R}}
\end{equation}

where $\sqrt{\mu/R}$ is the characteristic orbital velocity at radius $R$. These variables form the input feature vector

\begin{equation}
\mathbf{x} = (x,u)
\end{equation}

The rotation period is normalized using the characteristic orbital timescale

\begin{equation}
T_{\mathrm{scale}} = \sqrt{\frac{R^{3}}{\mu}}
\end{equation}

Instead of predicting the rotation period directly, the regression model predicts the dimensionless quantity

\begin{equation}
\tau = \frac{T_{\mathrm{rot}}}{T_{\mathrm{scale}}}
\end{equation}

This normalization improves numerical stability and allows the regression model to capture the physical scaling of the system.

\subsection{Target Transformation}

The residual amplitude spans several orders of magnitude and may approach very small values in some regions of parameter space. To stabilize the regression, the amplitude is predicted in logarithmic form

\begin{equation}
y_A = \log(A)
\end{equation}

A small numerical floor is applied to avoid singular values in the logarithmic transformation.

\subsection{Local Polynomial Regression}

The regression model uses a local polynomial regression approach. Instead of fitting a single global polynomial across the entire parameter space, predictions are obtained by fitting a polynomial locally to nearby simulation samples.

For a query point $\mathbf{x}_0$, the $k$ nearest neighbors are identified in the normalized feature space using the Euclidean distance. A polynomial feature expansion of degree $d$ is then applied to the neighboring samples,

\begin{equation}
\Phi_d(\mathbf{x})
\end{equation}

which generates all monomials of the input variables up to order $d$.

A weighted least-squares problem is solved using these local samples,

\begin{equation}
\hat{y}(\mathbf{x}_0)
=
\arg\min_{\beta}
\sum_{i=1}^{k}
w_i
\left(
y_i - \beta^\top \Phi_d(\mathbf{x}_i)
\right)^2
\end{equation}

where the weights are inversely proportional to the distance from the query point,

\begin{equation}
w_i = \frac{1}{d_i + \varepsilon}
\end{equation}

with $d_i$ denoting the distance to the $i$-th neighbor and $\varepsilon$ a small constant ensuring numerical stability.

Separate regressions are performed for the logarithmic amplitude and the normalized rotation period.

\subsection{Training and Validation Procedure}

The model is trained using the dataset generated on the regular simulation grid within the feasible region defined in Section~\ref{subsec:tether-tension}. The training dataset contains 4718 simulations, each providing the parameters $(R,\ell,v_{\mathrm{rel}})$ together with the corresponding residual amplitude $A$ and rotation period $T_{\mathrm{rot}}$.

To evaluate the generalization performance of the regression model, an independent dataset of randomly sampled parameter combinations within the feasible region is generated. This dataset contains 241 samples drawn from the continuous parameter space rather than from the original simulation grid. It is split into a validation dataset of 120 samples and a test dataset of 121 samples.

A grid search is performed over the following hyperparameters:

\begin{equation}
d \in \{1,2,3,4\}, \qquad
k \in \{50,100,150,200,250,300\}
\end{equation}

where $d$ is the degree of the polynomial feature expansion used in the local regression and $k$ is the number of nearest neighbors used to construct the local training neighborhood. For each combination $(d,k)$, predictions are computed on the validation dataset and evaluated using a composite error metric combining the scaled amplitude error and the relative error of the rotation period. The hyperparameter pair yielding the lowest validation score is selected as the final model configuration. The validation score is defined as the sum of the mean scaled amplitude error and the mean relative error of the rotation period. Specifically,

\begin{equation}
\mathrm{Score} =
\frac{1}{N}\sum_{i=1}^{N}
\frac{|A_i-\hat{A}_i|}{A_i + A_{\mathrm{scale}}}
+
\frac{1}{N}\sum_{i=1}^{N}
\frac{|T_i-\hat{T}_i|}{T_i}
\end{equation}
\begin{table}[h]
\centering
\caption{Validation scores for the local polynomial regression model across the full hyperparameter grid. Lower values indicate better performance.}
\label{tab:regression_search}
\begin{tabular}{ccc}
\hline
Polynomial Degree & Neighbors ($k$) & Validation Score \\
\hline
1 & 50  & $1.1627\times10^{-2}$ \\
1 & 100 & $1.9569\times10^{-2}$ \\
1 & 150 & $2.8955\times10^{-2}$ \\
1 & 200 & $3.6167\times10^{-2}$ \\
1 & 250 & $4.2084\times10^{-2}$ \\
1 & 300 & $4.8865\times10^{-2}$ \\

2 & 50  & $1.8320\times10^{-3}$ \\
2 & 100 & $3.3408\times10^{-3}$ \\
2 & 150 & $5.8330\times10^{-3}$ \\
2 & 200 & $7.2468\times10^{-3}$ \\
2 & 250 & $7.4638\times10^{-3}$ \\
2 & 300 & $9.9628\times10^{-3}$ \\

3 & 50  & $7.6396\times10^{-4}$ \\
3 & 100 & $1.3676\times10^{-3}$ \\
3 & 150 & $2.8900\times10^{-3}$ \\
3 & 200 & $4.2441\times10^{-3}$ \\
3 & 250 & $6.2644\times10^{-3}$ \\
3 & 300 & $8.7059\times10^{-3}$ \\

4 & 50  & $1.3131\times10^{-3}$ \\
4 & 100 & $\mathbf{6.1288\times10^{-4}}$ \\
4 & 150 & $1.4639\times10^{-3}$ \\
4 & 200 & $5.0174\times10^{-3}$ \\
4 & 250 & $3.9308\times10^{-3}$ \\
4 & 300 & $7.2826\times10^{-3}$ \\
\hline
\end{tabular}
\end{table}
where $A_i$ and $T_i$ denote the simulated values, $\hat{A}_i$ and $\hat{T}_i$ are the corresponding model predictions, and $A_{\mathrm{scale}}$ is a small stabilization constant defined as the first percentile of the amplitude values in the validation dataset. This normalization prevents numerical instability when the residual amplitude approaches zero. The resulting model is then evaluated on the independent test dataset to estimate predictive accuracy on previously unseen parameter combinations. The validation results are summarized in Table~\ref{tab:regression_search}. The best performance is obtained for a polynomial degree $d=4$ with $k=100$ nearest neighbors.

\subsection{Test Dataset Evaluation}

The selected model is evaluated on an independent test dataset consisting of randomly sampled parameter combinations that are not part of the original simulation grid. The prediction accuracy is summarized in Table~\ref{tab:regression_test}.

\begin{table}[h]
\centering
\caption{Prediction accuracy of the selected regression model on the independent test dataset.}
\label{tab:regression_test}
\begin{tabular}{lcccc}
\hline
Target & MAE & Mean Relative Error & Median Relative Error & $R^2$ \\
\hline
Residual amplitude $A$ & $3.46\times10^{-6}$ & $1.14\times10^{-2}$ & $1.12\times10^{-8}$ & 0.999895 \\
Rotation period $T_{\mathrm{rot}}$ & $0.986$ s & $1.67\times10^{-4}$ & $2.36\times10^{-9}$ & 0.999993 \\
\hline
\end{tabular}
\end{table}
\subsection{Angular Trajectory Reconstruction}

The regression model predicts only the residual amplitude $A$ and the rotation period $T_{\mathrm{rot}}$. The full angular trajectory is reconstructed using a normalized residual template derived from the simulation dataset.

For each simulation, the tether orientation angle $\phi(t)$ is defined with respect to the local radial direction of the orbit. The trajectory can therefore be decomposed into a uniform rotation component in the orbital frame and a residual oscillatory term,
\begin{equation}
\phi(t) = \phi_{\mathrm{lin}}(t) + \phi_{\mathrm{res}}(t),
\end{equation}
where
\begin{equation}
\phi_{\mathrm{lin}}(t) = \phi(t_0) + \omega_{\mathrm{rot}} (t - t_0),
\qquad
\omega_{\mathrm{rot}} = \frac{2\pi}{T_{\mathrm{rot}}},
\end{equation}
represents the mean rotation of the tether relative to the orbital frame, and $\phi_{\mathrm{res}}(t)$ describes the deviation from uniform rotation caused by gravitational torques and dynamical coupling with the orbit.

The residual trajectories from the training dataset are normalized by their amplitudes,
\begin{equation}
\tilde{\phi}_{\mathrm{res}}^{(i)}(t)
=
\frac{\phi_{\mathrm{res}}^{(i)}(t)}{A^{(i)}},
\end{equation}
and averaged to construct a mean residual template
\begin{equation}
\phi_{\mathrm{template}}(\tau)
=
\frac{1}{N}
\sum_{i=1}^{N}
\tilde{\phi}_{\mathrm{res}}^{(i)}(\tau),
\qquad
\tau = \frac{t - t_0}{T_{\mathrm{rot}}}.
\end{equation}

For a new parameter set $(R,\ell,v_{\mathrm{rel}})$, the regression model predicts $\hat{A}$ and $\hat{T}_{\mathrm{rot}}$. The residual signal is reconstructed as
\begin{equation}
\hat{\phi}_{\mathrm{res}}(t)
=
\hat{A}\,
\phi_{\mathrm{template}}
\left(
\frac{t - t_0}{\hat{T}_{\mathrm{rot}}}
\right),
\end{equation}
and the full angular trajectory in the orbital frame is obtained as
\begin{equation}
\hat{\phi}(t)
=
\phi(t_0) +
\hat{\omega}_{\mathrm{rot}} (t - t_0) +
\hat{\phi}_{\mathrm{res}}(t),
\qquad
\hat{\omega}_{\mathrm{rot}} = \frac{2\pi}{\hat{T}_{\mathrm{rot}}}.
\end{equation}

If the inertial orientation of the tether is required, the orbital rotation must also be included through the orbital angle $\theta(t)$,
\begin{equation}
\hat{\phi}_{\mathrm{inert}}(t) = \theta(t) + \hat{\phi}(t).
\end{equation}

The reconstruction accuracy is evaluated on the independent random test dataset described above. The resulting errors are summarized in Table~\ref{tab:phi_reconstruction}.

\begin{table}[h]
\centering
\caption{Prediction accuracy of the reconstructed angular trajectory and average evaluation time of the surrogate model.}
\label{tab:phi_reconstruction}
\begin{tabular}{lccc}
\hline
Signal & Mean Relative Error & Mean Max Absolute Error & Avg. prediction Time [s] \\
\hline
Residual angle $\phi_{\mathrm{res}}(t)$ & $3.84\times10^{-5}$ & $1.72\times10^{-5}$ & $6.69\times10^{-4}$ \\
Full angle $\phi(t)$ & $6.57\times10^{-6}$ & $1.94\times10^{-5}$ & $\sim 10^{-8} $ (overhead) \\
\hline
\end{tabular}
\end{table}

\begin{figure}[h]
\centering

\begin{subfigure}{0.48\textwidth}
\centering
\includegraphics[width=\linewidth]{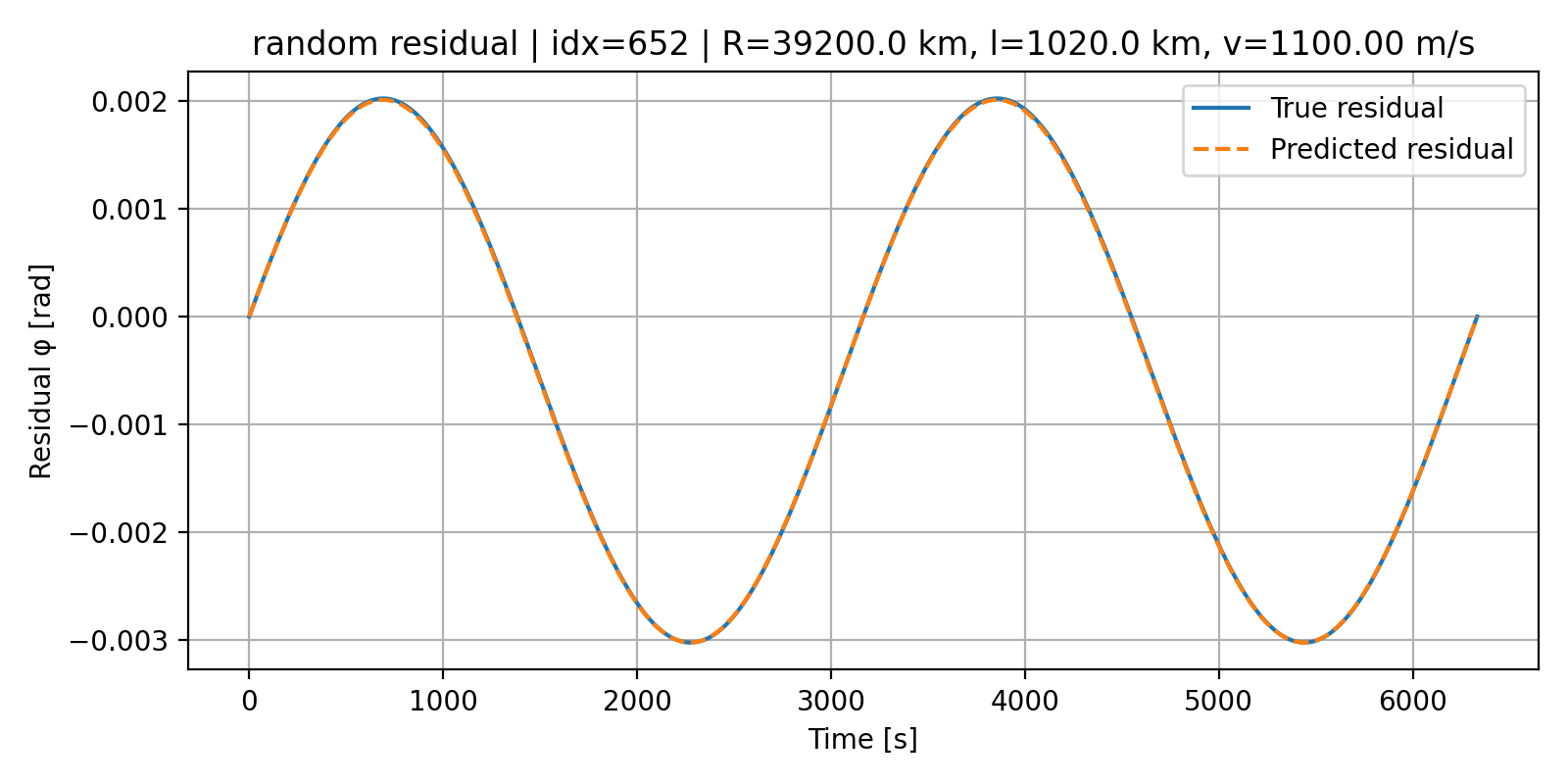}
\caption{Predicted residual signal compared with numerical simulation.}
\label{fig:phi_residual_prediction}
\end{subfigure}
\hfill
\begin{subfigure}{0.48\textwidth}
\centering
\includegraphics[width=\linewidth]{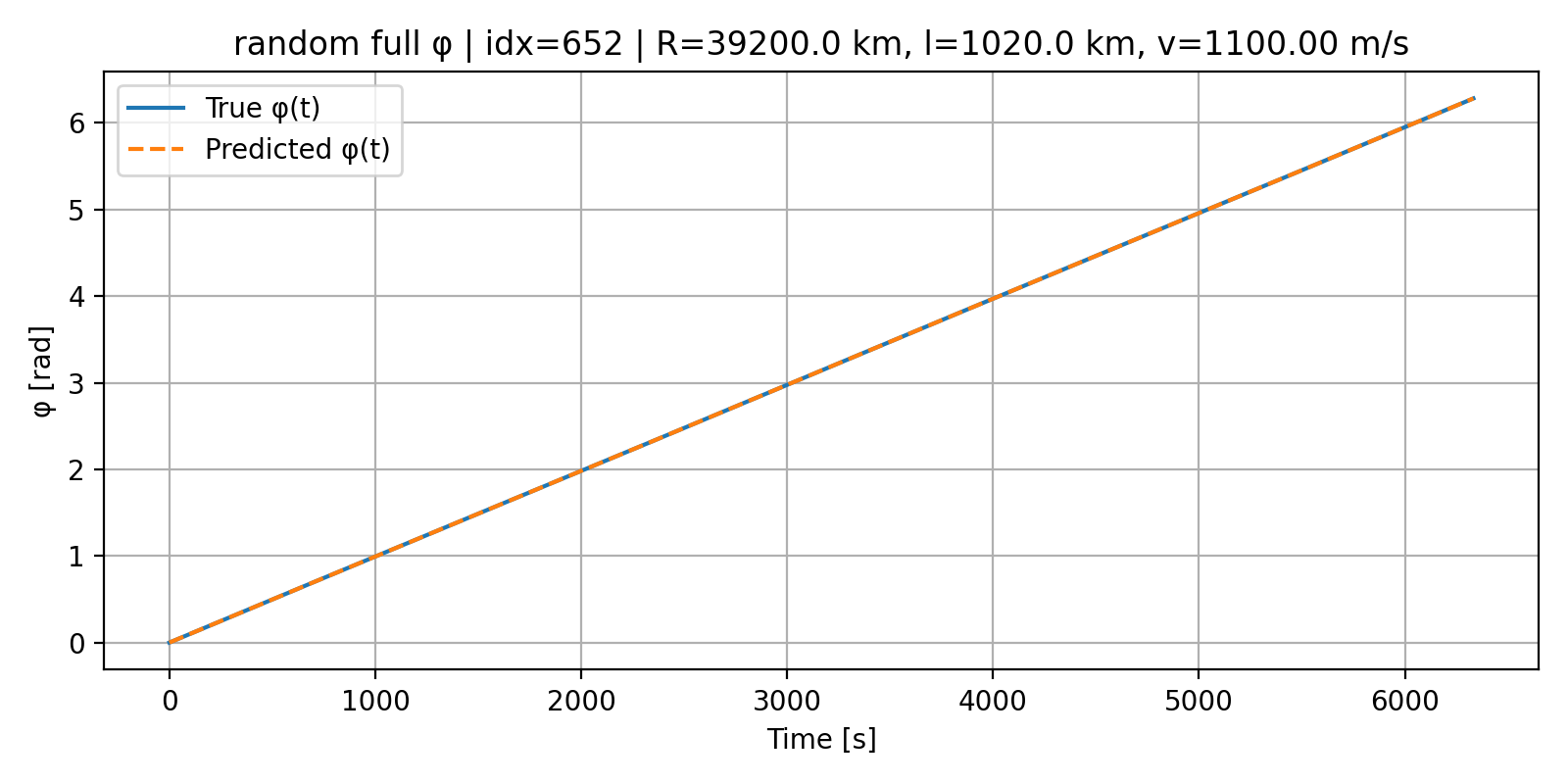}
\caption{Reconstructed angular trajectory $\phi(t)$ compared with numerical}
\label{fig:phi_full_prediction}
\end{subfigure}
\

\caption{Comparison between surrogate model predictions and numerical simulations.}
\label{fig:phi_prediction}
\end{figure}

Examples of the predicted residual signal and the reconstructed angular trajectory compared with the numerical simulations are shown in Figures~\ref{fig:phi_residual_prediction}-\ref{fig:phi_full_prediction}.

\subsection*{B. Multi-tether synchronization algorithm}

Algorithm 1 generates the initial database of feasible tether configurations at the base orbital altitude. The algorithm scans the admissible tether length range and azimuthal phase angles to determine all physically feasible elliptical states. For each configuration, the surrogate tether dynamics model is used to estimate the instantaneous tether spin and corresponding endpoint velocity. Only configurations satisfying the structural and dynamical feasibility constraints are retained. 

\begin{algorithm}[H]
\caption{Initial tether grid construction (Stage 0)}
\footnotesize
\begin{algorithmic}[1]

\Require
$\mu$, $R_\oplus$;
$R_{\mathrm{safe}}=R_\oplus+\mathrm{ALT\_OFFSET}$;
$r\in[r_{\min},r_{\max}]$ with step $\Delta r$;
$N_\phi$ azimuth samples;
feasibility function $\mathcal{F}(R_{\mathrm{cm}},r,v_{\mathrm{rel}})$.

\Ensure
Initial feasible tether set $\mathcal{T}_0$

\State $\mathcal{T}_0 \gets \emptyset$

\ForAll{$r \in [r_{\min},r_{\max}]$ and $\phi_j = 2\pi j/N_\phi$}
    \State $R_{\mathrm{cm}} \gets R_{\mathrm{safe}} + r$
    \State $\Omega \gets \sqrt{\mu/R_{\mathrm{cm}}^3}$
    \State compute corrected tether spin $w(\phi_j)$ from the surrogate motion
    \State $v_{\mathrm{rel}} \gets |w(\phi_j)|\,r$

    \If{$\mathcal{F}(R_{\mathrm{cm}},r,v_{\mathrm{rel}})$}
        \State store $(r,\phi_j,R_{\mathrm{cm}},\Omega,w(\phi_j))$ in $\mathcal{T}_0$
    \EndIf
\EndFor

\end{algorithmic}

\end{algorithm}

Algorithm 2 constructs candidate ballistic transfer trajectories originating from feasible tether releases and identifies those compatible with resonant synchronization conditions. For each tether configuration, the surrogate model reconstructs the release state and the resulting Keplerian orbit. Trajectories violating safety constraints, such as perigee below the safe altitude or excessive apoapsis, are discarded. The remaining trajectories are analyzed for orbital–tether resonance by comparing the ratio between the orbital mean motion and the inertial tether rotation rate to nearby rational ratios. For each resonance pair $(p,q)$ only the best candidate node is retained (in terms of time propagation), forming the resonant node set $\mathcal{C}_k$.
\begin{algorithm}[H]
\caption{Elliptical node construction and resonance detection}
\footnotesize
\begin{algorithmic}[1]

\Require
Tether set $\mathcal{T}_k$,
ratio map $\{(p,q)\}$,
$\Delta s_{\mathrm{tol}}$

\Ensure
Resonant node set $\mathcal{C}_k$

\State $\mathcal{N}_k\gets\emptyset$

\ForAll{$(r,\phi,R_{\mathrm{cm}},\Omega,w)$ in $\mathcal{T}_k$}

\State Reconstruct the surrogate tether trajectory and compute the release state $(x,y,v_x,v_y)$ at phase $\phi$

\State Compute Kepler orbit $(a,e,r_p,r_a,n,T)$ from the release state

\If{$r_p < R_{\mathrm{safe}}$}
    \State \textbf{continue}
\EndIf

\If{$\mathrm{cap\_apoapsis}$ and $r_a > R_{\mathrm{target}}$}
    \State \textbf{continue}
\EndIf

\State Store node $(x_0,y_0,v_{x0},v_{y0},n)$

\EndFor


\State $\mathcal{C}_k\gets\emptyset$

\ForAll{node candidates}

\State $w_{\mathrm{inert}}\gets \Omega + w$

\State $x\gets n/w_{\mathrm{inert}}$

\State $(p,q)\gets\mathrm{ClosestPQ}(x)$

\State $\Delta s=r_0\,2\pi q \left|x-\frac{p}{q}\right|$

\If{$\Delta s>\Delta s_{\mathrm{tol}}$}
    \State \textbf{continue}
\EndIf

\State Store candidate resonant node for pair $(p,q)$

\EndFor

\State For each rational pair $(p,q)$ keep only the node
with the smallest ratio error $\left|x-\frac{p}{q}\right|$

\end{algorithmic}

\end{algorithm}

Algorithm 3 performs the stage-to-stage propagation of the multi-tether transfer chain. Each resonant node is numerically propagated along its Keplerian trajectory, and candidate capture points are evaluated. At each sampled trajectory point an inverse capture solver determines whether a new tether configuration can reproduce the local release state while satisfying feasibility constraints. Additional synchronization checks enforce resonance between the new tether rotation and the orbital motion. Valid solutions generate the next tether set $\mathcal{T}_{k+1}$, which becomes the starting point for the subsequent stage of the outward transfer.
\begin{algorithm}[H]
\caption{Resonant tether chaining with inverse capture}
\footnotesize
\begin{algorithmic}[1]

\Require
Resonant node set $\mathcal{C}_k$

\Ensure
Next-stage tether set $\mathcal{T}_{k+1}$

\State $\mathcal{T}_{k+1}\gets\emptyset$

\ForAll{resonant nodes}

\State Numerically propagate the orbit for one period using
adaptive time steps and a Verlet integrator

\ForAll{sampled trajectory points $(x,y,v_x,v_y)$}

\State Solve inverse capture to find $(r',R'_{\mathrm{cm}},w')$
such that the tether release state matches $(x,y,v_x,v_y)$

\If{solver fails}
    \State \textbf{continue}
\EndIf

\If{not $\mathcal{F}(R'_{\mathrm{cm}},r',|w'r'|)$}
    \State \textbf{continue}
\EndIf

\State $\Delta R=(R'_{\mathrm{cm}}-r')-(R_{\mathrm{cm}}-r)$

\If{$\Delta R\le0$}
    \State \textbf{continue}
\EndIf

\State $\Omega' \gets \sqrt{\mu/(R'_{\mathrm{cm}})^3}$

\State $w'_{\mathrm{inert}} \gets \Omega' + w'$

\State $y_{\mathrm{rat}} \gets n_{\mathrm{ell}} / w'_{\mathrm{inert}}$

\State $(l,m)\gets\mathrm{ClosestPQ}(y_{\mathrm{rat}})$

\State $\Delta s = r_{\mathrm{attach}}\,2\pi m
\left| y_{\mathrm{rat}} - \frac{l}{m} \right|$

\If{$\Delta s>\Delta s_{\mathrm{tol}}$}
    \State \textbf{continue}
\EndIf

\For{$\phi_j=2\pi j/N_\phi$}
    \State store $(r',\phi_j,R'_{\mathrm{cm}},\Omega',w')$ in $\mathcal{T}_{k+1}$
\EndFor

\EndFor
\EndFor

\State Merge solutions with similar $(R'_{\mathrm{cm}}, r')$ to remove duplicates

\If{$\mathcal{T}_{k+1}=\emptyset$}
    \State \Return failure
\EndIf

\end{algorithmic}
\end{algorithm}

\subsection*{C. Numerical Architecture of the Constructed Tether Chain}
\label{chain}

The outward propagation of payloads in the Space-Clock Elevator architecture proceeds through a sequence of synchronized dynamical interactions between rotating tethers and elliptical transfer nodes. 
Each stage of the chain begins when a payload reaches the endpoint of a rotating tether and is released onto an elliptical trajectory whose orbital frequency is rationally synchronized with the inertial angular velocity of the next tether. 
The payload then follows this elliptical orbit until the synchronization condition produces a resonant encounter with the subsequent tether endpoint. 
At this point the payload is captured by the rotating tether and transferred to the next orbital stage.

Thus, the transport process follows the repeating sequence

\[
\text{tether} \;\rightarrow\; \text{elliptical node} \;\rightarrow\; \text{tether} \;\rightarrow\; ... \;\rightarrow\; \text{elliptical node}  \;\rightarrow\; \text{tether}
\]

Each elliptical node therefore acts as a dynamical bridge between two adjacent tethers, enabling stepwise outward motion without continuous propulsion. 

Tables~\ref{tab:tether_chain_a} and~\ref{tab:node_chain_b} summarize the parameters of the rotating tether chain and the corresponding synchronization nodes, respectively. 
The configuration is constructed under an increment limit of 5000 km between successive stages, and together these tables define the full dynamical setup used to generate the propagation and timing results discussed in this work.

\begin{table}[h]
\centering
\scriptsize
\caption{ Parameters of the rotating tethers forming the synchronized transfer chain (increment limit $\Delta R <  5000km$.).}
\label{tab:tether_chain_a}
\begin{tabular}{cccccccccc}
\hline
stage & $R_{cm}$ (km) & \shortstack{tether \\ radius (km)} & \shortstack{endpoint \\ inward (km)} & \shortstack{endpoint \\ outward (km)} & $\Omega$ (rad/s) & $w$ (rad/s) & $\omega_{\text{inert}}$ (rad/s) & $v_{\text{cm}}$ (m/s) & $v_{\text{earth}}$ (m/s) \\
\hline
0  & 7411.4 & 40.4  & 7371.0 & 7451.8 & 0.000989 & 0.019016 & 0.020005 & 768.25 & 8141.74 \\
1  & 11591.2 & 161.4 & 11429.8 & 11752.6 & 0.000506 & 0.004768 & 0.005274 & 769.64 & 6715.35 \\
2  & 12121.8 & 90.1  & 12031.7 & 12211.9 & 0.000473 & 0.005734 & 0.006207 & 516.86 & 6293.78 \\
3  & 16196.2 & 250.5 & 15945.7 & 16446.7 & 0.000306 & 0.003026 & 0.003333 & 758.00 & 5795.57 \\
4  & 19872.4 & 233.8 & 19638.6 & 20106.2 & 0.000225 & 0.002489 & 0.002714 & 581.93 & 5113.18 \\
5  & 21531.8 & 429.6 & 21102.2 & 21961.4 & 0.000200 & 0.001698 & 0.001898 & 729.57 & 5117.94 \\
6  & 23918.2 & 152.4 & 23765.8 & 24070.6 & 0.000171 & 0.001878 & 0.002048 & 286.07 & 4394.31 \\
7  & 28635.0 & 176.9 & 28458.1 & 28811.9 & 0.000130 & 0.001525 & 0.001655 & 269.69 & 4023.63 \\
8  & 33448.6 & 256.5 & 33192.1 & 33705.1 & 0.000103 & 0.001285 & 0.001389 & 329.68 & 3808.18 \\
9  & 37263.0 & 109.1 & 37153.9 & 37372.1 & 0.000088 & 0.001445 & 0.001533 & 157.67 & 3437.82 \\
10 & 38176.0 & 170.0 & 38006.0 & 38346.0 & 0.000081 & 0.001050 & 0.001131 & 178.77 & 3424.72 \\
11 & 41932.6 & 299.6 & 41633.0 & 42232.2 & 0.000073 & 0.001082 & 0.001155 & 324.03 & 3440.52 \\
\hline
\end{tabular}
\end{table}

\begin{table}[h]
\centering
\scriptsize
\caption{ Synchronization parameters of elliptical transfer nodes connecting successive tethers (increment limit $\Delta R <  5000km$.}
\label{tab:node_chain_b}
\begin{tabular}{cccccccc}
\hline
stage & radius (km) & $v_{prop}$ (m/s) & tether radius (km) & $p$ & $q$ & ratio error & $T_{sync}$ (s) [h] \\
\hline
0  & 7411.4 & 0.0000  & 40.4  & -- & --  & -- & 0 \,[0.00] \\
1  & 11591.2 & 39.1612 & 161.4 & 11 & 87  & 2.0e-06 & 103643.16 \,[28.79] \\
2  & 12121.8 & 5.6094  & 90.1  & 9  & 106 & 6.3e-05 & 210940.01 \,[58.59] \\
3  & 16196.2 & 71.5880 & 250.5 & 3  & 29  & 2.7e-05 & 265614.93 \,[73.78] \\
4  & 19872.4 & 30.0957 & 233.8 & 5  & 53  & 1.2e-04 & 388317.77 \,[107.87] \\
5  & 21531.8 & 7.0177  & 429.6 & 5  & 63  & 7.5e-05 & 596878.33 \,[165.80] \\
6  & 23918.2 & 12.7703 & 152.4 & 7  & 68  & 1.1e-05 & 805461.69 \,[223.74] \\
7  & 28635.0 & 36.3568 & 176.9 & 3  & 34  & 1.6e-04 & 934524.18 \,[259.59] \\
8  & 33448.6 & 11.8885 & 256.5 & 7  & 88  & 7.6e-05 & 1332720.00 \,[370.20] \\
9  & 37263.0 & 8.9498  & 109.1 & 7  & 108 & 1.1e-04 & 1775388.85 \,[493.16] \\
10 & 38176.0 & 2.1760  & 170.0 & 5  & 69  & 7.2e-05 & 2165158.11 \,[601.43] \\
11 & 41932.6 & 12.8051 & 299.6 & 3  & 34  & 1.1e-04 & 2436638.65 \,[676.84] \\
\hline
\end{tabular}
\end{table}

Tables~\ref{tab:tether_chain_unlim_a} and~\ref{tab:node_chain_unlim_b} summarize the parameters of the rotating tether chain and the corresponding synchronization nodes for the case without an increment limit. 
In this configuration, the spacing between successive stages is not constrained, and together these tables define the full dynamical setup used to generate the propagation and timing results.

\begin{table}[h]
\centering
\scriptsize
\caption{ Parameters of the rotating tethers forming the synchronized transfer chain (no increment limit).}
\label{tab:tether_chain_unlim_a}
\begin{tabular}{cccccccccc}
\hline
stage & $R_{cm}$ (km) & \shortstack{tether \\ radius (km)} & \shortstack{endpoint \\ inward (km)} & \shortstack{endpoint \\ outward (km)} & $\Omega$ (rad/s) & $w$ (rad/s) & $\omega_{\text{inert}}$ (rad/s) & $v_{\text{cm}}$ (m/s) & $v_{\text{earth}}$ (m/s) \\
\hline
0 & 7439.2  & 68.2  & 7371.0 & 7507.4 & 0.000984 & 0.019014 & 0.019998 & 1296.77 & 8683.68 \\
1 & 16731.8 & 458.6 & 16273.2 & 17190.4 & 0.000292 & 0.002553 & 0.002845 & 1170.61 & 6185.17 \\
2 & 29178.2 & 407.7 & 28770.5 & 29585.9 & 0.000127 & 0.001175 & 0.001302 & 479.27  & 4226.93 \\
3 & 41953.4 & 493.8 & 41459.6 & 42447.2 & 0.000073 & 0.001446 & 0.001519 & 714.01  & 3832.62 \\
\hline
\end{tabular}
\end{table}
\begin{table}[H]
\centering
\scriptsize
\caption{ Synchronization parameters of elliptical transfer nodes connecting successive tethers (no increment limit).}
\label{tab:node_chain_unlim_b}
\begin{tabular}{cccccccc}
\hline
stage & radius (km) & $v_{prop}$ (m/s) & tether radius (km) & $p$ & $q$ & ratio error & $T_{sync}$ (s) [h] \\
\hline
0 & 7439.2  & 0.0000  & 1244.5 & -- & -- & -- & 0 \,[0.00] \\
1 & 16731.8 & 161.2090 & 977.0  & 4  & 25 & 2.17e-04 & 55221.76 \,[15.34] \\
2 & 29178.2 & 47.0886  & 527.3  & 8  & 55 & 1.83e-04 & 265398.11 \,[73.72] \\
3 & 41953.4 & 54.7903  & 881.6  & 3  & 56 & 6.0e-06 & 231593.68 \,[64.33] \\
\hline
\end{tabular}
\end{table}
\end{document}